\documentclass[a4paper, 12pt]{article}
\usepackage{graphicx} 
\usepackage[left=2cm, right=2cm, top=2cm, bottom=2cm]{geometry}
\usepackage{amsfonts}
\usepackage{amsmath}
\usepackage{mathrsfs} 
\usepackage{color}
\usepackage[numbers,sort&compress]{natbib}
\usepackage{hyperref}
\usepackage{authblk}
\usepackage[per-mode=symbol]{siunitx}
\usepackage{relsize} 
\usepackage[percent]{overpic}
\usepackage{multirow}
\usepackage{pgfplots}
\usepackage{tikz}
\usetikzlibrary{shapes,arrows}
\usepackage{caption} 
\hypersetup{hyperindex, linkcolor=blue, colorlinks =true, citecolor=blue, urlcolor=magenta} 

\definecolor{darkgreen}{RGB}{0, 100, 0}

\newcommand{\II}{\mathbf{I}}
\newcommand{\uu}{\mathbf{u}}
\newcommand{\F}{\mathbf{F}}
\newcommand{\qq}{\mathbf{q}}
\newcommand{\XX}{\mathbf{X}}
\newcommand{\LLambda}{\mathbf{\Lambda}}
\newcommand{\ddelta}{\boldsymbol{\delta}}
\newcommand{\dLambda}{\delta \mkern-1mu \Lambda}
\newcommand{\ddLambda}{\boldsymbol{\delta}\mkern-2mu\mathbf{\Lambda}}
\newcommand{\TT}{\mathbf{T}}
\newcommand{\RR}{\mathcal{R}}

\newcommand{\Norm}[1]{\left\lVert#1\right\rVert}
\newcommand{\Det}[1]{\left| #1 \right|}
\newcommand{\Dev}[1]{\text{dev}\!\left(#1\right)}
\newcommand{\abs}[1]{\left| #1 \right|}

\newcommand{\detF}{J}

\newcommand{\added}[1]{#1}

\makeatletter
\DeclareRobustCommand{\svdots}
{\vbox{\baselineskip3\p@ \lineskiplimit\z@
       \hbox{.}\hbox{.}\hbox{.}}}
\makeatother

\title{Topology Optimization of Pneumatic Soft Actuators\\
       Based on Porohyperelasticity}
{}
\author[ ]{Sumit Mehta}
\author[*]{Konstantinos Poulios}

\affil[ ]{Department of Civil and Mechanical Engineering \protect\\
Technical University of Denmark, Denmark}

\date{}
\begin{document}

\maketitle

\vspace{-13mm}
\begin{abstract}
This paper introduces a new nonlinear topology optimization framework which employs porohyperelasticity for providing computational design of pneumatic soft actuators.
Density-based topology optimization is used with the objective of maximizing the bending response in a soft actuator made of an elastomer, for given actuation pressure and external resistance.
Pressurization of interconnected cavities is modeled via an extension of the Darcy flow theory that is valid for large deformations.
Essential for the good performance of the framework is a carefully chosen interpolation scheme for the permeability between void and solid regions, as well as a suitable definition of a drainage term in the solid regions.
Results are shown for a variety of actuation pressures and maximum allowable strain energy density levels, covering a wide range of system responses from small to rather large deformations.
\end{abstract}

\let\thefootnote\relax\footnote{\hskip -0.5cm\scriptsize $^{\rm *}$Corresponding author: kopo@dtu.dk}

\section{Introduction}

Soft robots, typically made of elastomers, are fundamentally different than rigid link robots, offering distinct advantages compared to the latter.
Their kinematic response upon actuation may not be as precisely controlled as in rigid robots, but their inherent adaptability to a less well-defined environment results in a more forgiving interaction with their surroundings.
They are typically made of gels and elastomers \citep{kim2013soft}, which in general have a compliance comparable to biological tissue, making soft robots a particularly attractive choice for interacting with the human body.

Common actuation mechanisms in soft robotics devices are pneumatic, hydraulic, piezoelectric, thermomechanical, or even purely mechanical by means of tendons/cables.
Among these options, pneumatic actuation is a very attractive one due to its simplicity, but at the same time it is inherently difficult to design and optimize for.
This difficulty stems from the fact that the provided gas pressure acts on structure walls, whose position and orientation change significantly upon activation.
The present work focuses on the design and optimization of pneumatically activated soft actuators, especially motivated by the bright outlook on possible uses for robot-human interaction, e.g., in the medical sector.

So far, due to the lack of automated approaches, the design of soft robotics actuators heavily relies on either the designer's intuition or biomimetic techniques \cite{kim2013soft, slesarenko2018strategies, mosadegh2014pneumatic}.
Inspiration from nature is certainly very useful for discovering new concepts, nevertheless, it is not sufficient for making quantitative decisions or for optimizing a certain design type.
Further parameter or shape optimization is necessary to reach a concrete high-performing design \cite{2023Ch-So-Ch-Gu-Zh}.
Topology optimization, on the other side, spans a design space which is typically large enough to contain the solutions found in nature and, at the same time, offers a systematic fully automated process.
It generates structural designs, without human intervention, by optimizing the material distribution in a given domain with respect to prescribed performance metrics, while fulfilling any imposed design constraints \cite{2013Be-Si}.
Significant efforts in using topology optimization for designing pneumatically activated soft robots do exist, but they either assume small strains~\cite{2021Ku-La}, or a fixed pressurized chamber geometry~\cite{2024Ko-Gh-Mo-Ta-Yu-Yu-Sa-Ka-Qi-No}.

Developing a general topology optimization framework for pneumatically activated soft actuators, entails certain difficulties.
The main difficulty stems from the fact that the wall surfaces of the structure, upon which the activation pressure is exerted, are a priori unknown.
An early intent to solve this problem, at least in 2D, was by extracting an explicitly defined surface from the implicit, i.e., density-based, description of the structure \cite{2000Ha-Ol,2004Du-Ol}. 
Despite more recent improvements to this approach, such as \cite{2012Le-Ma,2019Pi-Ne-Ki}, the two main drawbacks of this approach remain.
Extracting an explicit surface is relatively easy in 2D, but it becomes significantly less straightforward and less numerically robust in 3D.
Moreover, the problem of detecting the connectivity between a pressure source and potentially pressurized surfaces is still left untreated by this approach.
One approach that actually deals with the connectivity to the pressure source is the three-phase (solid/fluid/void) topology optimization model proposed in reference \citep{2007Si-Cl}, which employs a displacement-pressure mixed formulation and different interpolation functions for the bulk and shear moduli.
Apart from the usual solid and void regions, this method additionally relies on an incompressible fluid region which can transfer the applied pressure to the surrounding solid, without excessive deformation in the fluid.
Later, reference \cite{2009Br-Ci} improved on the numerical stability of the mixed finite element formulation for incompressibility.

Incompressibility, however, is not really essential for transferring pressure to a solid structure.
In fact, a compressible fluid like air can transfer pressure to a surrounding structure as effectively as a nearly incompressible fluid like water.
The essential difference between the two in a Lagrangian numerical setting is the extent of deformations observed in the fluid region.
For the case of a soft robotics actuator though, even if the actuation fluid is modeled as entirely incompressible, a quite large feedstock volume will still be necessary in order to effectuate the large deformations met in these types of applications.
This would inevitably result in severe displacements within the fluid region.
In order to avoid this, rather than treating the fluid as incompressible, it is advantageous from a modeling point of view, to use an expansible medium.
This was basically done already in reference \cite{2001Ch-Ki} for linearized elasticity and more recently for finite-strain elasticity in reference \cite{2020Ca-Po-Ni}.
Both works impose a prescribed pressure in the fluid region, which in turn results in volumetric expansion to the necessary extent for reaching the imposed pressure.
A further refinement of this approach, valid for large strains and supporting also self-contact by means of the third medium approach, can be found in \cite{2024Fa-Ho-Do-Ro}.

Nevertheless, the last three references lack a general procedure for identifying whether a cavity is connected to a given pressure source.
To allow for an expansible medium, unlike the incompressible condition from \citep{2007Si-Cl}, and at the same time track the connectivity to the pressure source, Kumar et al.~\citep{2020Ku-Fr-La,2021Ku-La} proposed to use poroelasticity theory, with the only limitation of the small strain assumption.
According to this approach, void regions in the design domain are treated as a very permeable and extremely low volume fraction sponge, while solid regions are treated as impermeable and stiff.
A mixed formulation is used with displacement and pressure fields in a coupled system of partial differential equations.
The propagation of pressure in highly permeable void regions is governed by the Darcy flow equations, and is through the theory of poroelasticity coupled to the mechanical equilibrium equations.
In addition to the use of standard poroelasticity, references \citep{2020Ku-Fr-La,2021Ku-La} propose a drainage term in solid regions, in order to avoid leaking of pressure to neighboring cavities through solid walls, which is otherwise inevitably independent of how low the solid permeability is set.
This drainage term does not have a physical world equivalent that is easy to imagine, but it is an essential component of the proposed model.
The main limitation of the topology optimization framework from \citep{2020Ku-Fr-La,2021Ku-La} is its small strain assumption.
Although the authors have demonstrated that designs optimized using the small strain approximation, perform also well when they are post-evaluated with finite-strain simulations, the limitation is still severe because soft actuators typically exhibit very large deformations.
These should certainly be accounted for during the optimization.

The present work extends the aforementioned topology optimization approach from poroelasticity to porohyperelasticity, thereby extending its validity to large deformations.
The proposed framework makes use of the drainage term from \citep{2021Ku-La}, it adopts the mixed Lagrangian porohyperelasticity theory by \cite{1996Si-Ka-Mc-Ba}, and incorporates both of these models into the nonlinear topology optimization framework by \citep{2021Bl-Si-Po}.
Regarding the porohyperelasticity formulation in particular, a two-field continuous Galerkin model is used, based only on displacement and pressure fields.
It is known that this simple two-field model might lead to oscillations across abrupt changes in permeability \cite{kadeethum2020finite}, however, the numerical examples included in the present work demonstrate that it is sufficiently accurate despite its simplicity.
Regarding the parametrization of the design, the same approach is followed as in \cite{2021Bl-Si-Po}, where a level-set function is optimized, which is mapped to a density field with a diffuse solid-void boundary.
RAMP interpolation is used for stiffness and a new interpolation scheme is proposed for the permeability and drainage coefficients, between void and solid.
Using the HuHu void regularization from \cite{2021Bl-Si-Po} turned out to be essential in order to avoid excessive mesh distortion in void regions.
Moreover, a p-norm strain energy density constraint is employed in order to control the amount of allowable elastomer deformation in the pneumatic soft actuators obtained by the optimization algorithm.

\section{Model foundations}

\subsection{Design parametrization}
\label{sec:design_parametrization}
Density-based topology optimization relies on the concept of varying material density field that determines the presence and absence of material in a fictitious design domain $\Omega$.
The goal is to find the best material distribution for a structure by minimizing a user-defined objective function, under certain geometric and deformation constraints. 
There are different ways for parametrizing the material distribution in $\Omega$.
In this work, following \cite{2021Bl-Si-Po,2023Bl-Si-Po}, a level-set function field $\chi \in (-\infty,\infty)$ is used as the primary design variable.
The field $\chi$ is then mapped to a physical density field $\rho$ through the logistic function
\begin{equation}
\rho{(\chi)} = \dfrac{1}{1 + e^{- \chi}}.
\label{eq:level_set_fnc}
\end{equation}

By its definition, $\rho$ remains between 0 to 1 in the design domain.
In this parametrization, no box constraints are required to enforce these limits on $\rho$.
Instead, a maximum slope constraint is applied on the level-set field $\chi$ in order to enforce a minimum interface width between void and solid regions.
This constraint is imposed by including the p-norm penalization term
\begin{equation}
C_i(\chi) =
\mathop{\mathlarger{\mathlarger{\int}}}_{\Omega}
  \dfrac{c_i}{6}
  \left\langle\Norm{\nabla\chi}-\dfrac{8}{L_i}\right\rangle^{\!6}
d\Omega,
\label{eq:C_x}
\end{equation}
in the objective function, with a sufficiently large weight constant $c_i$.
The Macaulay brackets notation $\langle x\rangle\!=\!\max(0,x)$ is used for compactness.
With the term $C_i(\chi)$ included in the objective, the spatial gradient (slope) of the level-set field $\chi$ will not exceed $8/L_i$ significantly.
As the optimization converges towards a black-white design, the interface between solid and void will attain an approximately constant width $L_i$, in the final design.
\added{
A minimum interface width $L_i$ should be chosen in the order of few finite elements to ensure that the solid-void transition will not be sharper than the discretization can represent.}

\subsection{Porohyperelastic model}
\label{sec:porohyperelasticity}
As explained in the introduction, poroelasticity based topology optimization elegantly implements design dependent pressure loads, ensuring at the same time connectivity to a pressurization source \cite{2020Ku-Fr-La,2021Ku-La}.
The only drawback of the standard Darcy flow theory, used in the aforementioned works, is that it is limited to small deformations, which is a too strict constraint when considering soft robotics.
Fortunately, a mathematically rigorous extension of poroelasticity to large deformations is readily available in the theory of porohyperelasticity, presented in detail in \cite{1996Si-Ka-Mc-Ba}.
Among the different formulations described in \cite{1996Si-Ka-Mc-Ba}, the present work adopts the two-field Lagrangian formulation, with pressure and displacement fields over the undeformed domain $\Omega$, respectively denoted as $p$ and $\uu$.

\added{Similar to references \cite{2020Ku-Fr-La,2021Ku-La}, only the steady state of the flow problem and isotropic permeability are required.
Hence, any time derivatives in the porohyperelasticity model from \cite{1996Si-Ka-Mc-Ba} are dropped, and the permeability tensor is replaced by a scalar value.
Moreover, the two-field formulation from \citep{1996Si-Ka-Mc-Ba} was adapted to the very common Neo-Hookean constitutive law from \cite{1985Si-Ta-Pi}, also used in \cite{2021Bl-Si-Po}. These simplifications and adaptations lead after a few algebraic manipulations to} the following weak form equations
\begin{align}
&\RR_p(\chi,p,\uu)\!\left[\delta p\right]=\nonumber\\
&~\int_{\Omega} k(\chi) \detF
                \left(\F^{-1}\F^{-T}\nabla p \right)
                \!\cdot\! \nabla\delta p
                +\big(Q_{\mathrm{in}}\!(\XX)\left(p\!-\!p_{\mathrm{in}}\right)
                      +Q_{\mathrm{out}}\!(\chi)\left(p\!-\!p_{\mathrm{out}}\right)
                 \big)\delta p
                ~d\Omega&&
                \!\!\!\!\!\!= 0~~~\forall~\delta p,
\label{eq:Darcy_flow}
\\[10pt]
&\RR_\uu(\chi,p,\uu,\qq)[\ddelta\uu]=\nonumber\\
&~\int_{\Omega} \left(\left(K(\chi) \ln{\detF}\!-\!\detF p\right) \II
                            +G(\chi) \detF^{-2/3}\,\Dev{\F\F^T}\right)
                  \!:\!
                  \left(\nabla\ddelta\uu\,\F^{-1}\right)
                  +
                  c_r ~\mathbb{H} \uu  \svdots \mathbb{H} \ddelta\uu
                  ~d\Omega&& \nonumber\\
&~~+\int_{\Gamma} \qq \cdot \ddelta\uu~d\Gamma&&
                  \!\!\!\!\!\!= 0~~~\forall~\ddelta\uu,
\label{eq:mech_equilibrium}
\end{align}
where $\F\!=\!\II+\nabla\uu$ is the deformation gradient for the displacement field $\uu$, and $\detF\!=\!\Det{\F}$.
The spatially varying coefficients $k(\chi)$, $K(\chi)$, and $G(\chi)$ are respectively, the Eulerian permeability, and the initial bulk and shear moduli of the porohyperelastic ersatz material model.
Choosing suitable definitions for these three coefficients as a function of the design variable field $\chi$ is an important modeling task discussed in the next subsection in detail.
In the context of topology optimization, $K(\chi)$ and $G(\chi)$ are typically 6 orders of magnitude smaller in the void than in the solid, while for the permeability $k(\chi)$ the present work will use a value in the solid which is 6 orders of magnitude smaller than in the void.
The vector field $\qq$ appearing in Eq.~\eqref{eq:mech_equilibrium} is a traction acting on part $\Gamma$ of the boundary of the design domain $\Omega$, to be defined in more detail later.
\added{If necessary, homogeneous Dirichlet conditions can also be imposed on part of the domain boundary by restricting the solution space for fields $\uu$ and $\delta\uu$ accordingly.}
In addition to the terms adopted from \cite{1996Si-Ka-Mc-Ba}, Eqs.~\eqref{eq:Darcy_flow} and~\eqref{eq:mech_equilibrium} contain three terms, multiplied with $Q_{\mathrm{in}}(\XX)$, $Q_{\mathrm{out}}(\chi)$, and $c_r$ respectively, which require some further clarifications.

\subsubsection*{Source and drainage terms}
The coefficient $Q_\mathrm{in}$, is a fixed function of the position vector $\XX$ in the undeformed domain $\Omega$.
In most of the domain it is zero, apart from a small region where a pressurization source is defined with a high pressure $p_\mathrm{in}$.
A sufficiently large value of $Q_\mathrm{in}$ within the pressurization source region, means that fluid volume is generated within this region until the target source pressure $p_\mathrm{in}$ is approached.
One could alternatively define a source at the boundary of $\Omega$ as is the standard case in \cite{1996Si-Ka-Mc-Ba}, however here we prefer to define a volumetric source, because it leads to smoother flow velocities close to the source region.

Similarly, the term multiplied with $Q_\mathrm{out}$ corresponds to the drainage term proposed in \cite{2020Ku-Fr-La}.
Unlike $Q_\mathrm{in}$, the distribution of $Q_\mathrm{out}$ over the domain $\Omega$ is a priori unknown, i.e., it is design-dependent in the context of topology optimization.
Following \cite{2020Ku-Fr-La}, the drainage term is added to all regions occupied by solid material.
A zero pressure $p_\mathrm{out}\!=\!0$ used in this term, means that fluid will vanish as soon as it reaches the solid walls of a pressurized cavity.
This is an elegant solution for the problem of leaking of pressure between neighboring cavities, that was proposed and treated in detail in \cite{2020Ku-Fr-La}.
The exact definition of $Q_\mathrm{out}(\chi)$ follows \cite{2020Ku-Fr-La} and will be provided in the next subsection.

\subsubsection*{Void regularization}
The HuHu void regularization term in Eq.~\eqref{eq:mech_equilibrium}, involving the Hessian $\mathbb{H} \uu$ of the displacement field and scaled with $c_r$, is known from \citep{2021Bl-Si-Po}, where it was introduced in order to regularize void elements for modeling frictionless contact by the third medium approach.
It turns out that the same approach is very useful for regularizing the deformation of the extremely high porosity void in the proposed porohyperelastic formulation.

The regularization term, corresponds to the higher-order strain energy density $\frac{1}{2}\,c_r\,\mathbb{H} \uu \svdots \mathbb{H} \uu$, added to the hyperelastic potential.
However, a deliberate exploitation of notation should be pointed out.
After discretization of the displacements field with standard Lagrange elements, the term $c_r\,\mathbb{H} \uu \svdots \mathbb{H} \ddelta\uu$ is not resolved consistently anymore, due to lack of C¹ continuity in the $\uu$ field.
In that sense, the method should not be considered as a regularization of the void continuum, but a regularization of the finite elements in the void region.
Following \citep{2021Bl-Si-Po}, a value of $c_r\!=\!10^{-6} L^2 K_\mathrm{s}$ is used throughout this work, where $L$ is a characteristic dimension of the domain introduced later, and $K_\mathrm{s}\!=\!K(\infty)$ is the initial bulk modulus in the solid.
Due to the relatively small value of the $c_r$ constant, the effect of the respective term is insignificant in the solid regions where the terms involving the stiffness parameters $K$ and $G$ become dominant.

\subsection{Material interpolation scheme}
Essential for the topology optimization model is the definition of the ersatz material properties $k(\chi)$, $K(\chi)$, and $G(\chi)$, as well as the drainage term $Q_\mathrm{out}(\chi)$, as a function of the design field variable $\chi$.

\begin{figure}[t]
\centering
\begin{overpic}[width=0.65\textwidth]
               {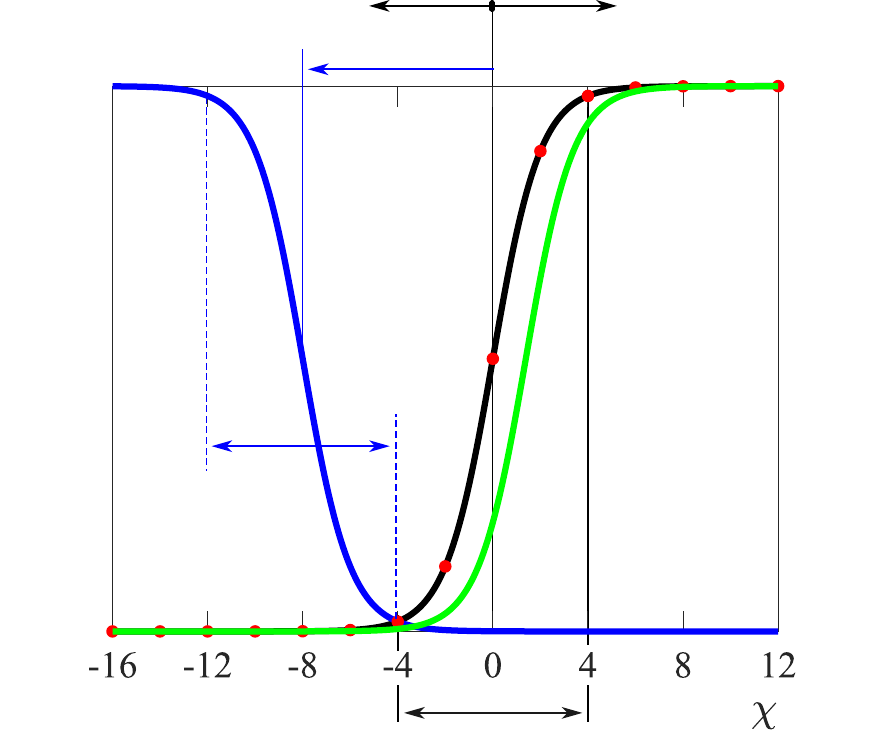}
  \put (35,83){Void}
  \put (72,83){Solid}
  \put (45,79){\color{blue}{$\ell_k$}}
  \put (15,71){\color{blue}{$k/k_\mathrm{v}$}}
  \put (32,30){\color{blue}{$L_i$}}
  \put (53,43){$\rho$}
  \put (62,45){\color{green}{$K/K_\mathrm{s}$}}
  \put (62,40){\color{green}{$G/G_\mathrm{s}$}}
  \put (49,68){\color{red}{$\dfrac{Q_\mathrm{out}}{Q_s}$}}
  \put (55,-0.5){$L_i$}
\end{overpic}
\caption{Material interpolation for the density $\rho(\chi)$, stiffness parameters $K(\chi), G(\chi)$, permeability $k(\chi)$, and drainage term $Q_\mathrm{out}(\chi)$ as a function of the level-set variable $\chi$.}
\label{fig:mat_interp}
\end{figure}
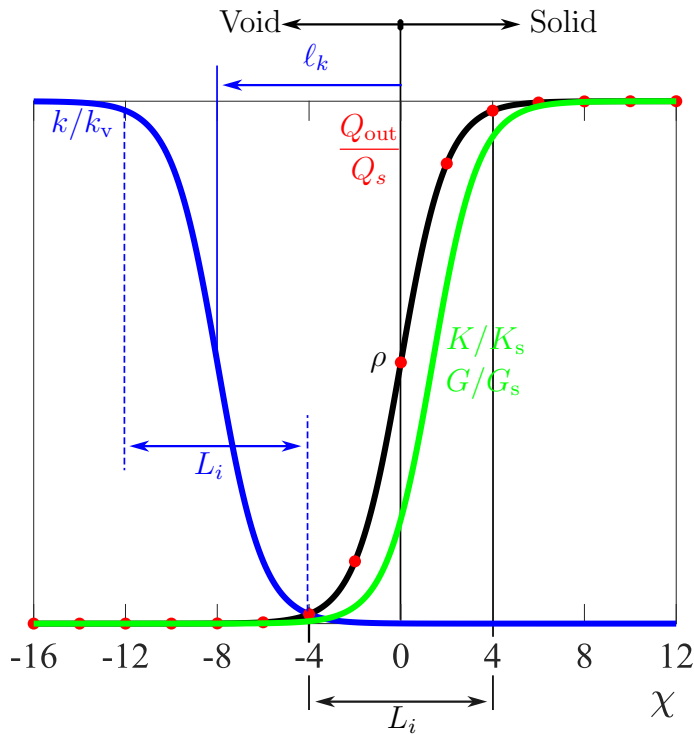

For the stiffness parameters $K(\chi)$ and $G(\chi)$, the present work uses the standard RAMP interpolation scheme \citep{2001St-Sv, 2021Bl-Si-Po}, so that
\begin{equation}
\begin{Bmatrix}K(\chi)\\G(\chi)\end{Bmatrix}
=
\begin{Bmatrix}K_\mathrm{s}\\G_\mathrm{s}\end{Bmatrix}
\left(
\mathcal{E}_0 + (1 - \mathcal{E}_0) \dfrac{\rho(\chi)}{1 + \hat{p} (1 - \rho(\chi))}
\right),
\label{eq:ramp_interp}
\end{equation}
where $K_\mathrm{s}$ and $G_\mathrm{s}$ are the initial bulk and shear moduli in the solid, $\hat{p}$ is the RAMP penalization parameter, and $\mathcal{E}_0$ is the small positive void-to-solid stiffness contrast value.
The rather common values, $\mathcal{E}_0\!=\!10^{-6}$ and $\hat{p}\!=\!3$ are used in all numerical examples throughout the present work.
The RAMP interpolation is illustrated in Figure~\ref{fig:mat_interp} as a function of the level-set value $\chi$ by means of the green color curve.

For the permeability parameter $k(\chi)$, in units of length\textsuperscript{4}$\times$force\textsuperscript{-1}$\times$time\textsuperscript{-1}, the following interpolation between a large void permeability $k_\mathrm{v}$ and a tiny solid permeability $k_\mathrm{s}$ is used,
\begin{equation}
k(\chi)
= k_\mathrm{v}
+ (k_\mathrm{s}-k_\mathrm{v})\,
\rho\!\left(\chi + \dfrac{8}{L_i} \ell_k\right).
\label{eq:perm_interpolation}
\end{equation}

As illustrated in Figure~\ref{fig:mat_interp}, offsetting the level-set value with $8 \ell_k/L_i$ shifts the permeability transition by $\ell_k$ length units into the void, i.e., $\ell_k$ can be interpreted as a dilation dimension.
The choice of this new parameter $\ell_k$ is crucial for an accurate modeling of the design-dependent pressure load.
Parametric studies have shown that the choice
\begin{equation}
k_\mathrm{s}=10^{-6}k_\mathrm{v}~~~\text{and}~~~\ell_k=L_i,
\end{equation}
results in the pressure drop from the void into the solid occurring approximately at the conventional void-solid interface contour for $\rho\!=\!0.5$.
These values for the permeability contrast of $10^{-6}$ and the permeability dilation equal to $L_i$ are used for all examples in the present work.
Changing the permeability contrast would require adapting $\ell_k$ as well, i.e., these two parameters cannot be defined independently.

A rather simple choice is made for the drainage term function $Q_\mathrm{out}(\chi)$.
As shown in Figure~\ref{fig:mat_interp}, $Q_\mathrm{out}$ has been made proportional to the density $\rho(\chi)$.
Its scaling with a constant $Q_s$ follows~\cite{2020Ku-Fr-La}, leading to the definition
\begin{equation}
Q_\mathrm{out}(\chi) =
\underbrace{\left(\dfrac{-\ln(0.1)}{L_{p0.1}}\right)^2 k_\mathrm{s}}
           _{\displaystyle Q_s}
~ \rho\left(\chi\right),
\label{eq:Q_out}
\end{equation}
where $L_{p0.1}$ is a length parameter, which corresponds to a penetration depth into the solid material.
From the definition of $Q_s$ in Eq.~\eqref{eq:Q_out} follows that the quantities $Q_\mathrm{out}$ and $Q_s$ have units in length\textsuperscript{2}$\times$force\textsuperscript{-1}$\times$time\textsuperscript{-1}.

The penetration depth $L_{p0.1}$ is defined as the distance from the pressurized surface of the solid to the point where the pressure has dropped to 10\% of the external pressure.
\added{
The choice of this parameter is essential for the performance of the model.
A too large value will lead to a slow decrease of the pressure from the surface into the solid, and in the case of a thin walled structure this can even lead to pressure leakage.
On the contrary, a too small value will lead to a transition which is too sharp to capture with a certain discretization.
Reference~\cite{2020Ku-Fr-La} recommends a large penetration depth equal to the size of two finite elements.
In the present work we define this depth as a fraction of the minimum interface width $L_i$, in particular
}
$$L_{p0.1}=0.02\,L_i.$$
\added{This is normally considerably smaller than an element size. Results are not as sensitive to this parameter as for the parameters $k_s$ amd $\ell_k$ discussed above, but higher $L_{p0.1}$ values than this recommendation will shift the pressure transition slightly towards the interior of the solid.}

\subsection{Strain energy density constraint}
The strain energy density for the Neo-Hookean material from Eq.~\eqref{eq:mech_equilibrium}, is a function of the design variable $\chi$ and the displacements field $\uu$, defined as
\begin{equation}
\Psi(\chi,\uu)=
  \dfrac{K(\chi)}{2} \left(\ln{\detF}\right)^2
  +\dfrac{G(\chi)}{2} \left(\detF^{-2/3}\,\Norm{\F}^2-3\right),
\label{eq:strain_energy_density}
\end{equation}
where the two elasticity moduli $K(\chi)$ and $G(\chi)$ are based on the RAMP interpolation Eq.~\eqref{eq:ramp_interp}.

In order to formulate a well-posed optimization problem, it is essential to exclude mechanism-like structures with infinitesimal hinges, either by a geometrical length scale constraint, or by limiting the allowable stresses or strains in the design.
In the present work, a limit is imposed on the strain level in the solid as well as in some portion of the gray interface between solid and void, by means of a strain energy constraint on a dilated version of the design.
Because of the level-set-like nature of the field $\chi$, it is easy to obtain a dilated version of the solid, simply by adding a constant offset to $\chi$.
A dilation of the converged design by a length $\ell_\Psi$ corresponds to adding the constant $8\ell_\Psi/L_i$ to the level-set variable $\chi$.
Based on these arguments, the following dimensionless p-norm term is added to the objective function
\begin{equation}
C_\Psi(\chi, \uu) =
\mathop{\mathlarger{\mathlarger{\int}}}_{\Omega}
  \dfrac{c_\Psi}{6}
  \Biggl\langle
    \dfrac{\Psi\!\left(\chi\!+\!\dfrac{8}{L_i}\ell_\Psi,\uu\right)}
          {\Psi_\mathrm{lim}}
    -1
  \Biggr\rangle^{\!6}
d\Omega,
\label{eq:strain_constraint}
\end{equation}
which penalizes any exceeding of the allowable strain energy density $\Psi_\mathrm{lim}$.
The penalization weight $c_\Psi$ is a positive dimensionless constant large enough for this term to be significant compared to the main term in the objective function.

\subsection{Leak pressure constraint}
When optimizing an internally pressurized structure, it is important to be able to detect leaking of the pressurized fluid and exclude leaky designs.
Assuming that there is an external boundary of the domain, denoted with $\Gamma_\mathrm{ext}$, that should remain at ambient pressure, it is easy to track leakage by monitoring the pressure on such a boundary.
Due to the nature of the porohyperelastic model, with the permeability of the solid being tiny but not zero, there will always be some degree of leaking.
Therefore, imposing a strict constraint on $\Gamma_\mathrm{ext}$ is not an option.
Instead, pressure values above the ambient pressure $p_\mathrm{out}$ can simply be penalized.
This is done in the present work by adding the penalization term
\begin{equation}
C_p(p)=\dfrac{1}{\Det{\Gamma_\mathrm{ext}}}
       \int_{\Gamma_\mathrm{ext}}
       \dfrac{c_p}{2}\left(\dfrac{p-p_\mathrm{out}}
                                 {p_\mathrm{in}-p_\mathrm{out}}
                     \right)^2
       d\Gamma,
\label{eq:ext_press_penalization}
\end{equation}
to the objective function, with a sufficiently large penalization factor $c_p$.
\added{In the numerical examples of the present work, the actuation pressure $p_\mathrm{in}$ will always be larger than the environment pressure $p_\mathrm{out}$, but this is not strictly necessary.
The presented formulation would still work for a negative pressure actuation, i.e. for $p_\mathrm{in}<p_\mathrm{out}$.}

\subsection{Design surface area}
The design parametrization according to Section~\ref{sec:design_parametrization}, allows for an easy evaluation of the total surface area of a fully converged design.
\added{
In general, surface area can be computed as the volume integral of a smeared Dirac function being 1 on the surface and 0 elsewhere.
Here, a volume integral of the quantity $\rho(1\!-\!\rho)$ will be considered instead, with an appropriate scaling factor.
In the fully converged design, the level-set field is expected to attain the maximum allowable slope of $8/L_i$, imposed by means of Eq.~\eqref{eq:C_x}.
In this converged situation, in direction normal to the surface, the transition of $\rho$ from 0 in the void to 1 in the solid, will have the form shown in Figure~\ref{fig:mat_interp}.
With $\xi$ in physical length units, denoting the signed distance perpendicular to the surface, the exact expression for a converged density field will be $\rho(\xi)\!=\!1/(1\!+\!e^{-8\xi/L_i})$.
Integrating along the surface normal direction, from far outside to far inside the surface, yields}
\begin{equation}
\int_{-\infty}^{\infty}\rho(\xi)\left(1-\rho(\xi)\right)d\xi
=\int_{-\infty}^{\infty}\dfrac{e^{-8\xi/L_i}}{\left(1+e^{-8\xi/L_i}\right)^2}d\xi
=\dfrac{L_i}{8}.
\end{equation}
\added{The result is not equal to 1, as it would be for the Dirac function.
Hence, in order to use $\rho(1\!-\!\rho)$ as a substitute for a smeared Dirac function, it is necessary to scale it with $8/L_i$.
This leads to the following estimate of the total surface area $A_\mathrm{surf}$ of a converged design as a functional of the level-set field $\chi$,}
\begin{equation}
A_\mathrm{surf}(\chi)=\dfrac{8}{L_i}\int_\Omega \rho(\chi)\left(1-\rho(\chi)\right) d\Omega.
\label{eq:surf_area}
\end{equation}
This result can be seen as a physical reinterpretation of the measure of non-discreteness quantity, often found in the topology optimization literature.

\subsection{Demonstration by examples}

\begin{figure}[!b]
\centering
\setlength{\tabcolsep}{1pt}
\begin{tabular}{c c c}
\begin{overpic}[abs,height=0.315\linewidth,trim=1cm 0cm 18.2cm 0cm,clip]
               {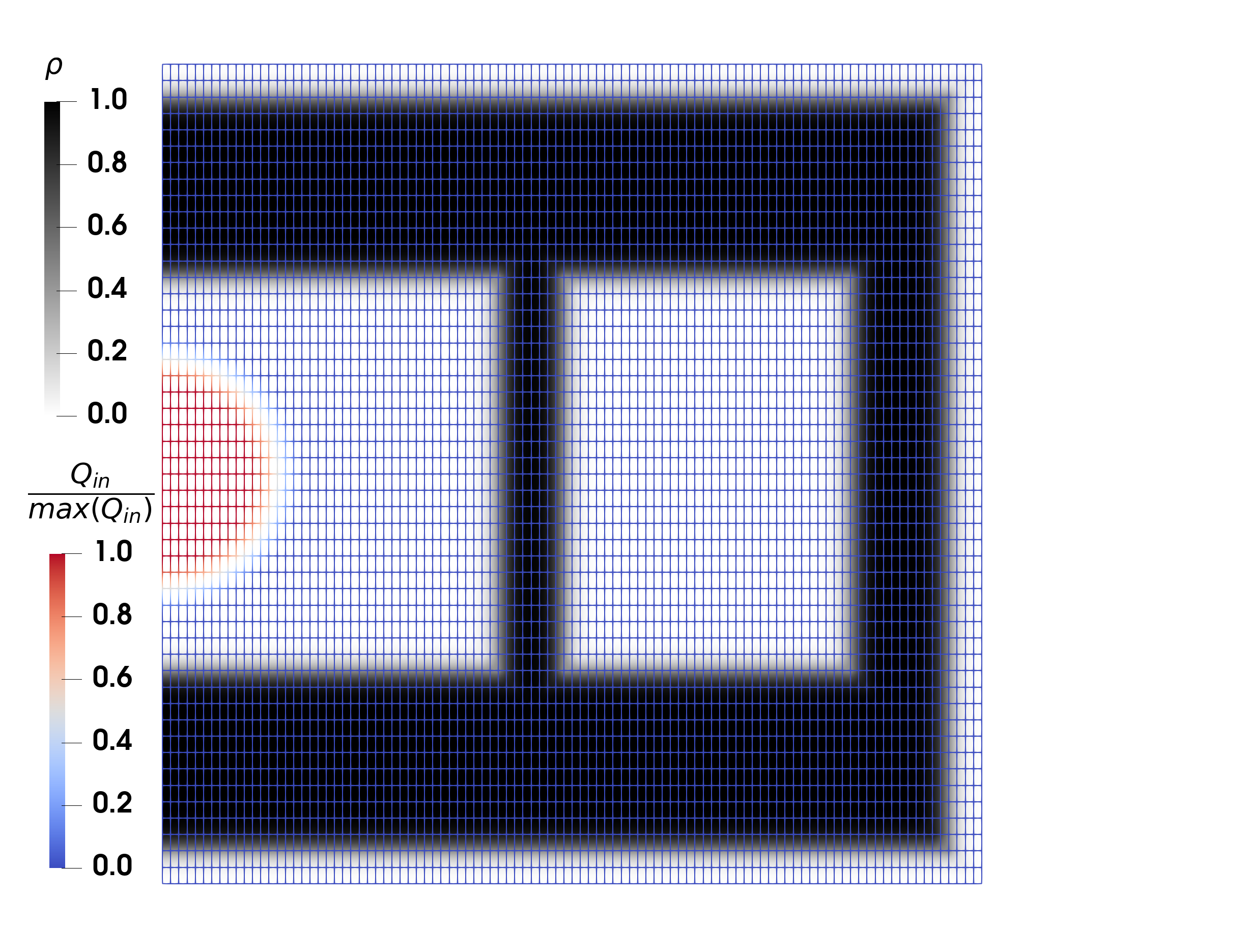}
\put(-2,2){a)}
\end{overpic}
&
\begin{overpic}[abs,height=0.315\linewidth,trim=1.9cm 0cm 11.4cm 0cm,clip]
               {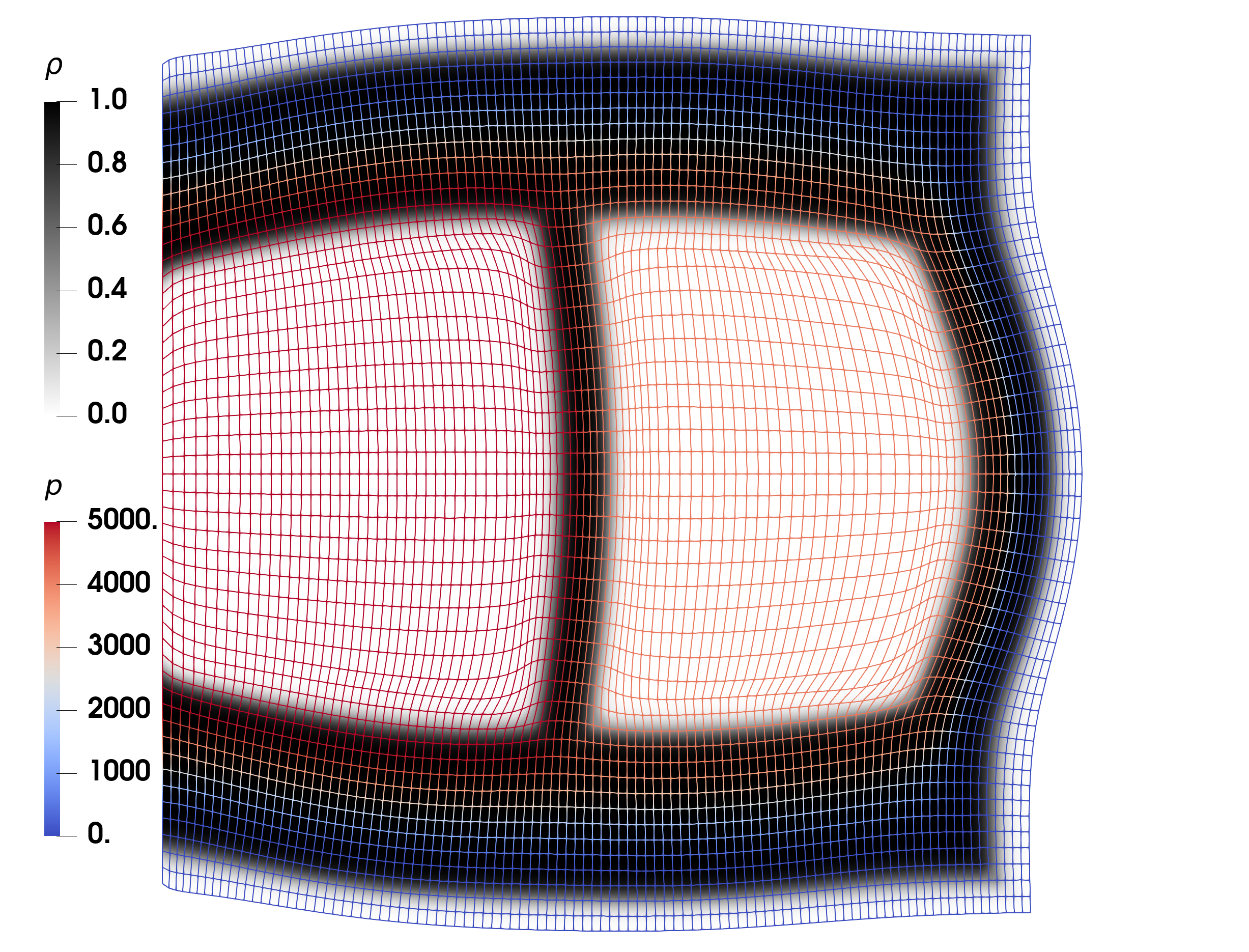}
\put(2,2){b)}
\end{overpic}
&
\begin{overpic}[abs,height=0.315\linewidth,trim=10.5cm 0cm 15.4cm 0cm,clip]
               {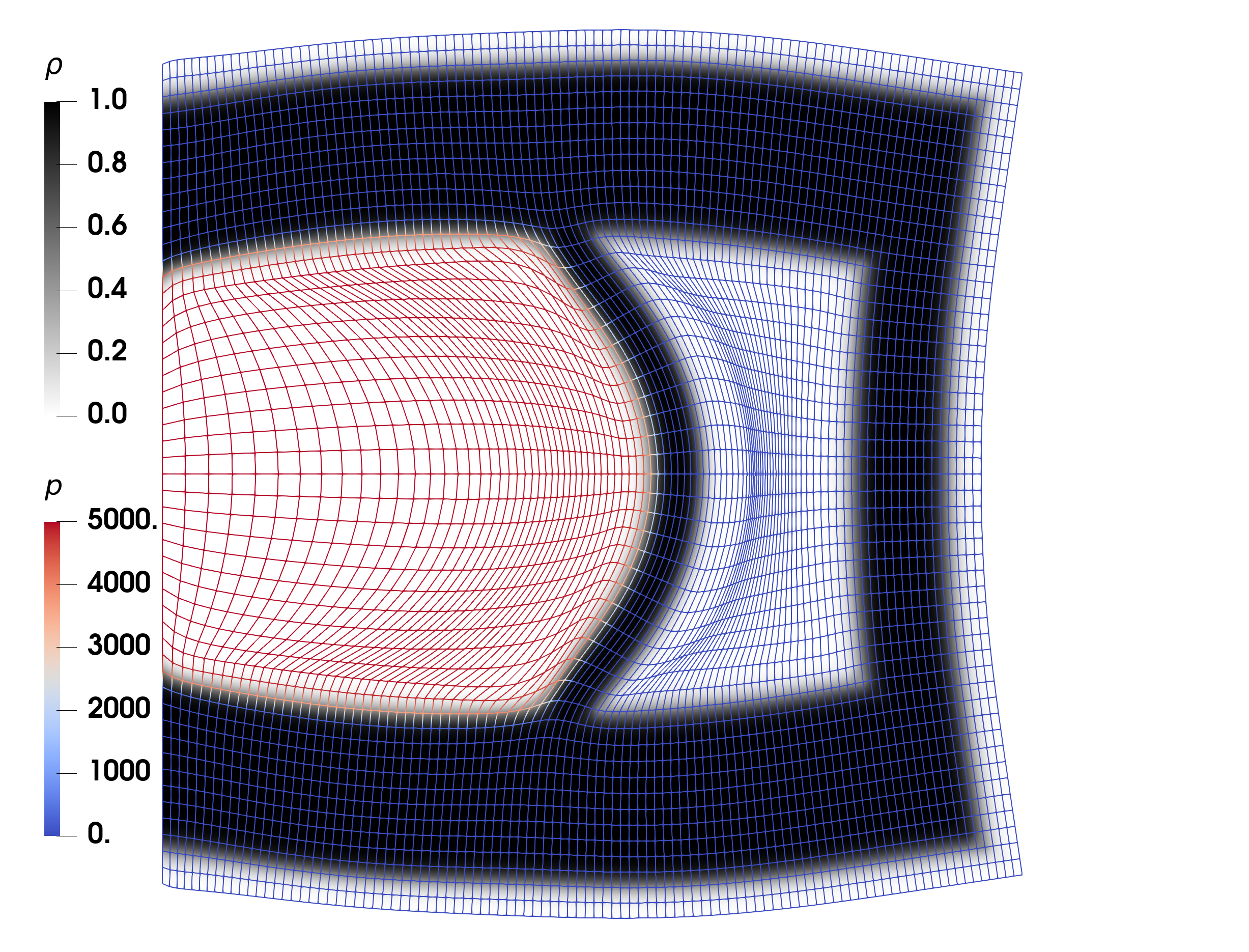}
\put(-9,2){c)}
\end{overpic}
\end{tabular}
\caption{Demo of structure with multiple cavities. a)~Undeformed configuration with the pressurized region. b)~Deformed configuration and pressure distribution obtained  without drainage in the solid, with $p\!=\!0$ enforced at the bottom, right, and top sides. c)~Deformed configuration and pressure distribution obtained with $p\!=\!0$ enforced in the solid, using a drainage term.}
\label{fig:leak_demo}
\end{figure}

To fix ideas, a couple of illustrative examples will be presented in this section.
The first \added{2D plane strain} example, shown in Figure~\ref{fig:leak_demo}, demonstrates the importance of the drainage term in Eq.~\eqref{eq:Darcy_flow}, with $Q_\mathrm{out}$ from Eq.~\eqref{eq:Q_out}.
\added{Pressurizing the left hand side cavity of the structure shown in Figure~\ref{fig:leak_demo}a, without the drainage term in the solid, but with the condition $p\!=\!p_\mathrm{out}\!=\!0$ enforced on the external boundary, leads to the response shown in Figure~\ref{fig:leak_demo}b.}
Two issues can be observed in this response.
The isolated second cavity is also pressurized, which indicates leaking through the intermediate wall, despite its low permeability.
The other issue is that the pressure in the solid drops linearly through the solid wall thickness, which does not really represent the design-dependent load acting on the interface between the void and the solid.
\added{The response shown in Figure~\ref{fig:leak_demo}c clearly demonstrates that both aforementioned issues are addressed after the inclusion of the drainage term in the solid.
A comparison of the solution shown in Figure~\ref{fig:leak_demo}c with a reference solution obtained using a pressure follower load on a body fitted mesh, can be found in \hyperref[appendixA]{Appendix~A}.
}

\begin{figure}[!t]
\centering
\begin{overpic}[abs,width=0.65\textwidth,trim=1cm 10.4cm 6cm 8.5cm,clip]
               {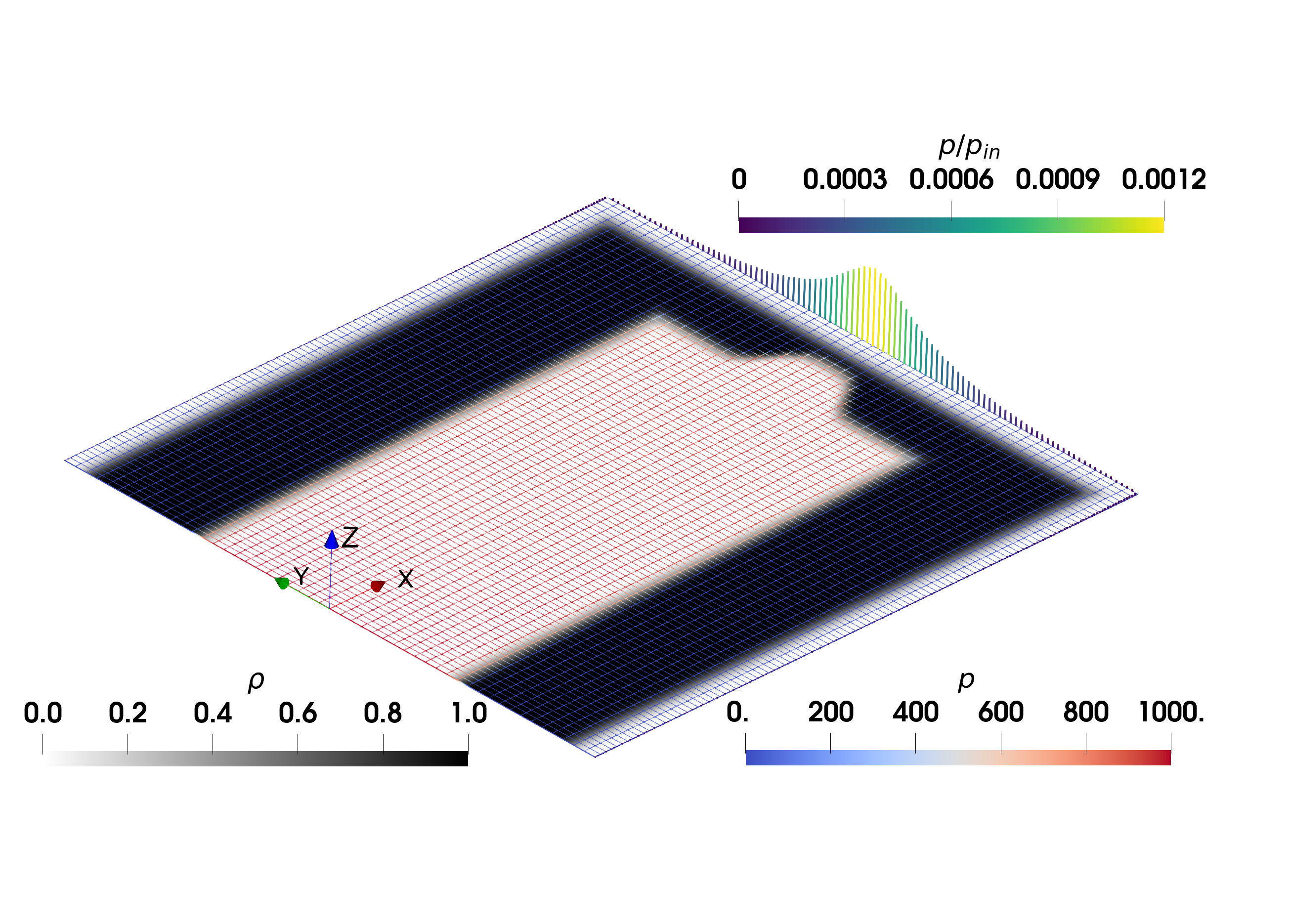}
\put(0,175){a)}
\end{overpic}
\begin{overpic}[abs,width=0.33\textwidth,trim=22cm 4.5cm 22cm 5cm,clip]
               {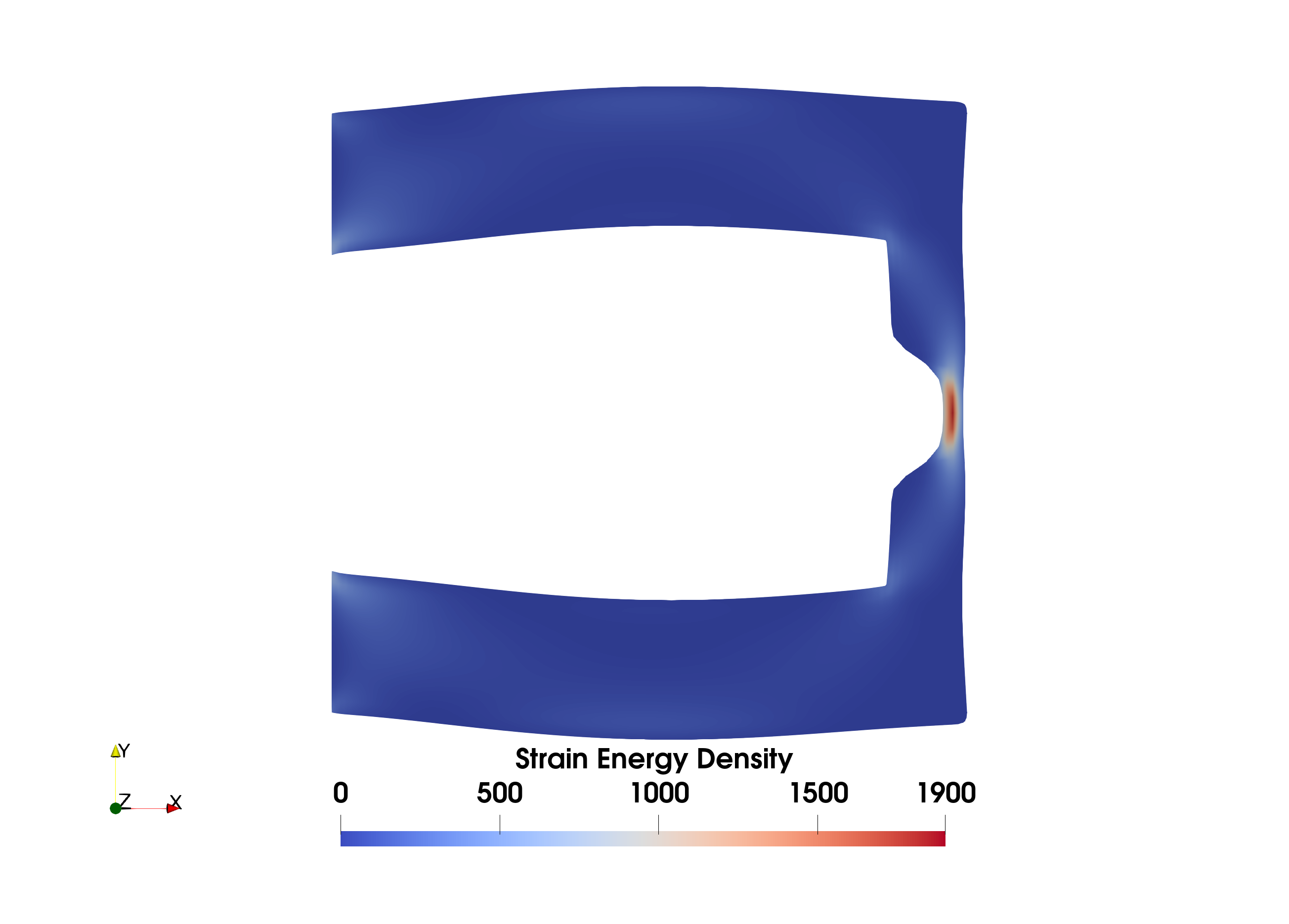}
\put(-7,175){b)}
\end{overpic}
\caption{Demo of single cavity structure with locally weakened wall. The same pressurized region is used as in Figure~\ref{fig:leak_demo}a, but with a lower source pressure. The bar plot in (a) illustrates the pressure at the external boundary, which needs to be penalized in order to avoid leaky designs. The strain energy density is illustrated on the deformed structure in~(b).}
\label{fig:leak_demo_hole}
\end{figure}

The same example can be used for assessing the accuracy of Eq.~\eqref{eq:surf_area} in estimating the total surface area of a design.
\added{
Assuming a $30\!\times\!30$ domain and out-of-plane thickness equal to 1 length unit, for the design shown in Figure~\ref{fig:leak_demo}a, the functional $A_\mathbf{surf}(\chi)$ according to Eq.~\eqref{eq:surf_area}, evaluates to 173.397 area units.}
Evaluating the actual area on a discrete version of the design results in \added{173.4} area units.
This demonstrates the level of accuracy of Eq.~\eqref{eq:surf_area}, under the condition that the level-set field has reached the slope constraint set in Eq.~\eqref{eq:C_x}.

The second example, shown in Figure~\ref{fig:leak_demo_hole}, demonstrates the role of the term $C_p$ from Eq.~\eqref{eq:ext_press_penalization} in detecting and avoiding leaky designs, as well as the role of the strain energy density from Eq.~\eqref{eq:strain_energy_density} in detecting overstretched thin membranes.
A single cavity structure is considered in this case, \added{also under a plane strain condition and} with the same pressurized region as before, but with a locally thinned wall at the opposite end.
Figure~\ref{fig:leak_demo_hole}a illustrates the computed pressure field plotted in the undeformed domain, and the pressure at the bottom, right, and top sides of the domain plotted in the out-of-plane direction as a ratio of the source pressure $p_\mathrm{in}$.
This ratio $p/p_\mathrm{in}$, which basically corresponds to the squared quantity in the integrand of Eq.~\eqref{eq:ext_press_penalization} for $p_\mathrm{out}\!=\!0$, attains larger values close to the thinned region of the structure wall.
Hence, by including the integral $C_p$ in the optimization, holes at structure walls that lead to leaking, can be excluded from the final design.

Figure~\ref{fig:leak_demo_hole}b shows the distribution of strain energy density in the deformed structure.
This result demonstrates how thinner membrane-like walls stretch excessively upon pressurization, and how such design features can be avoided by including a strain energy density constraint in the optimization, like with the p-norm integral $C_\Psi$ from Eq.~\eqref{eq:strain_constraint}.

\section{Topology optimization scheme} \label{sec:Top_Opt_scheme}

After having introduced the more general parts of the theory in the preceded section, this section will introduce a specific objective function that will be used in the soft robotics optimization examples to follow, as well as the actual optimization scheme.
Part of this optimization scheme is a continuous adjoint sensitivity analysis, also to be presented in this section.

\subsection{Output of a soft robotics actuator}
\label{subsec:rigid_arm}
In order to introduce an objective function that can be used for a class of soft robotics applications, we need to slightly extend the mechanical model presented in Section~\ref{sec:porohyperelasticity}.
It will be assumed that the boundary $\Gamma$ of the design domain $\Omega$, is connected to a rigid arm that represents the actuated output of the soft robotics system.
The placement, i.e. translation and rotation, of this rigid body will be defined by a global variable $\TT$.
In 3D, this variable would have 6 components, but the present work will be limited to 2D \added{plane strain} examples, hence the definition of $\TT\!=\!\left\{T_x,T_y,T_\theta\right\}$ with only two translational and one rotational components.

In the extended mechanical model, the coupling between the rigid body described by $\TT$ and the boundary $\Gamma$ of the deformable domain $\Omega$, is represented by means of the Lagrangian
\begin{equation}
\mathscr{L}(\uu, \qq, T_x,T_y,T_\theta)
=\int_\Gamma 
\left(\uu-\bar{\uu}\left(T_x,T_y,T_\theta\right)\right)\cdot\qq
~d\Gamma,
\label{eq:Largangian_rigid_body}
\end{equation}
where $\bar{\uu}$ is the rigid body displacements of any point on $\Gamma$ due to $T_x$, $T_y$, and $T_\theta$.
The variation of this Lagrangian with respect to $\uu$ leads to the $\qq\cdot\ddelta\uu$ integrand, already included in Eq.~\eqref{eq:mech_equilibrium}, while its variation with respect to $\qq$ provides a new equilibrium equation,
\begin{equation}
\int_\Gamma
\left(\uu-\bar{\uu}\left(T_x,T_y,T_\theta\right)\right)\cdot\ddelta\qq
~d\Gamma=0~~~\forall~\ddelta\qq.
\label{eq:q_multiplier_weak_form}
\end{equation}
The rigid arm modeled through the placement variable $\TT$, is assumed to be elastically supported on a foundation, with this support described by the elastic potential
\begin{equation}
W_\mathrm{sp}\!\left(T_x,T_y,T_\theta\right),
\end{equation}
which is to be specified for a particular numerical example.

Based on $W_\mathrm{sp}$ and $\mathscr{L}$, the equilibrium conditions for the rigid arm are provided by the equation
\begin{align}
W_{\mathrm{sp},T_x}\left[\delta T_x\right]&+
W_{\mathrm{sp},T_y}\left[\delta T_y\right]+
W_{\mathrm{sp},T_\theta}\left[\delta T_\theta\right]\nonumber\\
&-
\int_\Gamma
\left(
\bar{\uu}_{,T_x}\left[\delta T_x\right]+
\bar{\uu}_{,T_y}\left[\delta T_y\right]+
\bar{\uu}_{,T_\theta}\left[\delta T_\theta\right]
\right)\cdot\qq
~d\Gamma=0~~~\forall~\delta T_x,\delta T_y,\delta T_\theta,
\label{eq:rigid_body_equilirbium}
\end{align}
where all involved directional derivatives of $W_\mathrm{sp}$ and $\bar{\uu}$ will be straightforward to obtain in closed form.

In total, with this extension included, the overall system of mechanical equilibrium equations can be written as
\begin{equation}
\RR(\chi,p,\uu,\qq,T_x,T_y,T_\theta)\!\left[\delta p,\ddelta\uu,\ddelta\qq,\delta T_x,\delta T_y,\delta T_\theta\right]=0
~~~~\forall~\delta p,\ddelta\uu,\ddelta\qq,\delta T_x,\delta T_y,\delta T_\theta,
\label{eq:mech_eqs}
\end{equation}
which simply represents the sum of the weak form Eqs.~\eqref{eq:Darcy_flow}, \eqref{eq:mech_equilibrium}, \eqref{eq:q_multiplier_weak_form}, and~\eqref{eq:rigid_body_equilirbium}.
In the following, for the sake of brevity, the three components $T_x$, $T_y$, and $T_\theta$ will often be written in the vector form $\TT$.

\subsection{Objective function}
\label{sec:objective_function}
After having defined the actuated robot output as a rigid body and its environment as an elastic foundation, the performance of the robot can often be evaluated by means of an objective function $C_0$ that only depends on the rigid body placement variables, i.e.,
\begin{equation}
C_0(T_x,T_y,T_\theta)
\end{equation}
in closed form.
To ensure a general definition of the optimization problem, $C_0$ should be dimensionless.

In contrast to the case of lightweight design, for soft robotic actuators, the amount of material used is not an essential problem parameter.
However, when performing topology optimization without a material volume constraint, the optimization problem is ill-posed.
Material can be added arbitrarily, disconnected from the mechanically loaded structure, without affecting the design performance.
One way to recover a well-defined optimization problem, is by penalizing the total surface area of the design.
In that spirit, we will introduce the following contribution in the overall objective function
\begin{equation}
C_A(\chi)=c_A \dfrac{A_\mathrm{surf}(\chi)}{\Det{\Omega}^{2/3}}
~~~\text{in 3D, or}~~~
C_A(\chi)=c_A \dfrac{A_\mathrm{surf}(\chi)}{\Det{\Omega}^{1/2}}
~~~\text{in 2D},
\end{equation}
where $\Det{\Omega}$ is the volume of the design domain.
The functional $C_A(\chi)$ is a dimensionless measure of geometric complexity for the design described by the level-set $\chi$.

In total, accounting for the main objective function $C_0$, a penalization of geometric complexity with $C_A$, the limit on the level-set slope, weakly imposed with the p-norm $C_i$, the leak pressure constraint, weakly imposed with the 2-norm $C_p$, and the strain energy density constraint, weakly imposed with the p-norm $C_\Psi$, the overall objective function to be minimized with respect to $\chi$ is
\begin{equation}
C = C_0(T_x,T_y,T_\theta)+C_A(\chi)+C_i(\chi)+C_p(p)+C_\Psi(\chi,\uu).
\label{eq:obj_func}
\end{equation}
All terms involved are dimensionless. In their definitions, $C_A$, $C_i$, $C_p$, and $C_\Psi$ are all weighted with dimensionless coefficients $c_A$, $c_i$, $c_p$, and $c_\Psi$, respectively, but there is a difference between $C_A$ and the remaining three contributions.
The three contributions $C_i$, $C_p$, and $C_\Psi$ represent constraints, which once fulfilled within a certain level of precision, they will all have diminishing returns upon further increasing the respective weights $c_i$, $c_p$ and $c_\Psi$.
Hence, the choice of these weights, as long as they are large enough, is in general not essential for the final result of the optimization.
On the contrary, the surface area penalization $C_A$ does not exhibit diminishing returns.
Increasing the weight $c_A$ will always lead to designs with less total surface area, and hence less geometrical complexity, at the cost of worse performance in terms of the main objective $C_0$.
The Pareto front with respect to the $c_A$ weight in the multi-objective function $C$ is not expected to flatten out.
This suggests a careful and justified choice of this weight, as this choice is essential for the resulting optimized design.

The optimality condition with regard to the objective function $C$ from Eq.~\eqref{eq:obj_func} requires the variation $\delta C$ to vanish, i.e.,
\begin{equation}
\delta C = 
C_{0,\TT}      \!\left[ \ddelta\TT  \right]
+ C_{A,\chi}   \!\left[ \delta\chi \right]
+ C_{i,\chi}   \!\left[ \delta\chi \right]
+ C_{p,p}      \!\left[ \delta p   \right]
+ C_{\Psi,\chi}\!\left[ \delta\chi \right]
+ C_{\Psi,\uu} \!\left[ \ddelta\uu \right] = 0.
\label{eq:optimality_eq}
\end{equation}
In this equation, the variations $\delta p$, $\ddelta\uu$, and $\ddelta\TT$ cannot be considered as independent, because they are actually coupled to $\delta\chi$ by means of the overall mechanical equilibrium condition according to Eq.~\eqref{eq:mech_eqs}.
This implicit dependence will be treated by means of the adjoint method.

\subsection{Damped design evolution}
Adopting the approach from \cite{2021Bl-Si-Po}, the optimality Eq.~\eqref{eq:optimality_eq} is included in a monolithic Newton-Raphson loop, together with the physics equations, and the adjoint equations, to be presented later.
This implies a fully consistent linearization of Eq.~\eqref{eq:optimality_eq}.
Moreover, in order to deal with the excessively non-convex system, the system is augmented, also according to \cite{2021Bl-Si-Po}, with a damping of the design variable, due to the following transient term added to the objective function,
\begin{equation}
C_t = \int_{\Omega}
      \dfrac{\dot{\chi}^2 + L_t^2 \Norm{\nabla \dot{\chi}}^2}{2}
      ~d\Omega,
\label{eq:design_damping}
\end{equation}
where $\dot{\chi}$ is the pseudo-time derivative of the level-set field, and $L_t$ is a design diffusivity length scale.
Including this term to the overall objective function, converts the one-shot solution of the optimality condition into a multi-step design evolution procedure which corresponds to pseudo-time integration from design time $t\!=\!0$ to infinite design time.

Backward Euler discretization of the pseudo-time derivatives in Eq.~\eqref{eq:design_damping}, and differentiation with respect to $\chi$, leads to the semi-discrete term
\begin{equation}
C_{t,\chi}(\chi,t)[\delta \chi]=
\int_{\Omega}
\dfrac{\chi-\chi_\mathrm{old}}
      {\Delta t^2}
\delta\chi
+ L_t^2 \dfrac{\nabla\chi- \nabla\chi_\mathrm{old}}
              {\Delta t^2}
        \cdot \nabla \delta \chi
~d\Omega,
\label{eq:design_damping_discr}
\end{equation}
to be added to the optimality condition, Eq.~\eqref{eq:optimality_eq}.
Note the corrected denominator $\Delta t^2$ in Eq.~\eqref{eq:design_damping_discr} compared to \cite{2021Bl-Si-Po}, which fixes the erroneous definition of the pseudo-time $t$ from this previous work.

With the term from Eq.~\eqref{eq:design_damping_discr} included in the optimality condition, every design update depends on the previous design $\chi_\mathrm{old}$ and the current time step $\Delta t$.
In order to reach infinite pseudo-time within a finite number of design updates, an adaptive time stepping is necessary.
Eventually, $\Delta t$ grows towards infinity, the contribution from Eq.~\eqref{eq:design_damping_discr} vanishes, and the original optimality condition is recovered.

\subsection{Adjoint sensitivity analysis}
The adjoint method is used in order to eliminate from the optimality condition Eq.~\eqref{eq:optimality_eq} all dependent variations $\delta p$, $\ddelta\uu$, and $\ddelta\TT$, which are coupled to the independent variation $\delta\chi$ due to Eq.~\eqref{eq:mech_eqs}.
To begin with, a set of adjoint variables, $\Lambda_p$, $\LLambda_\uu$, $\LLambda_\qq$, and $\LLambda_\TT$ are introduced with the same sizes as the respective physics variables $p$, $\uu$, $\qq$, and $\TT$.
Then, the augmented objective function is defined as
\begin{equation}
C^*=C_0(\TT)+C_A(\chi)+C_i(\chi)+C_p(p)+C_\Psi(\chi,\uu)+C_t(\dot{\chi})
   +\RR(\chi,p,\uu,\qq,\TT)\left[\Lambda_p,\LLambda_\uu,\LLambda_\qq,\LLambda_\TT\right],
\label{eq:aug_func}
\end{equation}
which leads to the new optimality condition, free of any dependent variations,
\begin{align}
  C_{A,\chi}   \!\left[ \delta\chi \right]
 +C_{i,\chi}   \!\left[ \delta\chi \right]
 +C_{\Psi,\chi}\!\left[ \delta\chi \right]
 +C_{t,\chi}   \!\left[ \delta\chi \right]
&+\RR_{,\chi}  \!\left[\Lambda_p,\LLambda_\uu{\color{gray},\LLambda_\qq,\LLambda_\TT}\right]
                 \left[ \delta\chi \right]
 \nonumber\\
 +~C_{p,p}     \!\left[ \dLambda_p \right]
&+\RR_{,p}     \!\left[ \Lambda_p,\LLambda_\uu{\color{gray},\LLambda_\qq,\LLambda_\TT}\right]
                 \left[ \dLambda_p \right]
 \nonumber\\
 +~C_{\Psi,\uu}\!\left[ \ddLambda_\uu \right]
&+\RR_{,\uu}   \!\left[\Lambda_p,\LLambda_\uu,\LLambda_\qq{\color{gray},\LLambda_\TT}\right]
                 \left[ \ddLambda_\uu \right]
 \nonumber\\
&+\RR_{,\qq}   \!\left[ {\color{gray}\Lambda_p,}\,\LLambda_\uu,
                       {\color{gray}\LLambda_\qq,}\,\LLambda_\TT\right]
                 \left[ \ddLambda_\qq \right]
 \nonumber\\
 +~C_{0,\TT}   \!\left[ \ddLambda_\TT \right]
&+\RR_{,\TT}   \!\left[ {\color{gray}\Lambda_p,\LLambda_\uu,}\,\LLambda_\qq,\LLambda_\TT\right]
                 \left[ \ddLambda_\TT \right]
               = 0.
\label{eq:optimality_adjoint_eqs}
\end{align}
The first row of Eq.~\eqref{eq:optimality_adjoint_eqs} corresponds to the optimality conditions, while the remaining four rows correspond to the adjoint equations that are solved for $\Lambda_p$, $\LLambda_\uu$, $\LLambda_\qq$, and $\LLambda_\TT$.
These equations are per definition linear with respect to the adjoint variables, but they include nonlinear dependencies on the design variable and the primary physics variables.
The particular mathematical expressions for any non-previously defined terms in Eq.~\eqref{eq:optimality_adjoint_eqs} are provided in \hyperref[appendixB]{Appendix~B}.
Eliminated dependencies on adjoint variables, upon differentiation of $\RR$ with respect to different quantities, are noted in Eq.~\eqref{eq:optimality_adjoint_eqs} with gray font color.
Formulating the adjoint problem in a discretization agnostic weak form implies that this is a so-called continuous adjoint approach.

\subsection{Optimization algorithm}
\label{sec:sol_alg}
The same optimization algorithm is used as in \cite{2021Bl-Si-Po}.
It consists of solving the optimality and adjoint conditions from Eq.~\eqref{eq:optimality_adjoint_eqs} simultaneously with the mechanical equilibrium conditions, from Eq.~\eqref{eq:mech_eqs}, in a monolithic system treated with the Newton-Raphson method.
This is in contrast to a staggered solution approach \cite{2023Fr-Si-Po}, which is in general more common in the literature.
The monolithic approach implies linearization of the adjoint Eq.~\eqref{eq:optimality_adjoint_eqs}, which provides second order, i.e. Hessian, information to the optimization.
Compared to first gradient-based methods, the monolithic approach is computationally more expensive but it exhibits quadratic convergence close to the optimality point.
In practice this means that close to the optimality point, the design damping term from Eq.~\eqref{eq:design_damping_discr} can be removed, and the exact optimality point is found just by means of the Newton-Raphson algorithm in the final design iteration.

\begin{figure}[b]
\tikzstyle{block} = [rectangle, rounded corners, fill=black!20,
                     minimum width=1cm, minimum height=0.5cm,
                     text centered, draw]
\tikzstyle{decision}= [diamond, aspect=5,
                       minimum width=2cm, minimum height=0.1cm,
                       text centered, draw, fill=black!20]
\tikzstyle{line}    = [draw, -latex']
\begin{center}
\begin{tikzpicture}[node distance = 1cm]
\node[rectangle,rounded corners,fill=black!20,draw,text centered,
      text width=9cm, inner ysep=4pt]
  (NR)
  {\small
   Newton-Raphson solution of Eqs.~\eqref{eq:mech_eqs}
   and~\eqref{eq:optimality_adjoint_eqs}\\
   for $\chi$, $p$, $\uu$, $\qq$, $\TT$,
   $\Lambda_p$, $\LLambda_\uu$, $\LLambda_\qq$, $\LLambda_\TT$
  };
\node[diamond,fill=black!20,draw,text centered,aspect=3,inner sep=0pt,
      below of=NR, node distance=1.4cm, text width=2cm]
  (conv)
  {\small Converged?};
\node (conv_y) at ([shift={(0.3,-0.7)}]conv){\footnotesize yes};
\node (conv_n) at ([shift={(1.9,0.15)}]conv){\footnotesize no};
\node[diamond,fill=black!20,draw,text centered,aspect=3,inner sep=0pt,
      below of=conv, node distance=1.5cm, text width=2.5cm]
  (finalcheck)
  {\small $\Delta t\!>\!\Delta t_{\max}$?};
\node[rectangle,rounded corners,fill=black!20,draw,text centered,
      right of=conv, node distance=3.5cm]
  (decr_dt)
  {\small $\Delta t \leftarrow \Delta t/2$};
\node (finalcheck_y) at ([shift={(2,0.15)}]finalcheck){\footnotesize yes};
\node (finalcheck_n) at ([shift={(0.3,-0.75)}]finalcheck){\footnotesize no};
\node[rectangle,rounded corners,fill=black!20,draw,text centered,
      below of=finalcheck, node distance=1.5cm, text width=2.3cm]
  (update)
  {\small
   $\chi_\mathrm{old}\leftarrow \chi$\\
   $t\leftarrow t\!+\!\Delta t$
  };
\node[diamond,fill=black!20,draw,text centered,aspect=3,inner sep=0pt,
      left of=update, node distance=3.5cm, text width=2.5cm]
  (NRitercheck)
  {\small N-R iterations};
\node (NRlt5) at ([shift={(0.3,0.8)}]NRitercheck){\footnotesize $<\!6$};
\node (NRgeq5) at ([shift={(-1.9,0.25)}]NRitercheck){\footnotesize $\geq\!6$};
\node[rectangle,rounded corners,fill=black!20,draw,text centered,
      above of=NRitercheck, node distance=1.5cm]
  (incr_dt)
  {\small $\Delta t \leftarrow 2\Delta t$};
\node [rectangle,rounded corners,fill=black!20,draw,text centered,
       right of=finalcheck, node distance=5cm, text width=4.5cm]
  (final)
  {\small Optimization completed};

  \path [line] (NR) -- (conv);
  \path [line] (conv) -- (finalcheck);
  \path [line] (finalcheck) -- (update);
  \path [line] (finalcheck) -- (final);
  \path [line] (conv) -- (decr_dt);
  \path [line] (decr_dt.east) -| (5,0) |- (NR.east);
  \path [line] (update) -- (NRitercheck);
  \path [line] (NRitercheck) -- (incr_dt);
  \path [line] (NRitercheck.west) -| (-6.,0) |- (NR.west);
  \path [line] (incr_dt.west) -- ([shift={(-1.5,0)}]incr_dt.west);
\end{tikzpicture}
\end{center}
\caption{Adaptive pseudo-time step update scheme for the monolithic optimization algorithm.}
\label{fig:dt_update_scheme}
\end{figure}
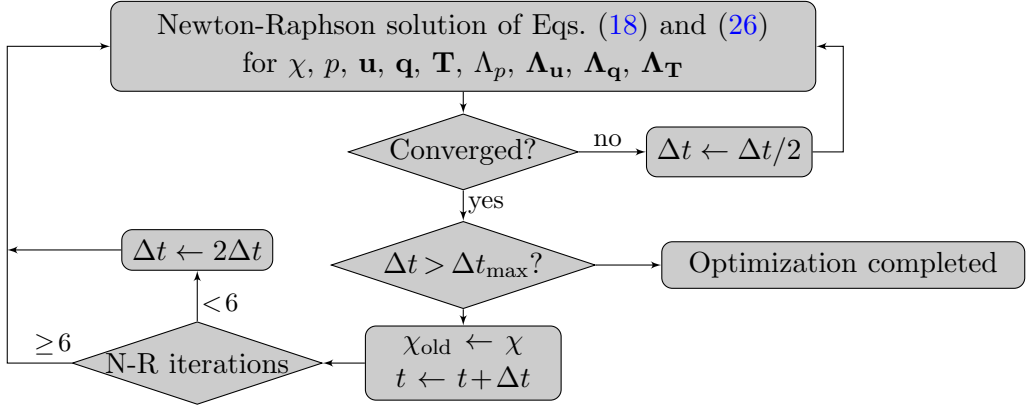

To kickstart the optimization, an initial design $\chi(t\!=\!0)$ is necessary, as well as the equilibrium solution $p$, $\uu$, $\qq$, and~$\TT$ for this design, according to equilibrium Eq.~\eqref{eq:mech_eqs}.
Due to large deformations, the mechanical problem is itself highly nonlinear; hence, finding such equilibrium will typically require incrementation of the source pressure $p_\mathrm{in}$, cf. Eq.~\eqref{eq:Darcy_flow}, in several steps toward its desired value.
Once an equilibrium solution has been determined for the initial design~$\chi$, a small enough initial pseudo-time step $\Delta t$ can be chosen, and the optimization can be started by solving the system of Eqs.~\eqref{eq:mech_eqs} and~\eqref{eq:optimality_adjoint_eqs} repeatedly, after substituting $\chi$ into $\chi_\mathrm{old}$ for each pseudo-time step.
The number of necessary Newton-Raphson iterations for each solution provides a good measure of the nonlinearity in the design increment problem. 
If $\Delta t$ is small, the added damping term dominates and the design converges fast within a few Newton-Raphson iterations indicating that larger time steps can be performed to get the next design.
In total, in order to reach infinite pseudo-time, where the optimization is completed, in a finite number of pseudo-time steps, requires an adaptive pseudo-time stepping scheme like the one from Figure~\ref{fig:dt_update_scheme}, used in the present work.

The topology optimization model described here has been implemented using the Python API of the \verb!C++! library GetFEM, leveraging the general weak form language available in this software \cite{2020Re-Po}.
The linearization of all nonlinear PDEs contained in Eqs.~\eqref{eq:mech_eqs}
and~\eqref{eq:optimality_adjoint_eqs} is performed in GetFEM symbolically.
GetFEM is also responsible for the assembly of the necessary residual vector and tangent matrix of the monolithic system treated in the algorithm from Figure~\ref{fig:dt_update_scheme}, while all linear system solutions are performed using the linear solver MUMPS~\cite{MUMPS}.
Linear 4-node Lagrange elements are used for the level-set field $\chi$ and the pressure field $p$.
Incomplete quadratic, 8-node Lagrange elements are used for the discretization of the displacement field $\uu$ and the multiplier field $\qq$, with the latter limited within the boundary $\Gamma$.
Gauss integration with $3\!\times\!3$ integration points per quadrilateral is used for the assembly of all terms.

\section{Results and discussion}
\label{sec:numerical_model}

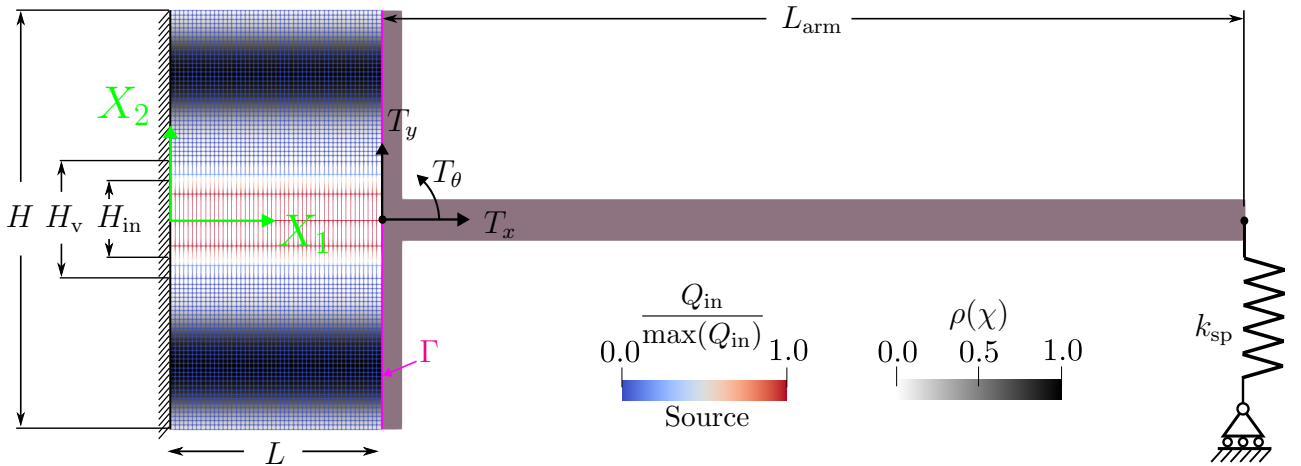
\begin{figure}[b!]
\centering
\begin{overpic}[width=\textwidth,trim=-1cm 0cm 0cm 0cm]
               {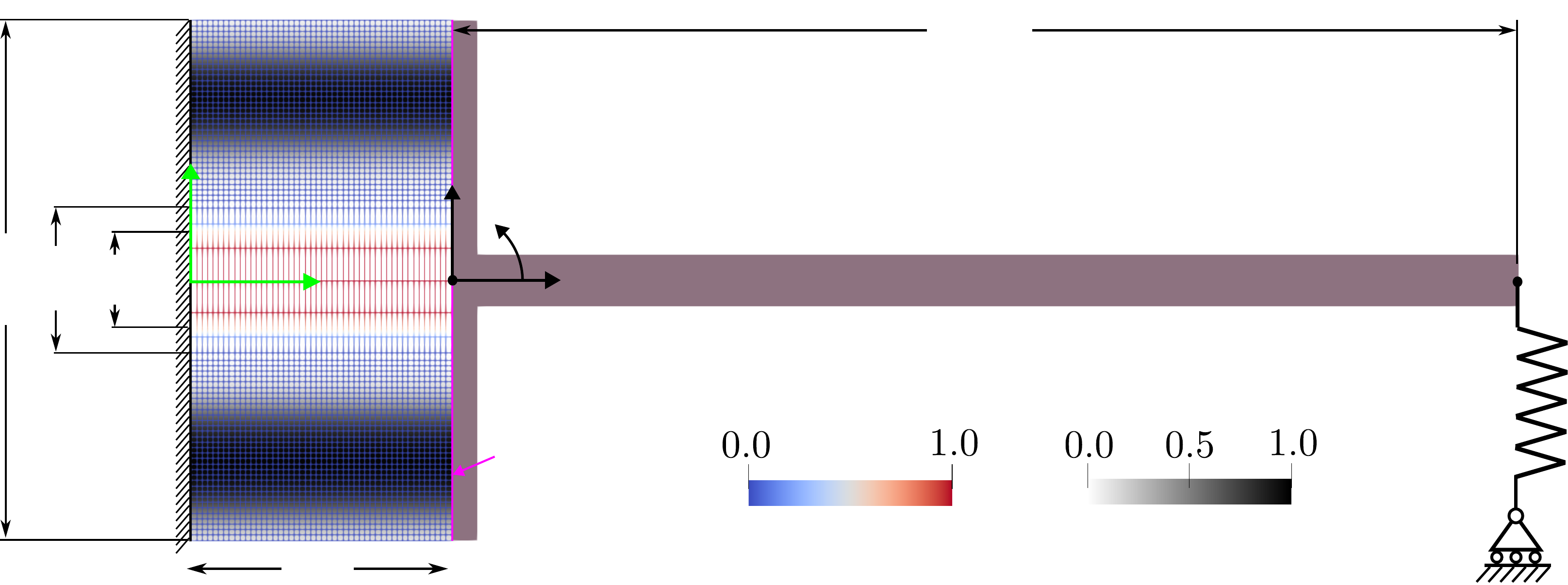}
  \put(74,11){$\rho(\chi)$}
  \put(52,3){\small{Source}}
  \put(50,11){\small
              $\dfrac{Q_\mathrm{in}}
                     {\max(Q_\mathrm{in})}$}
  \put(61,34){$L_\mathrm{arm}$}
  \put(21,0){$L$}
  \put(1,18.5){$H$}
  \put(4,18.5){$H_\mathrm{v}$}
  \put(8,18.5){$H_\mathrm{in}$}
  \put(22,17){\color{green}{\Large $X_1$}}
  \put(8,27){\color{green}{\Large $X_2$}}
  \put(34,22){$T_\theta$}
  \put(38,18){$T_x$}
  \put(30.5,26){$T_y$}
  \put(93,10){$k_\mathrm{sp}$}
  \put(33,8){\color{magenta}{$\Gamma$}}
\end{overpic}
\caption{Initial design in undeformed configuration, with position vector $\XX\!=\!(X_1,X_2)$.}
\label{fig:initial_design}
\end{figure}

To demonstrate the robustness of the porohyperelasticity-based topology optimization framework introduced in the previous sections, a simple 2D soft pneumatic actuator, shown in Figure~\ref{fig:initial_design}, will serve as a benchmark.
Extensive parametric studies will be presented and discussed, both with regard to physical and numerical parameters of the benchmark problem.

Figure~\ref{fig:initial_design} defines the design domain $\Omega$ of the benchmark problem, as an $L\!\times\!H$ rectangle.
It also shows the initial design in terms of the density field $\rho$, corresponding to the level-set function
\begin{equation}
\chi(t\!=\!0) =
  \dfrac{8}{L_i}\min
  \left(\dfrac{0.45H-\abs{X_2}}{3},~
        \dfrac{\abs{X_2}-0.25H}{3},~
        \abs{X_2}\!-\!H_\mathrm{v}/2\right)
        \!-\!0.05 \langle X_2 \rangle.
\label{eq:init_chi_2D}
\end{equation}
This initial design consists of two horizontal plates at a vertical distance of $0.7H$ with a thickness of approximately $0.2H$ and a rather diffuse transition between the void and solid regions.
Subtraction of the term $0.05\langle X_2 \rangle$ introduces an asymmetry in the initial design, making the upper wall somewhat thinner.
The passive void region height $H_\mathrm{v}$ will be defined and explained below.

The left edge of the design domain $\Omega$ is fixed, while its right edge $\Gamma$ is attached to the rigid arm.
The placement of the actuated arm is controlled by the placement variables $T_x$, $T_y$, and $T_\theta$, introduced in Section~\ref{subsec:rigid_arm}.
Now, for the specific geometry of the actuated rigid arm, the displacement field $\bar{\uu}$ at the right edge $\Gamma$ of the domain, can be specified as the following nonlinear function of $T_x$, $T_y$, and $T_\theta$
\begin{equation}
\bar{\uu}
=\begin{pmatrix}T_x - X_2\sin(T_\theta)\\
                T_y - X_2\left(1-\cos(T_\theta)\right)
 \end{pmatrix},
\label{eq:rigid_body_displacement_Gamma}
\end{equation}
which can be substituted into Eqs.~\eqref{eq:q_multiplier_weak_form}, and~\eqref{eq:rigid_body_equilirbium}.

\added{With the displacement of the end point of the arm given as $T_y\!+\!L_\mathrm{arm}\sin(T_\theta)$, the elastic potential energy for the attached spring can be specified, for use in Eq.~\eqref{eq:rigid_body_equilirbium}, as}
\begin{equation}
W_\mathrm{sp}
= \dfrac{1}{2} k_\mathrm{sp} \left(T_y + L_\mathrm{arm} \sin(T_\theta)\right)^2.
\label{eq:spring_energy}
\end{equation}
A linear spring with stiffness  $k_\mathrm{sp}$ is considered for simplicity, but any other rate-independent mechanical resistance from the environment could, \added{as easily, be implemented for the quasi static state captured by the physical model.}

Another term to be specified is the pressurization source term $Q_\mathrm{in}(\XX)$ present in Eq.~\eqref{eq:Darcy_flow}.
In this benchmark example, the pressurized region extends over a height $H_\mathrm{in}\!=\!0.25 H$ across the center of the design domain, with the respective term defined as
\begin{equation}
Q_\mathrm{in}(\XX)
= 10\,Q_s
  \left(1-\sin^2\left(\dfrac{\pi}{2}
                     \Biggl\langle
                     \dfrac{5}{2}\dfrac{\abs{X_2}}{H_\mathrm{in}}
                     -\dfrac{3}{2}
                     \Biggr \rangle_{\!\!0\dots 1}
                \right)
  \right),
\end{equation}
where $Q_s$ is the drainage intensity in the solid, defined previously in Eq.~\eqref{eq:Q_out}.
The notation $\langle x\rangle_{\!a\dots b}\!=\!\max(a,\min(x,b))$ is used for compactness \added{in the definition of a sinusoidal smooth Heaviside function attaining the value of one in the interior of the pressurized region.}
Scaling the inlet intensity in the pressurized region with a factor 10 compared to the drainage intensity in the solid, ensures an inflow which is strong enough in order for the target pressure $p_\mathrm{in}$ to be reached within good approximation.
At the same time, this moderate inflow intensity avoids a too strong fluid flow through the pressurized void regions, which could lead to excessive element deformations.
\added{With the void regularization described in Section~\ref{sec:porohyperelasticity}, increasing the scaling of the inlet intensity to $100 Q_s$, or even higher, should not be a problem.}

In order to avoid the presence of any solid material in the pressurized region of the domain, a passive void region with a height $H_\mathrm{v}\!=\!0.3 H$ across the center of the domain is \added{defined.
This is done by enforcing the condition $\chi\leq 8/L_i(\abs{X_2}\!-\!H_\mathrm{v}/2)$ to the level-set field, with the slope $8/L_i$ chosen for compatibility with Eq.~\eqref{eq:C_x}.
In a similar manner, in order to avoid boundary effects like the ones treated in \cite{2020Wa-Iv-Am-To}, the level-set field can be forced to be non-positive at the bottom and top of the domain, with the condition $\chi\leq 8/L_i(H/2\!-\!\abs{X_2})$.
Both of these constraints are enforced through the following penalization term, added} to the overall objective function $C^*$,
\begin{equation}
\dfrac{1000}{2}
\left\langle\chi-\dfrac{8}{L_i}
                 \min\left(H/2\!-\!\abs{X_2},~
                           \abs{X_2}\!-\!H_\mathrm{v}/2\right)
\right\rangle^2.
\label{eq:passive_domain}
\end{equation}
The \added{analytic} treatment of this term in the overall optimization scheme is fully equivalent to the terms $C_A(\chi)$ and $C_i(\chi)$ already present in $C^*$, cf. Eq.~\eqref{eq:aug_func}, and will therefore not be discussed further.
\added{Unlike $C_A(\chi)$, but similar to the penalized level-set slope constraint $C_i(\chi)$, the passive void region constraint from Eq.~\eqref{eq:passive_domain} is feasible.
Hence, once sufficiently enforced, further increase of its scaling should practically not affect the result of the optimization.
It should also} be noted that the term involving $H_\mathrm{v}$ in Eq.~\eqref{eq:init_chi_2D}, ensures that the initial design, defined in this equation, complies with the passive void region constraint.

The last and most important component necessary in order to complete the problem definition is the specification of the main objective function $C_0(\TT)$.
In general, it is desired to maximize the compression of the actuated spring, i.e., to minimize the function $T_y+L_\mathrm{arm}\sin(T_\theta)$, involving the two rigid arm placement components $T_y$ and $T_\theta$.
Since all other contributions in $C^*$ are dimensionless, the spring compression should also be expressed in dimensionless form.
This is achieved by dividing the aforestated expression with $L_\mathrm{arm}$.
Moreover, the optimized design should exhibit significant shear stiffness, in the sense that the origin of the rigid arm should not move upwards, i.e. the condition $T_y\!\leq\!0$ should be fulfilled, at least in a weak sense.
All these considerations together motivate a definition of the main objective function as
\begin{equation}
C_0(\TT) = \dfrac{1000}{2}
           \left\langle\dfrac{T_y}{L_\mathrm{arm}}\right\rangle^2
           +\dfrac{T_y}{L_\mathrm{arm}} + \sin(T_\theta).
\label{eq:main_obj_C0}
\end{equation}
The quadratic term penalizes any positive, i.e. upwards, displacement $T_y$ of the actuated rigid arm, while the remaining part maximizes the compression of the actuated spring.
The rigid arm length $L_\mathrm{arm}$ in the objective function is a rather essential parameter affecting the relative importance between shear and bending deformations.
The longer the rigid arm, the higher the ratio of bending-to-shear loading.

At this point, the optimization problem is fully defined.
All specific parameter values involved in the model equations, as well as in the discretization, are according to Table~\ref{tab:model_params} unless explicitly stated otherwise in the text.
The elasticity constants $E$ and $\nu$ of the solid material, listed in Table~\ref{tab:model_params}, correspond to a typical 3D-printed elastomer \cite{2023Wa-Br-Si}, and they were used to calculate the initial elastic moduli $K_\mathrm{s}$ and $G_\mathrm{s}$ appearing in Eq.~\eqref{eq:ramp_interp}.

\begin{table}[t!]
\centering
\caption{Default model and discretization parameters for optimized pneumatic soft actuator.}
\begin{tabular}{l| l|l}
    $\Det{\Omega}=L\!\times\!H = 15\!\times\!30$
          & Domain dimensions
          & \si{\milli\meter\squared} \\
    $N_L\!\times\!N_H = 48\!\times\!74$
          & Mesh size
          & \si{-} \\
    $E=2.736,~\nu=0.499$
          & Young's modulus and Poisson's ratio for the solid
          & \si{\newton\per\milli\meter\squared},~\si{-} \\
    $k_\mathrm{sp}=H E/2000$ & Spring stiffness
          & \si{\newton\per\milli\meter} \\
    $L_\mathrm{arm}=60$ & Length of actuated rigid arm
          & \si{mm} \\
    $p_\mathrm{in} = 0.02E$ & Source pressure
          & \si{\newton\per\milli\meter\squared} \\[2pt]
\hline&\\[-12pt]
    $c_A = 0.02$ & Penalization of estimated total surface area
          & \si{-} \\[2pt]
\hline&\\[-12pt]
    $L_i = 2$ & Void-solid interface width
          & \si{mm} \\
    $c_i = 1$ & Penalization of interface width constraint
          & \si{-} \\
    $L_t = 5 L_i$ & Design diffusivity length
          & \si{mm} \\[2pt]
\hline&\\[-12pt]
    $k_\mathrm{v}=\SI{1000}{\per\second}\times L^2/E$ & Void permeability
          & \si{\milli\meter^4\per\newton\per\second} \\
    $k_\mathrm{s} = 10^{-6}~k_\mathrm{v}$ & Solid permeability
          & \si{\milli\meter^4\per\newton\per\second} \\
    $\ell_k = L_i$ & Offset of solid-void permeability transition
          & \si{\milli\meter} \\[2pt]
\hline&\\[-12pt]
    $c_p = 1000$ & Penalization of leaking pressure constraint
          & \si{-} \\[2pt]
\hline&\\[-12pt]
    $\Psi_\mathrm{lim} = \tfrac{1}{2}\,0.17^2E$ & Max. allowable strain energy density
          & \si{\newton\per\milli\meter\squared} \\[2pt]
    $c_\Psi = 10^{6}$ & Penalization for strain energy density constraint
          & \si{-} \\
    $\ell_\Psi = 0.5$ & Design dilation for strain energy density constraint
          & \si{mm}\\
\end{tabular}
\label{tab:model_params}
\end{table}

\subsection{Parametric studies}

To show the efficacy and robustness of the proposed nonlinear topology optimization framework, the benchmark optimization problem introduced above, was solved for a wide range of parameters.
The optimized designs obtained, depend on various input parameters including the actuation pressure $p_\mathrm{in}$, spring stiffness $k_\mathrm{sp}$, rigid arm length $L_\mathrm{arm}$, allowable strain energy density $\Psi_\mathrm{lim}$, and surface area penalization $c_A$.
Starting with the same initial design in all cases considered, the effect of all these parameters on the optimized actuator design will be presented and discussed individually.

\subsubsection*{Actuation pressure and allowable strain energy density}

\begin{figure}[t!]
\setlength{\tabcolsep}{4pt}
\begin{tabular}{c| c c c c c c c} 
\multirow{2}{5mm}{$\dfrac{p_\mathrm{in}}{E}$} &
\multicolumn{6}{c}{$ \sqrt{2\Psi_\mathrm{lim}/E}$} \\
 & $0.03$ & $0.07$ & $0.11$ & $0.17$ & $0.23$ & $0.34$ & $0.68$ \\
\hline
  & \scriptsize -0.0024, 0.0371
  & \scriptsize -0.0024, 0.037
  & \scriptsize -0.0024, 0.037
  &
  &
  &
  &
  \\[-2pt]
\put(0,40){\rotatebox{90}{$0.001$}}
  & \includegraphics[width=0.11\textwidth,trim=36mm 21mm 38mm 21mm,clip]
                    {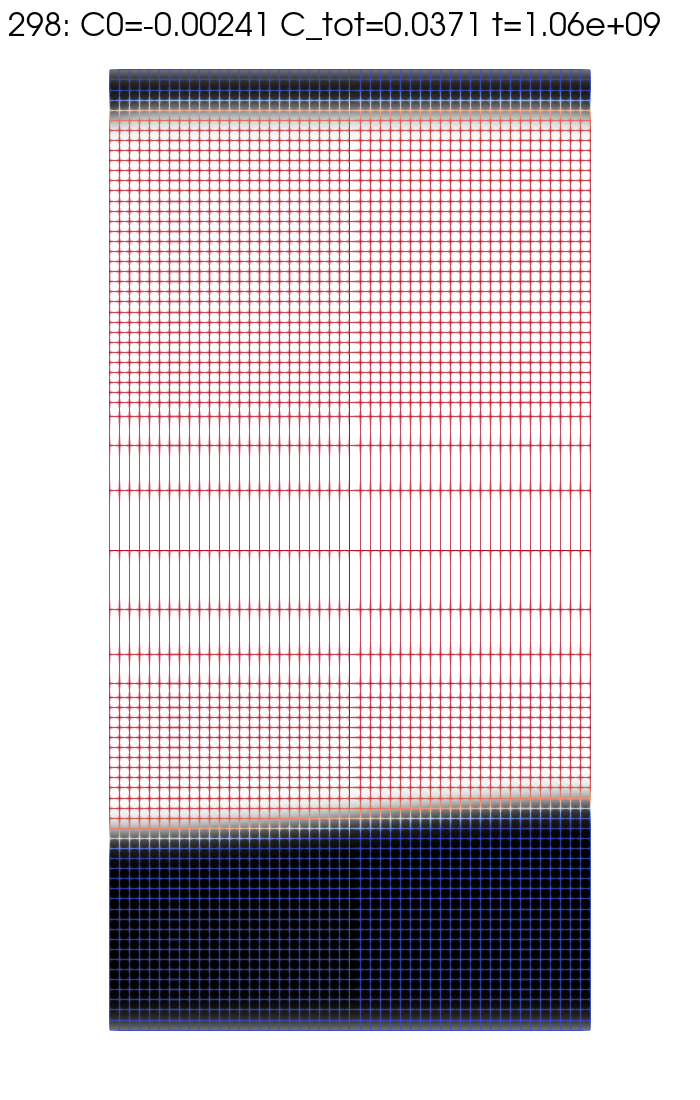}
  & \includegraphics[width=0.11\textwidth,trim=36mm 21mm 38mm 21mm,clip]
                    {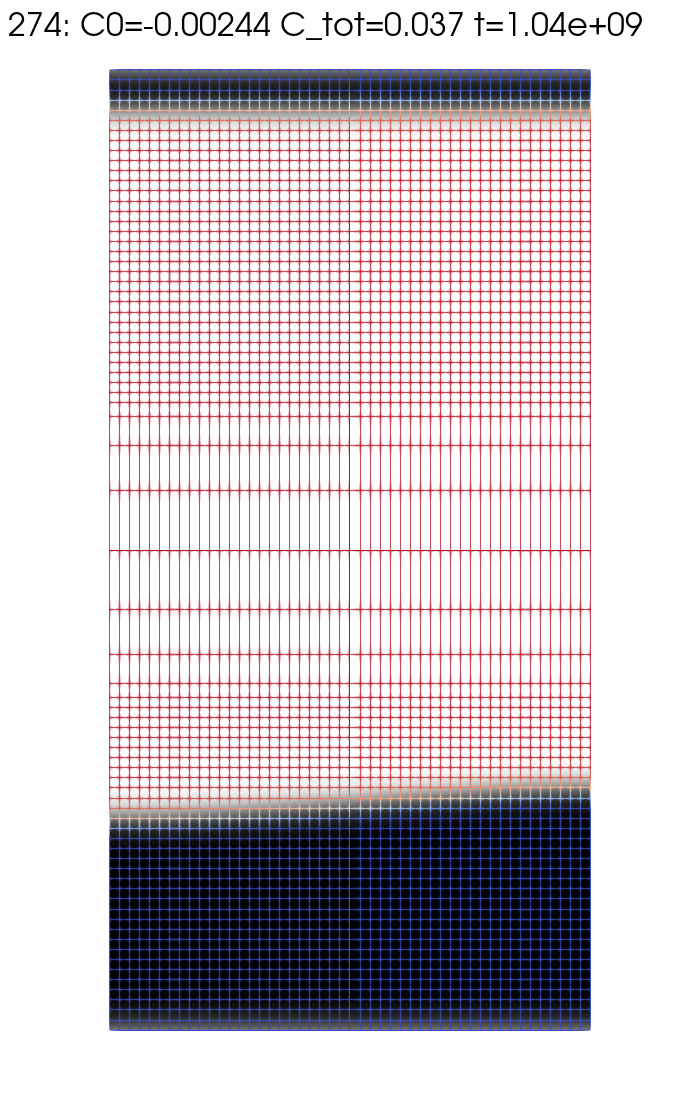}
  & \includegraphics[width=0.11\textwidth,trim=36mm 21mm 38mm 21mm,clip]
                    {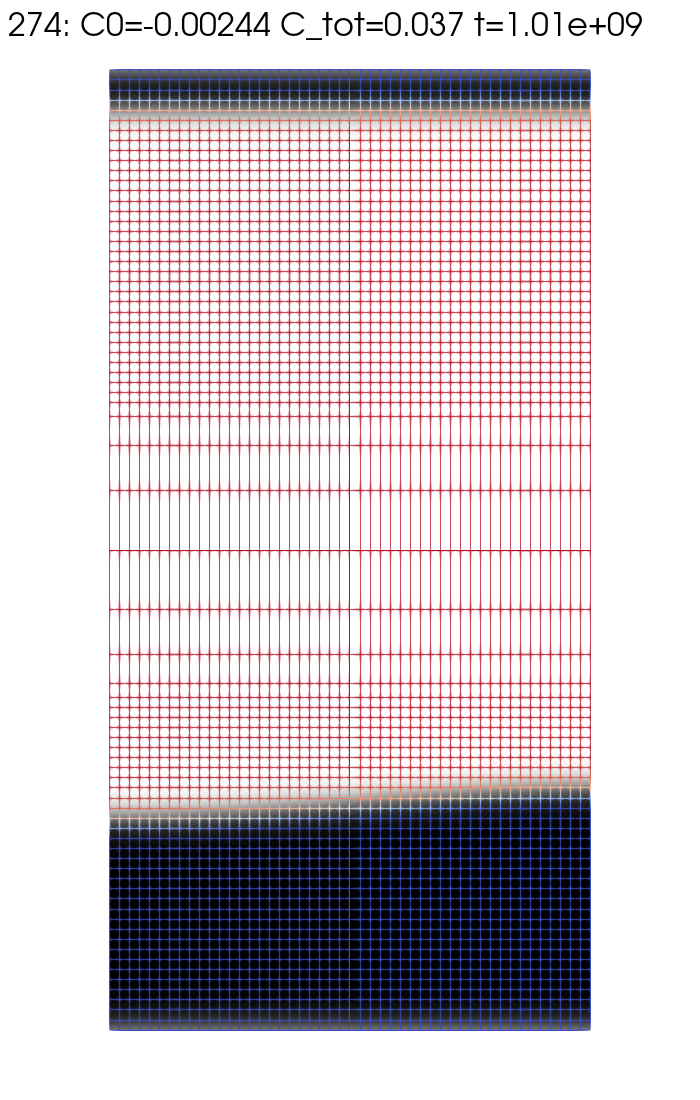}
  &  &  & &
  \\[0pt]
  & \scriptsize -0.0053, 0.0375
  & \scriptsize -0.0239, 0.0266
  & \scriptsize -0.0255, 0.024
  & \scriptsize -0.0259, 0.0234
  &
  & 
  & 
  \\[-2pt]
\put(0,40) {\rotatebox{90}{$0.005$}}
  & \includegraphics[width=0.11\textwidth,trim=36mm 21mm 38mm 21mm,clip]
                    {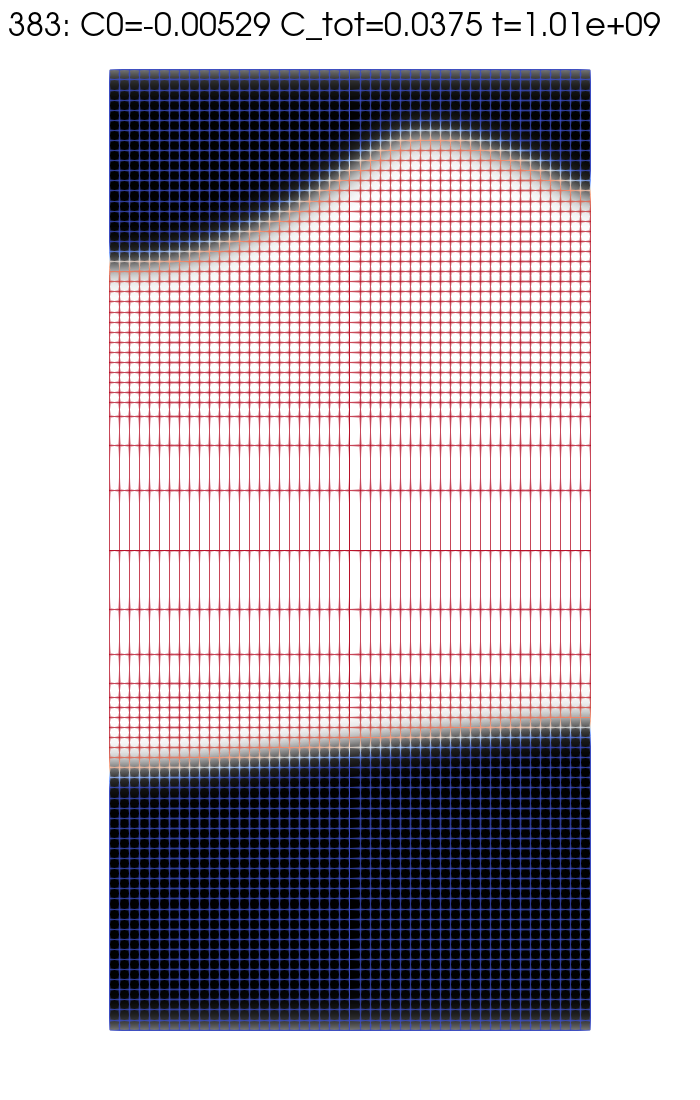} 
  & \includegraphics[width=0.11\textwidth,trim=36mm 21mm 38mm 21mm,clip]
                    {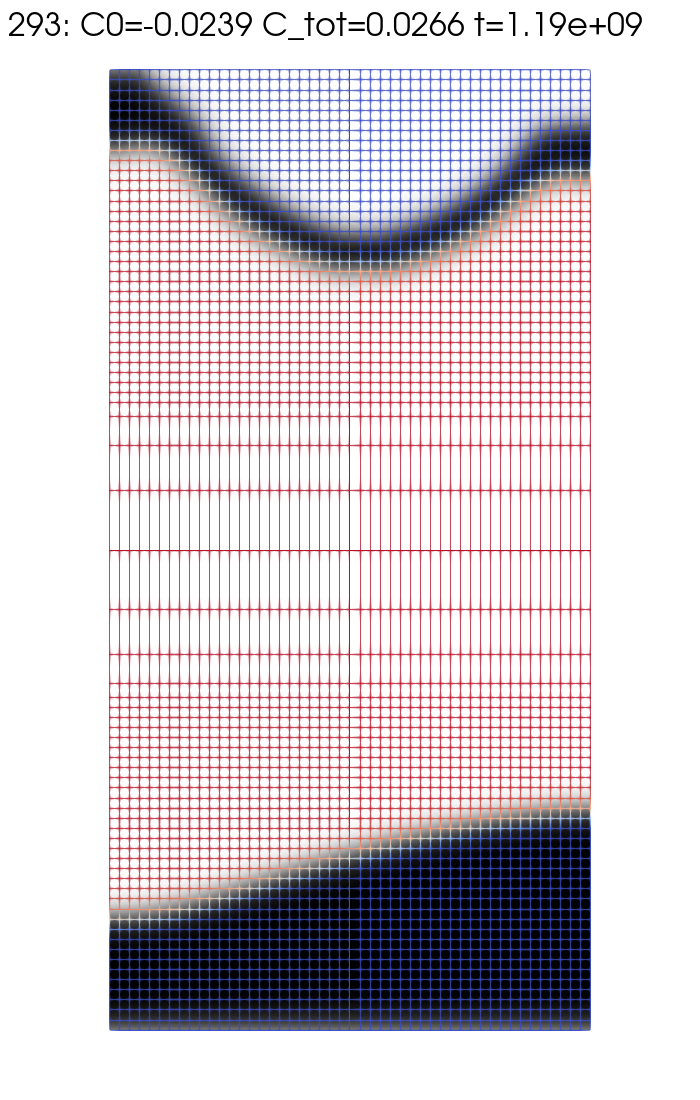}
  & \begin{overpic}[width=0.11\textwidth,trim=36mm 21mm 38mm 21mm,clip]
                   {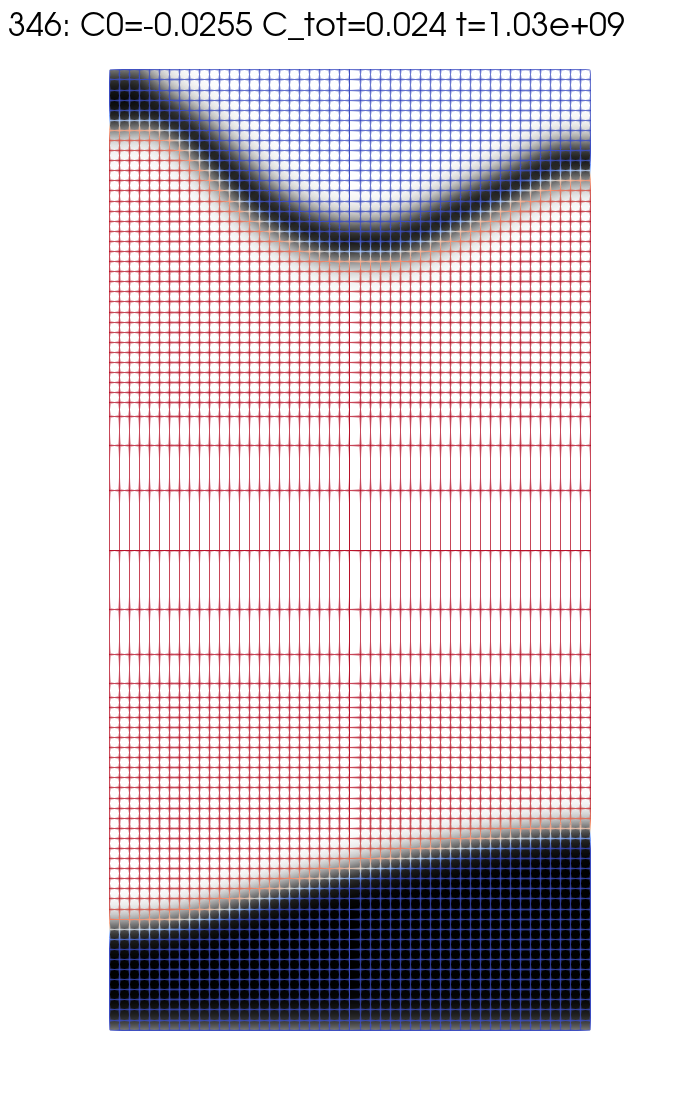}
      \put(25,90){\vector(3,2){100}}
      \put(120,100){
        \includegraphics[scale=0.14,trim=2.5cm 5.5cm 7.0cm 3.2cm,clip]
                        {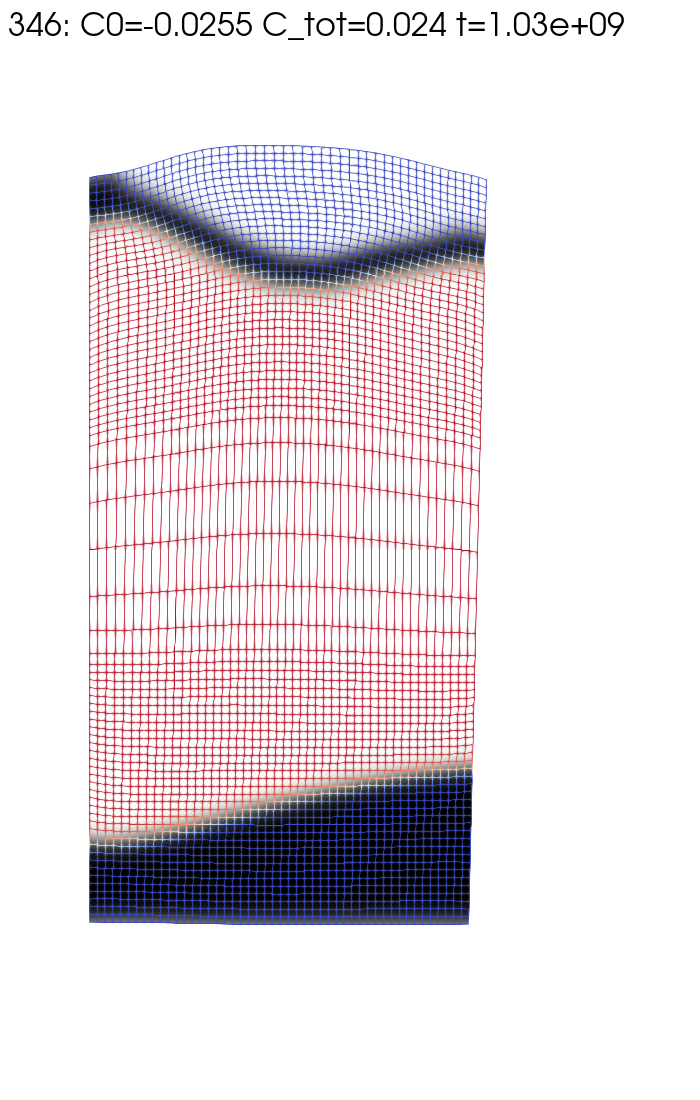}
      }
      \put(132,210){\small deformed}
    \end{overpic}
  & \includegraphics[width=0.11\textwidth,trim=36mm 21mm 38mm 21mm,clip]
                    {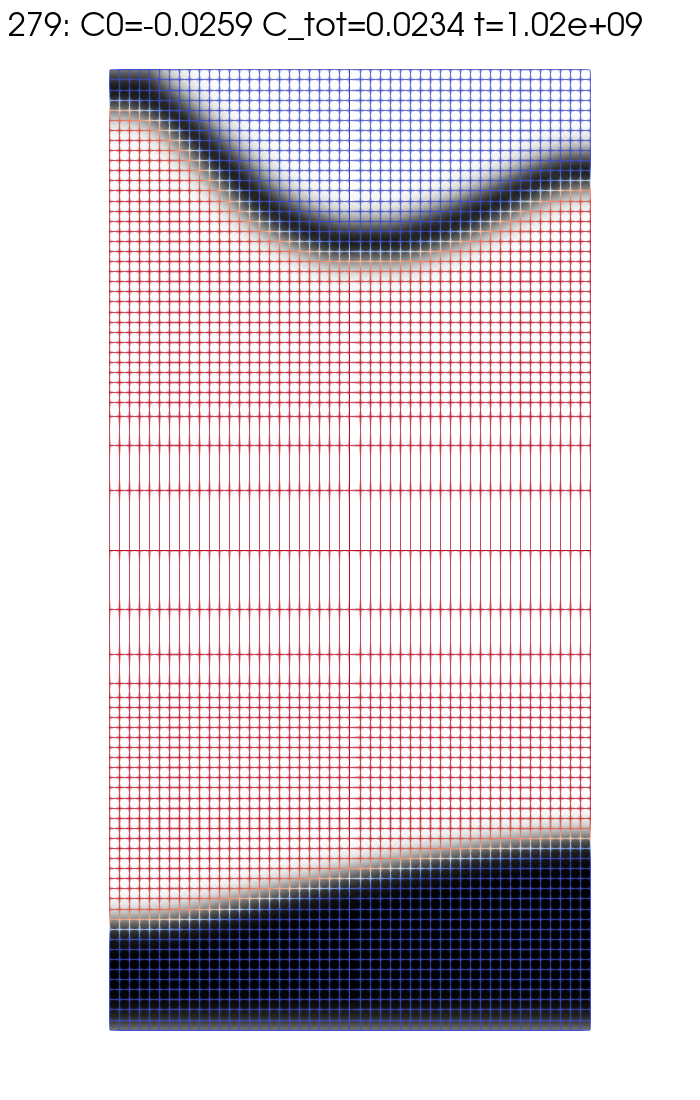}
  &
  & &
  \\[0pt]
  & \scriptsize -0.0047, 0.0397
  & \scriptsize -0.0393, 0.0151
  & \scriptsize -0.0501, 0.0023
  & \scriptsize -0.0586, -0.0056
  & \scriptsize -0.063, -0.0099
  &
  &
  \\[-2pt]
\put(0,40) {\rotatebox{90}{$0.01$}} 
  & \includegraphics[width=0.11\textwidth,trim=36mm 21mm 38mm 21mm,clip]
                    {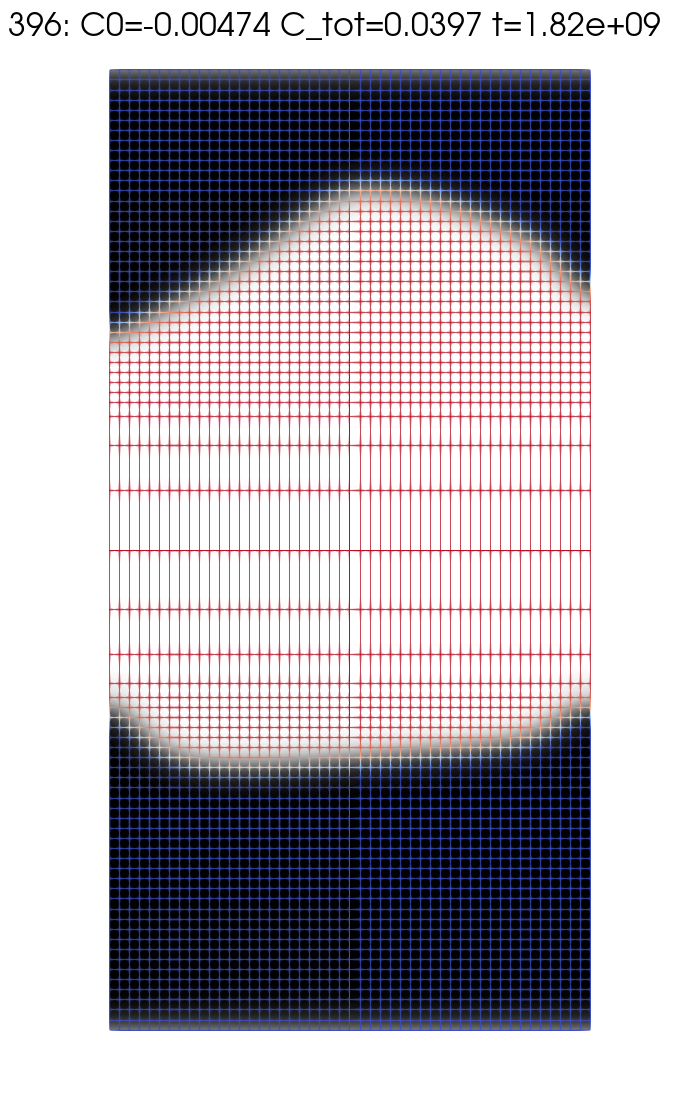}
  & \includegraphics[width=0.11\textwidth,trim=36mm 21mm 38mm 21mm,clip]
                    {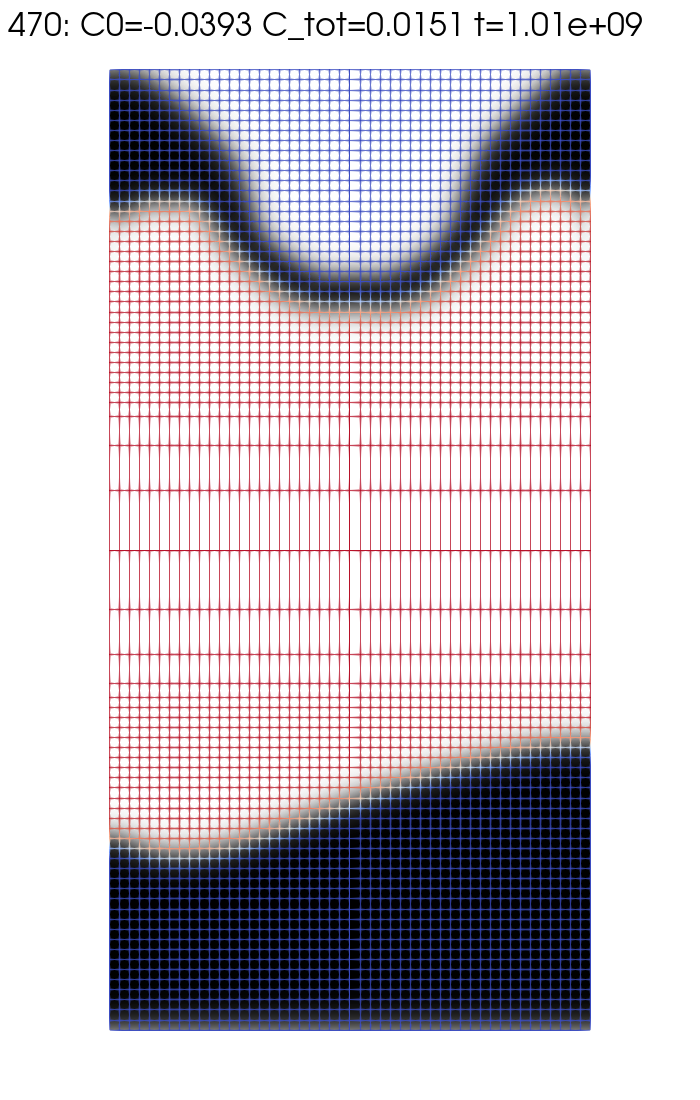} 
  & \includegraphics[width=0.11\textwidth,trim=36mm 21mm 38mm 21mm,clip]
                    {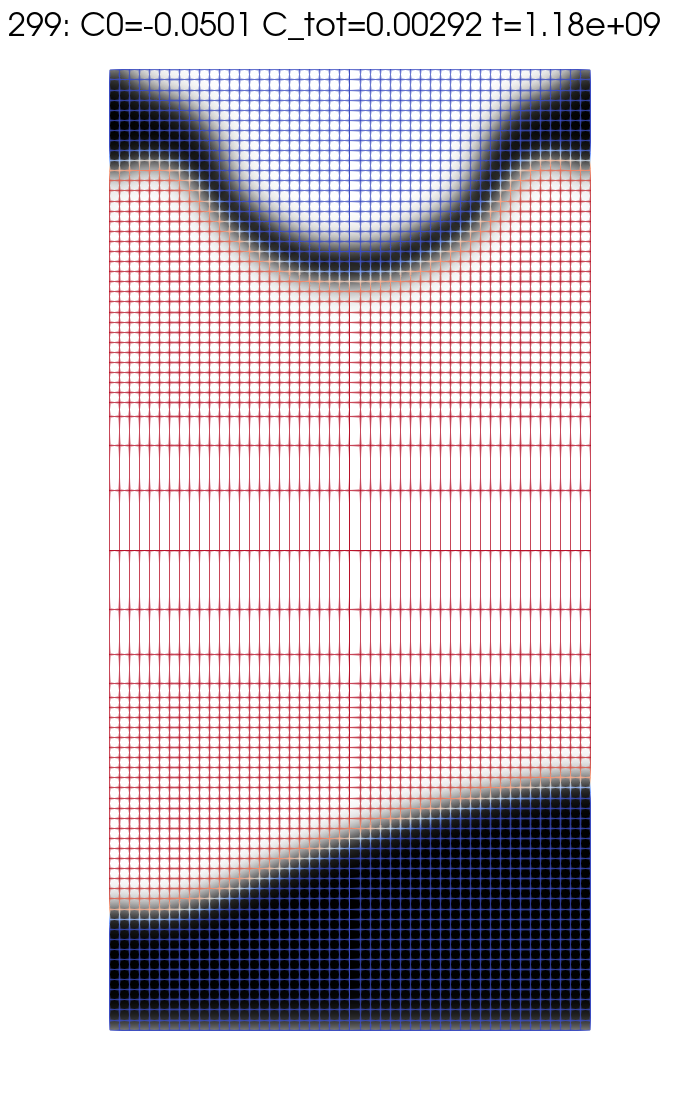}
  & \begin{overpic}[width=0.11\textwidth,trim=36mm 21mm 38mm 21mm,clip]
                   {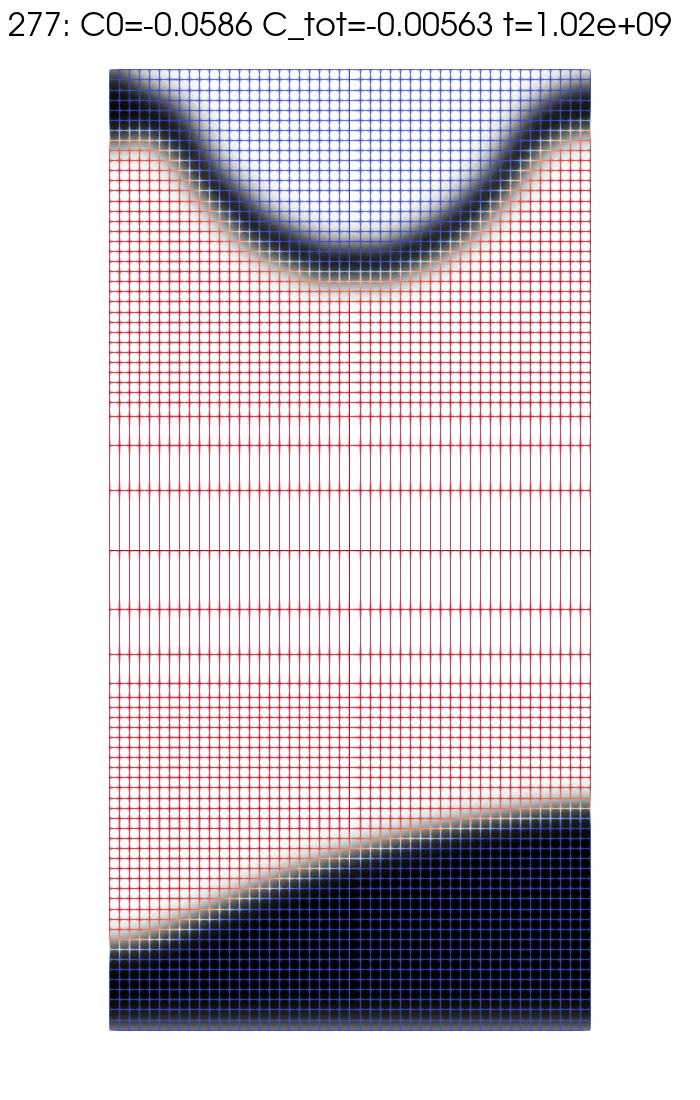}
        \put(25,90){\vector(3,2){100}}
        \put(120,97){
          \includegraphics[scale=0.14,trim=2.5cm 5.5cm 6.0cm 3.2cm,clip]
                          {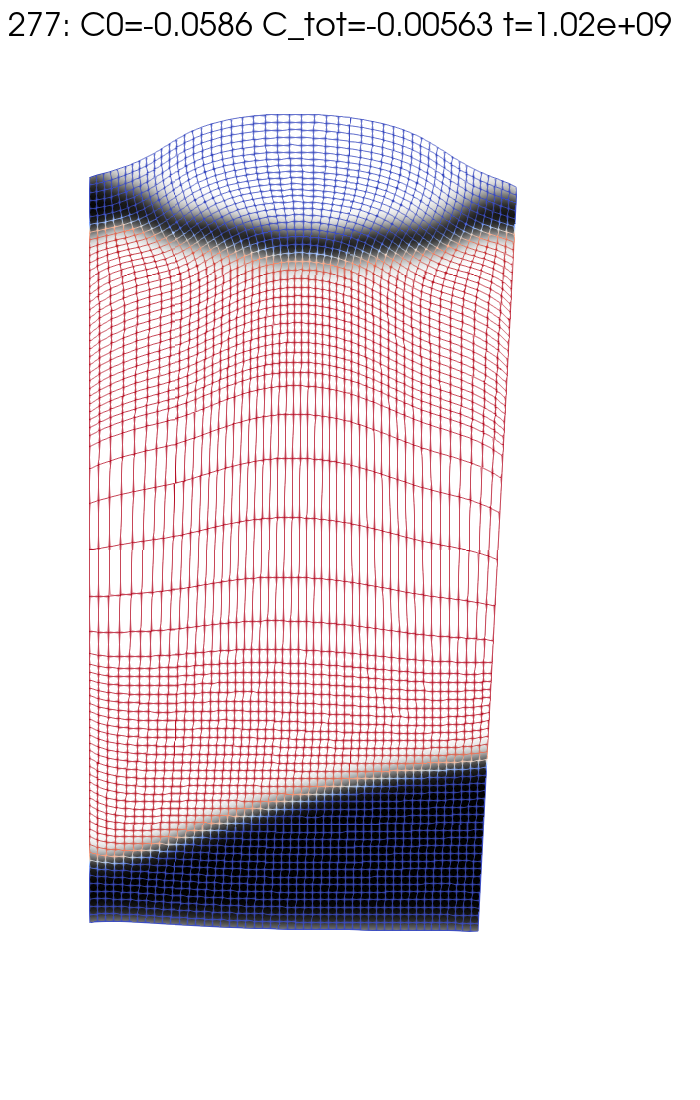}
        }
        \put(132,210){\small deformed}
      \end{overpic}
  & \includegraphics[width=0.11\textwidth,trim=36mm 21mm 38mm 21mm,clip]
                    {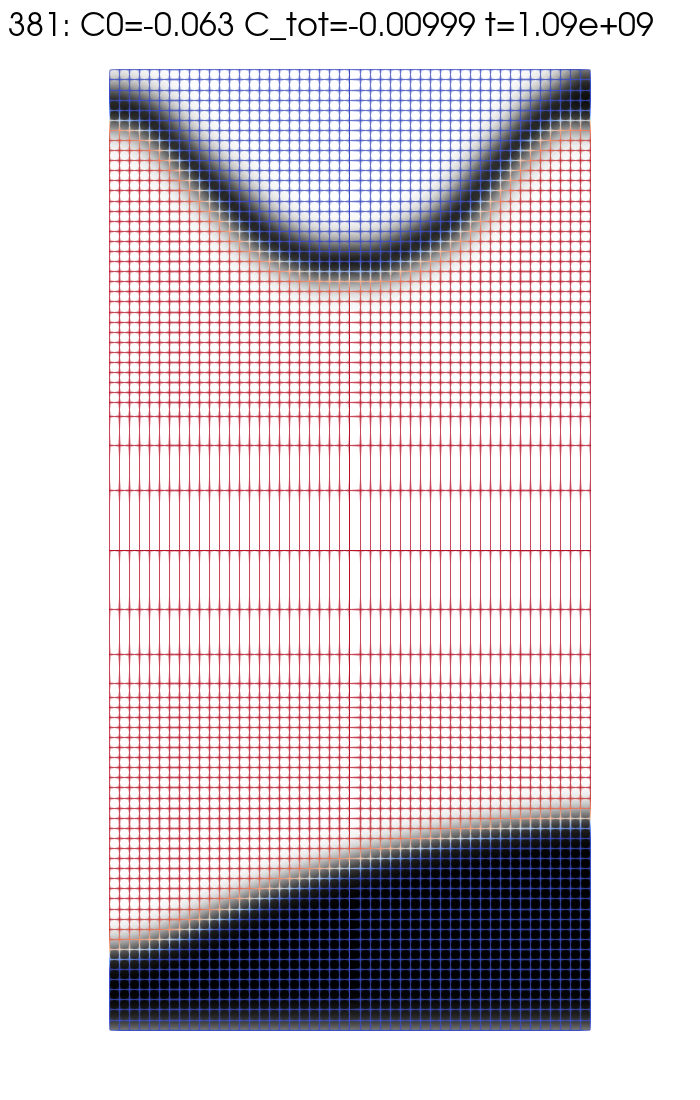}

  &
  &
  \\[0pt]
  &
  & \scriptsize -0.0131, 0.0311
  & \scriptsize -0.0561, 0.0061
  & \scriptsize -0.1016, -0.0422
  & \scriptsize -0.126, -0.0653
  & \scriptsize -0.168, -0.104
  &
  \\[-2pt]
\put(0,40) {\rotatebox{90}{$0.02$}} 
  &
  & \includegraphics[width=0.11\textwidth,trim=36mm 21mm 38mm 21mm,clip]
                    {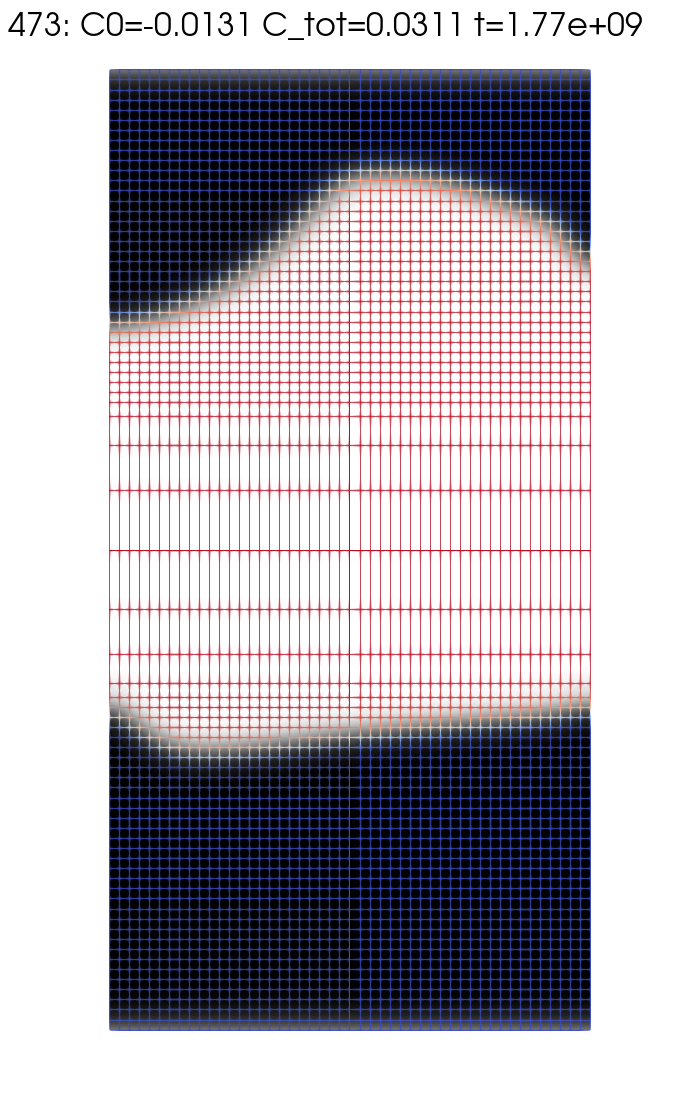} 
  & \begin{overpic}[width=0.11\textwidth,trim=36mm 21mm 38mm 21mm,clip]
                {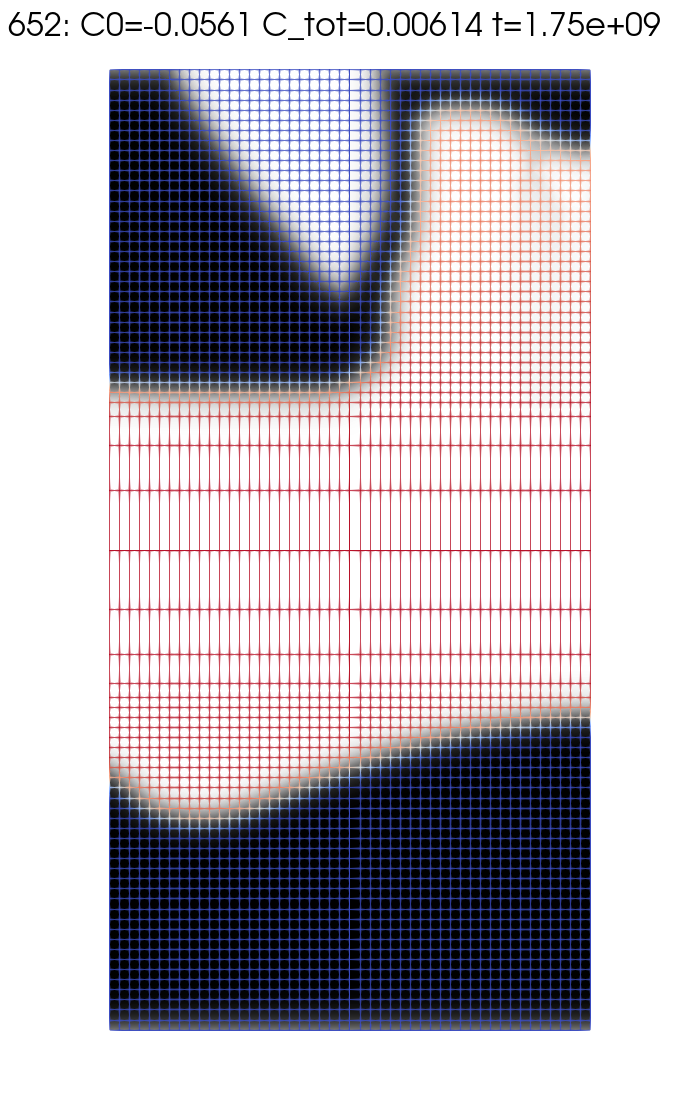}
      \put(25,50){\vector(-2,-3){65}}
      \put(-100,-120){
        \includegraphics[scale=0.14,trim=2.5cm 4.5cm 7cm 5.0cm,clip]
                        {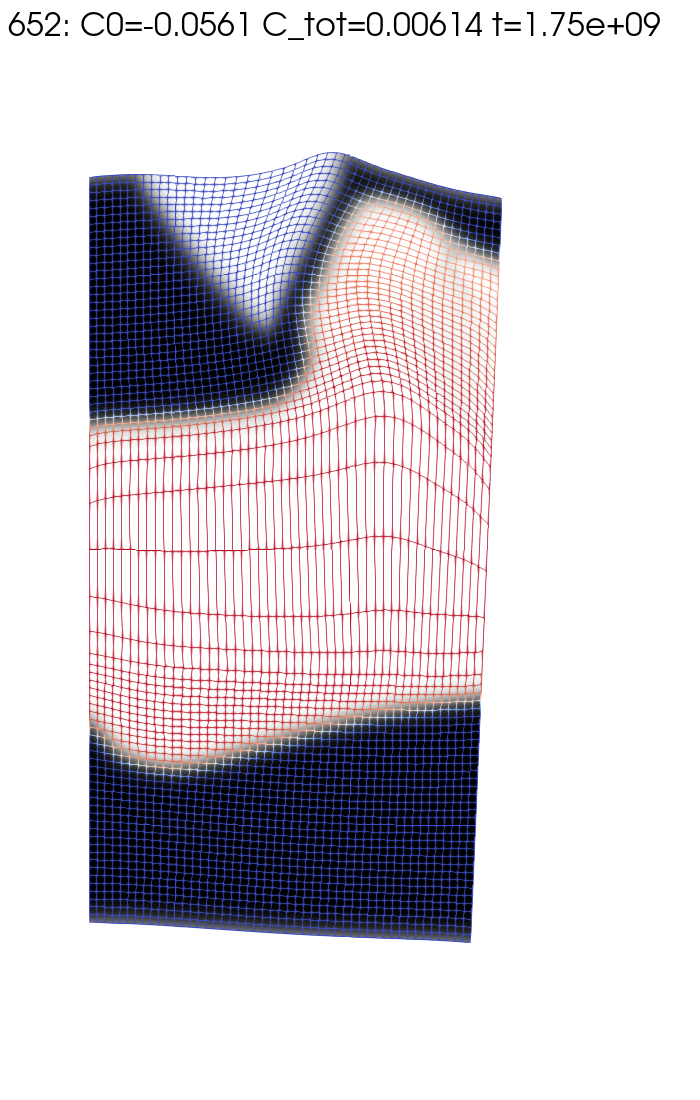}
      }
      \put(-90,-10){\small deformed}
    \end{overpic} 
  & \includegraphics[width=0.11\textwidth,trim=36mm 21mm 38mm 21mm,clip]
                    {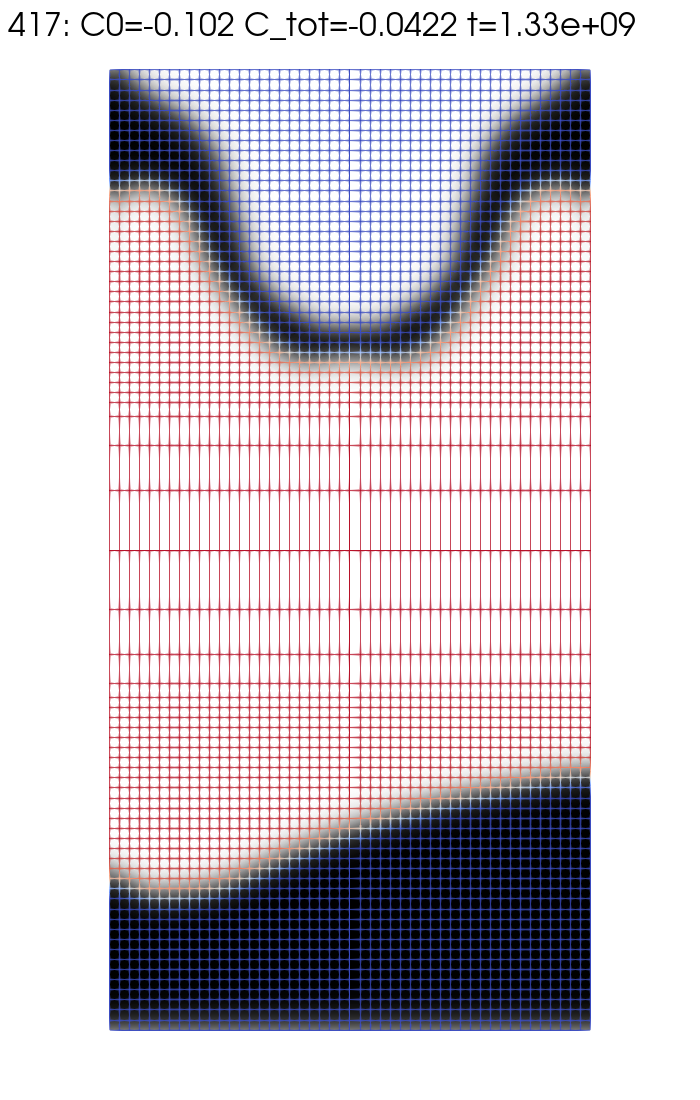} 
  & \includegraphics[width=0.11\textwidth,trim=36mm 21mm 38mm 21mm,clip]
                    {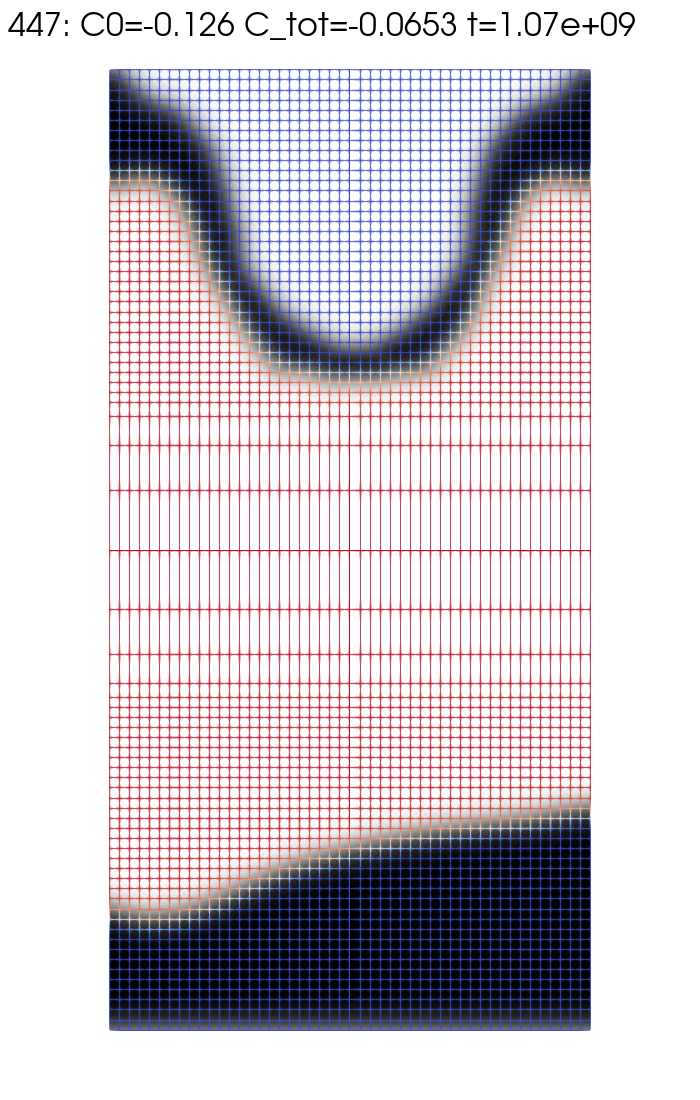}
  & \includegraphics[width=0.11\textwidth,trim=36mm 21mm 38mm 21mm,clip]
                    {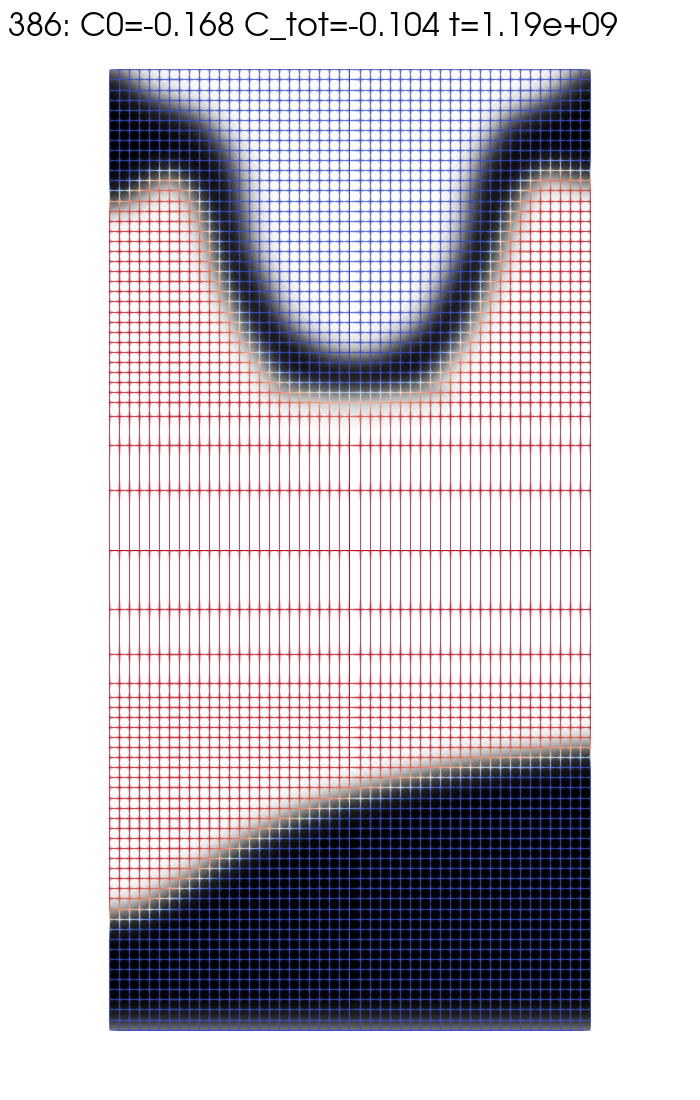} 
  &
  \\[0pt]
  & & 
  & \scriptsize -0.0185, 0.0287
  & \scriptsize -0.054, -0.0022
  & \scriptsize -0.158, -0.0923
  & \scriptsize -0.208, -0.14
  & \scriptsize -0.254, -0.181
  \\[-2pt]
\put(0,40) {\rotatebox{90}{$0.04$}} 
  & &
  & \includegraphics[width=0.11\textwidth,trim=36mm 21mm 38mm 21mm,clip]
                    {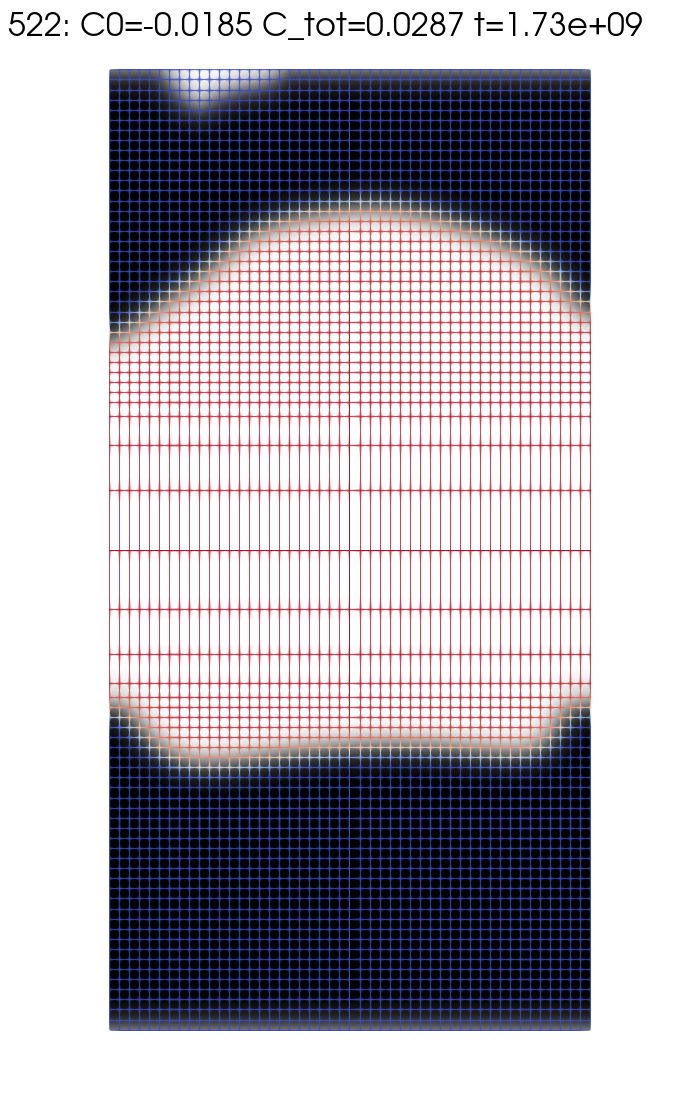}
  & \includegraphics[width=0.11\textwidth,trim=36mm 21mm 38mm 21mm,clip]
                    {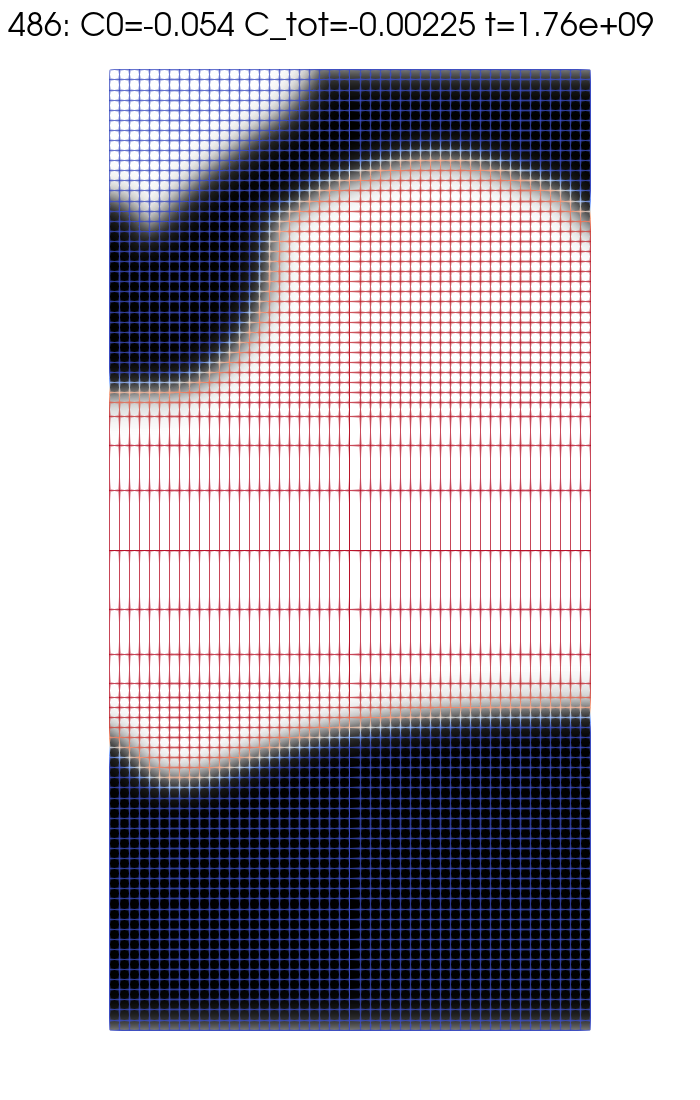} 
  & \includegraphics[width=0.11\textwidth,trim=36mm 21mm 38mm 21mm,clip]
                    {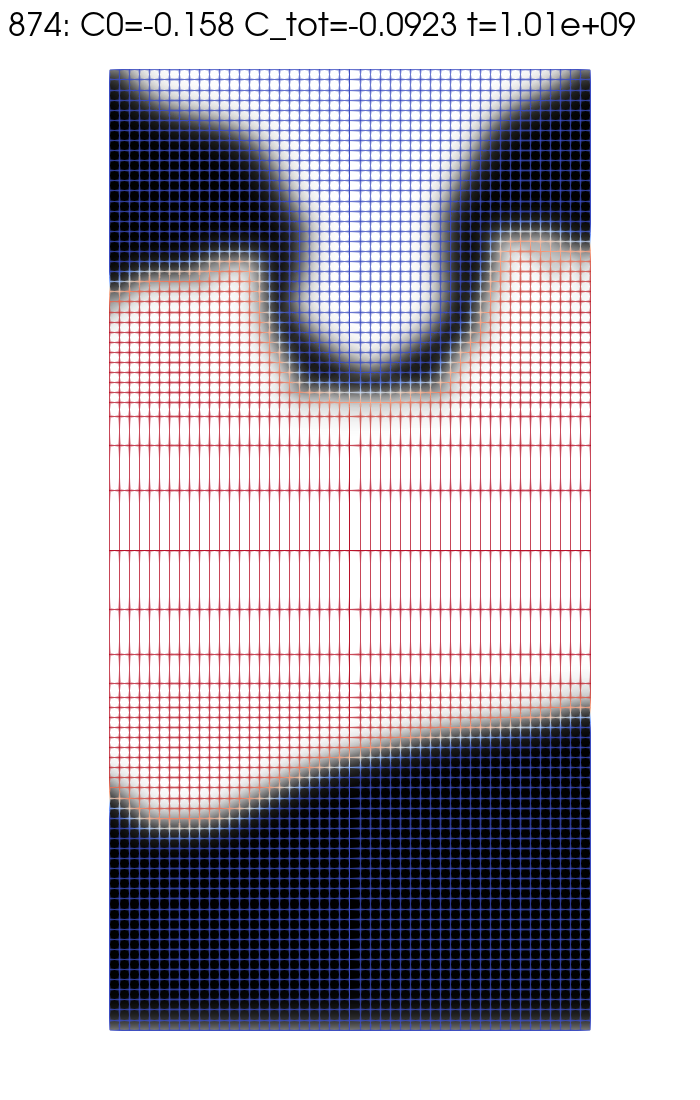}
  & \begin{overpic}[width=0.11\textwidth,trim=36mm 21mm 38mm 21mm,clip]
                  {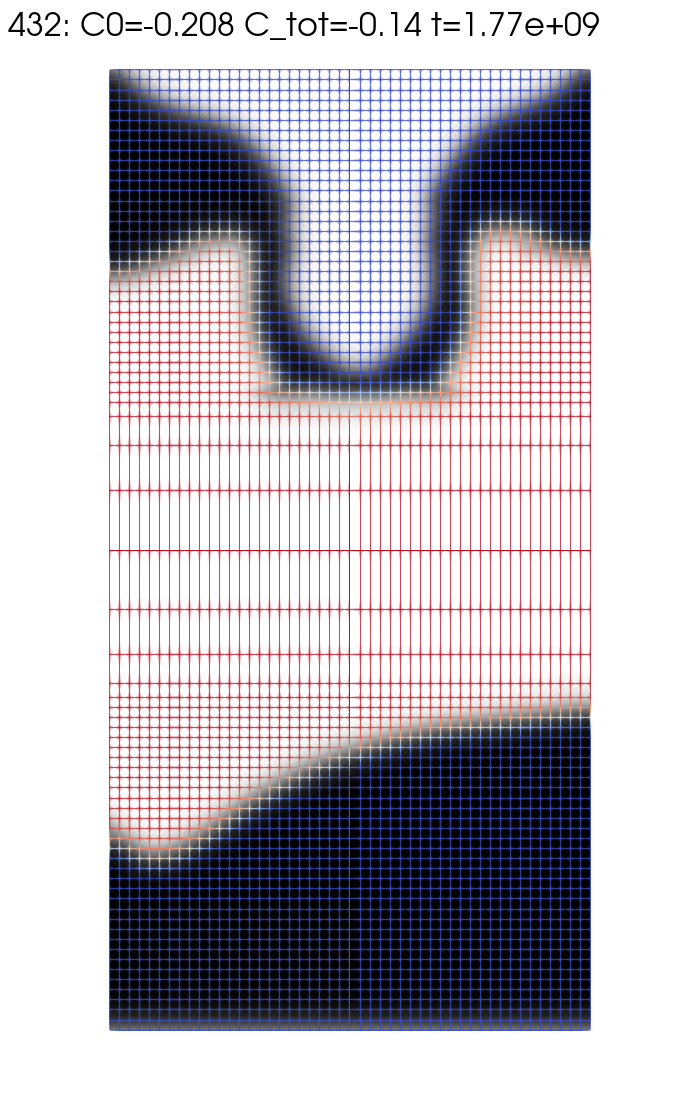}
      \put(25,75){\vector(2,3){57}}
      \put(55,160){
        \includegraphics[scale=0.14,trim=2.5cm 4.5cm 2.4cm 2.0cm,clip]
                          {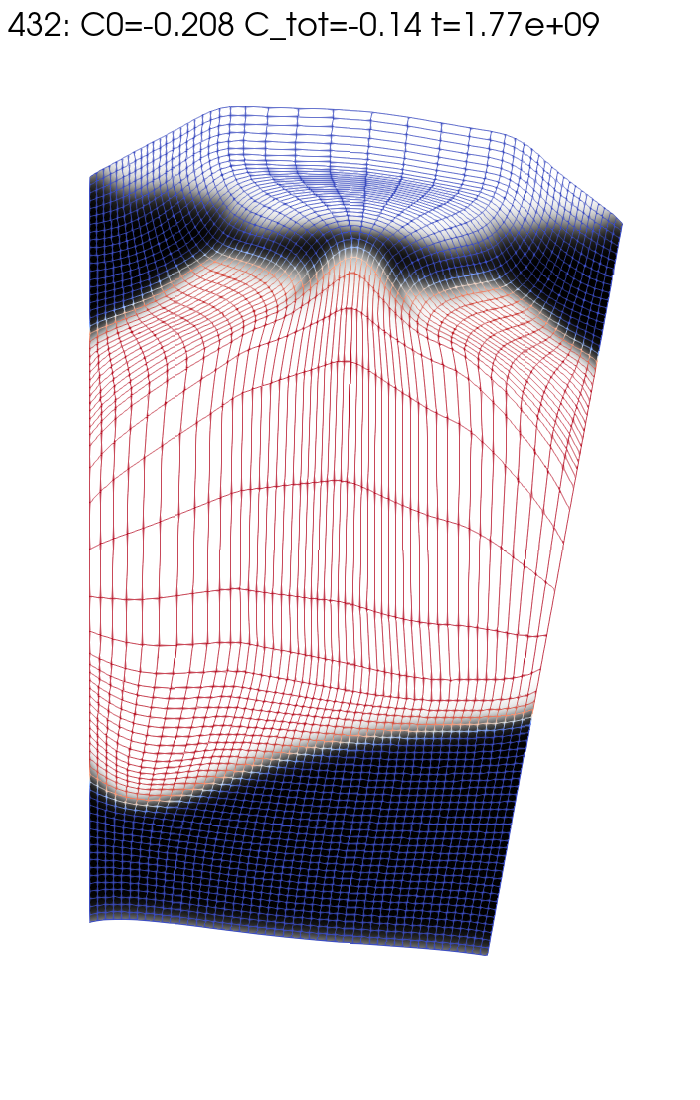}
      }
      \put(70,280){\small deformed}
    \end{overpic}
  & \includegraphics[width=0.11\textwidth,trim=36mm 21mm 38mm 21mm,clip]
                    {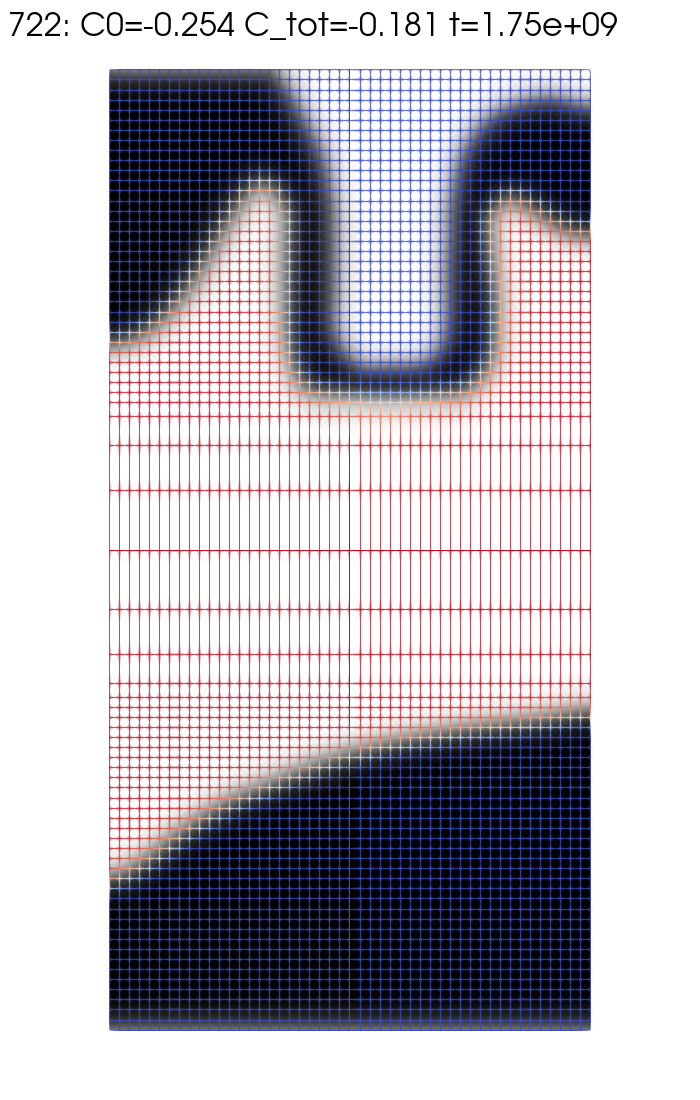}  
\end{tabular}
\caption{Optimized soft actuator designs for different levels of actuation pressure $p_\mathrm{in}$ and strain energy density constraint $\Psi_\mathrm{lim}$.
The values above each undeformed design correspond to the main objective $C_0$ and the overall objective function $C$, respectively.}
\label{fig:param_study_p_Wlim}
\end{figure}

Figure~\ref{fig:param_study_p_Wlim} presents results from a parametric study with respect to the actuation pressure $p_\mathrm{in}$ and the maximum allowable strain energy density $\Psi_\mathrm{lim}$.
The five rows of images cover a pressure range between $0.1\%$ and $4\%$ of the solid material initial Young's modulus~$E$, while the seven columns of images cover allowable strain energy densities corresponding to strain levels between 3\% and 68\%.
For the convenience of the reader, allowable strain energy densities $\Psi_\mathrm{lim}$ are reported in terms of the dimensionless quantity $\sqrt{2\Psi_\mathrm{lim}/E}$ which is easier to interpret as a linearly approximated allowable strain value.
All other parameters are kept fixed at their default values according to Table~\ref{tab:model_params}.

In total, Figure~\ref{fig:param_study_p_Wlim} reports 22 optimization results.
All respective designs are presented in their undeformed configuration, with an overlay mesh colored according to the pressure field~$p$, where blue corresponds to zero and red corresponds to $p\!=\!p_\mathrm{in}$.
Although, pressure fields are shown in the undeformed domain, they are still obtained based on the extended Darcy flow Eq.~\eqref{eq:Darcy_flow} which applies to the deformed configuration, as explained in the theory Section~\ref{sec:porohyperelasticity}.
For four selected cases, deformed configurations are also included in Figure~\ref{fig:param_study_p_Wlim} alongside the respective undeformed designs.

These results demonstrate how the finite strain capable optimization framework of this work yields distinct optimized designs depending on the actuation pressure $p_\mathrm{in}$, which is in contrast to small strain frameworks \cite{2020Ku-Fr-La,2021Ku-La}.
Moreover, the results also demonstrate how essential the inclusion of a strain energy density constraint is, especially for cases of relatively high actuation pressure.
For the highest pressure $p_\mathrm{in}\!=\!0.04E$, for example, the optimized designs exhibit very large deformations when the strain energy density constraint is relaxed by increasing $\Psi_\mathrm{lim}$.

\begin{figure}[!b]
\centering
\includegraphics[width=0.56\linewidth,trim=0cm 0.4cm 0cm 0.4cm,clip]
                {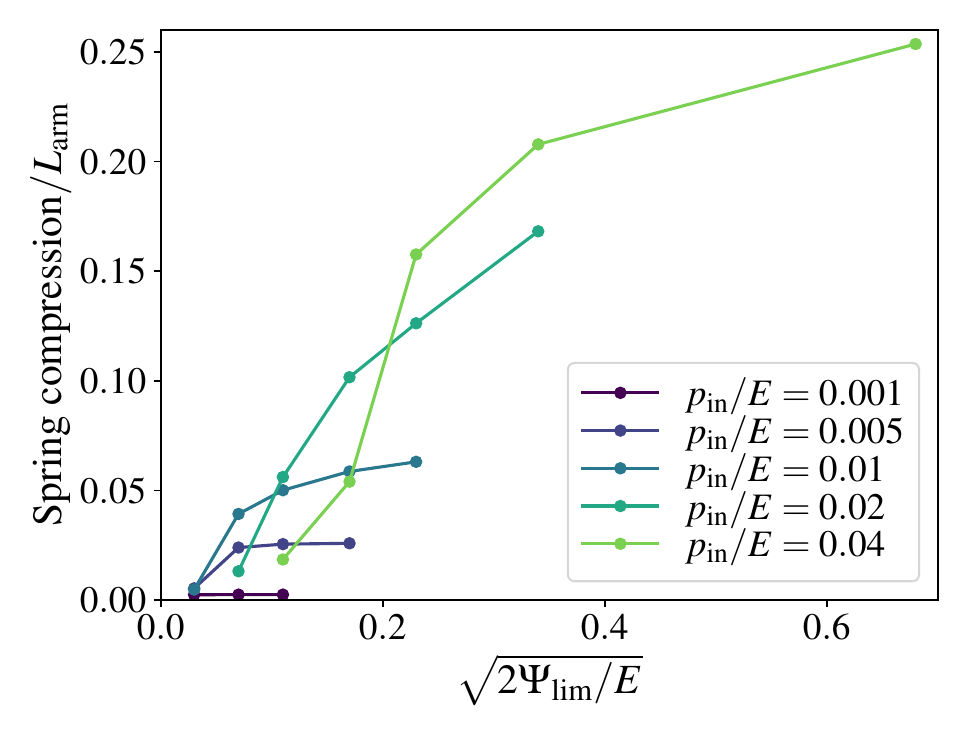}
\caption{Achieved spring compression for optimized designs under different actuation pressure and allowable strain energy density levels.}
\label{fig:spring_compression}
\end{figure}

The graph in Figure~\ref{fig:spring_compression} presents the achieved spring compression as a function of the allowable strain energy density $\Psi_\mathrm{lim}$, for different values of the actuation pressure $p_\mathrm{in}$.
Thereby it provides a good overview of the performance of all designs from Figure~\ref{fig:param_study_p_Wlim}.
As one would expect, for any given actuation pressure, increasing the allowable strain limit leads in general to larger spring compression.
All curves in Figure~\ref{fig:spring_compression} increase monotonically, apart from the entirely flat curve for the lowest actuation pressure, where the strain energy density constraint is simply not active at all.
Interestingly enough though, for a given level of allowable strain energy density, there seems to be some intermediate actuation pressure level that yields the largest spring compression.
For $\Psi_\mathrm{lim}\!=\!\tfrac{1}{2}\,0.07^2E$, for example, the structure designed to operate with $p_\mathrm{in}\!=\!0.01E$ performs best, while for $\Psi_\mathrm{lim}\!=\!\tfrac{1}{2}\,0.11^2E$ or $\Psi_\mathrm{lim}\!=\!\tfrac{1}{2}\,0.17^2E$, the best-performing actuation pressure level is $p_\mathrm{in}\!=\!0.02E$.
Of course, the optimized designs correspond most likely to local minima, nevertheless it is interesting to observe that the performance is not increasing monotonically with the applied actuation pressure.
Higher actuation pressure levels can only be exploited if correspondingly higher levels of deformation are allowed in the elastomer material.

\begin{figure}[!t]
\centering
\begin{tabular}{c c}
  \begin{overpic}[scale=0.27,trim=0mm -4mm 0mm -22mm,clip]
                 {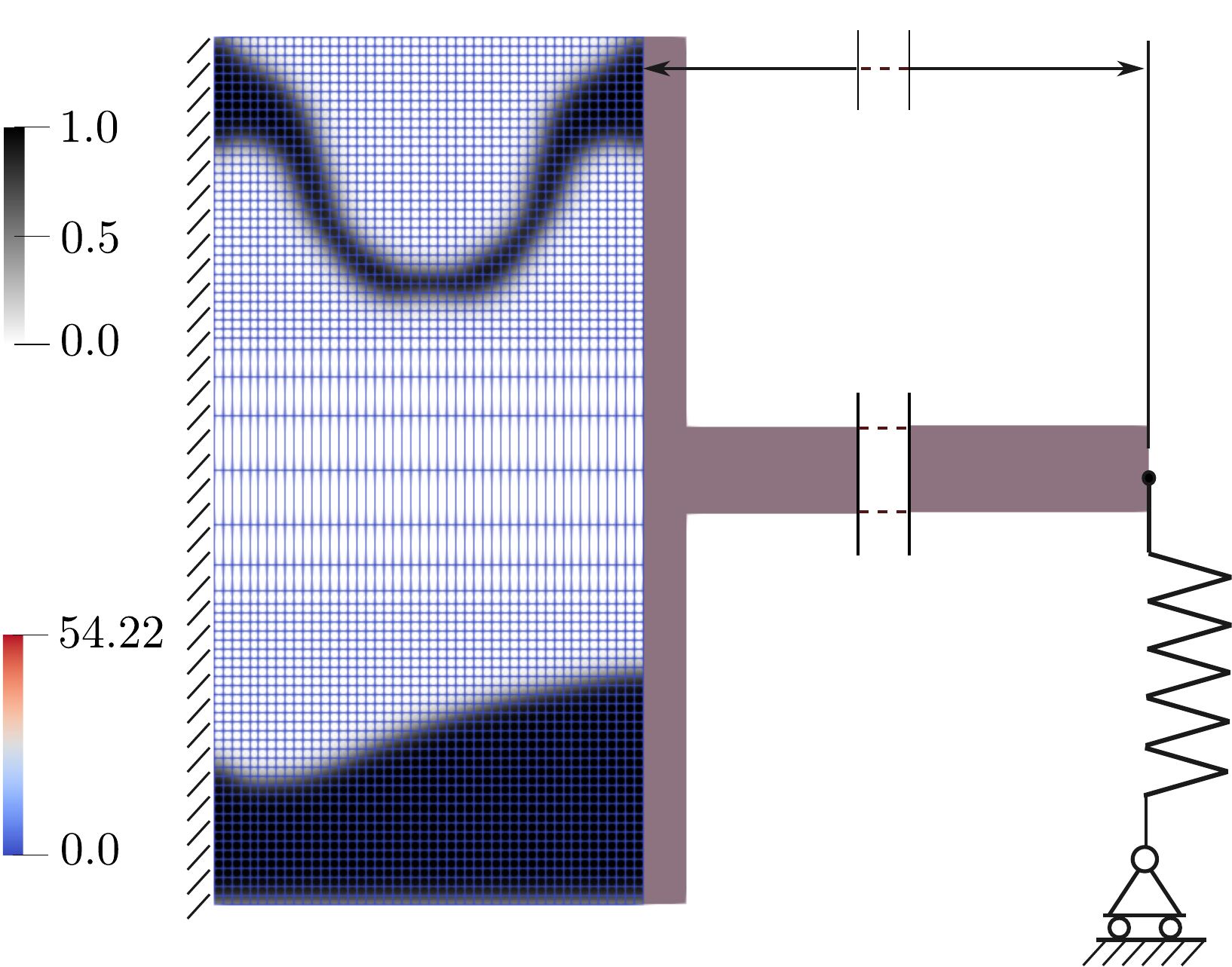}
    \put(66,79){$L_\mathrm{arm}$}
    \put(0,74){$\rho$}
    \put(-3,33){$p\,[\si{\kilo\pascal}]$}
  \end{overpic}
  & 
  \begin{overpic}[scale=0.27,trim=37mm 4mm 0cm 2mm,clip]
                 {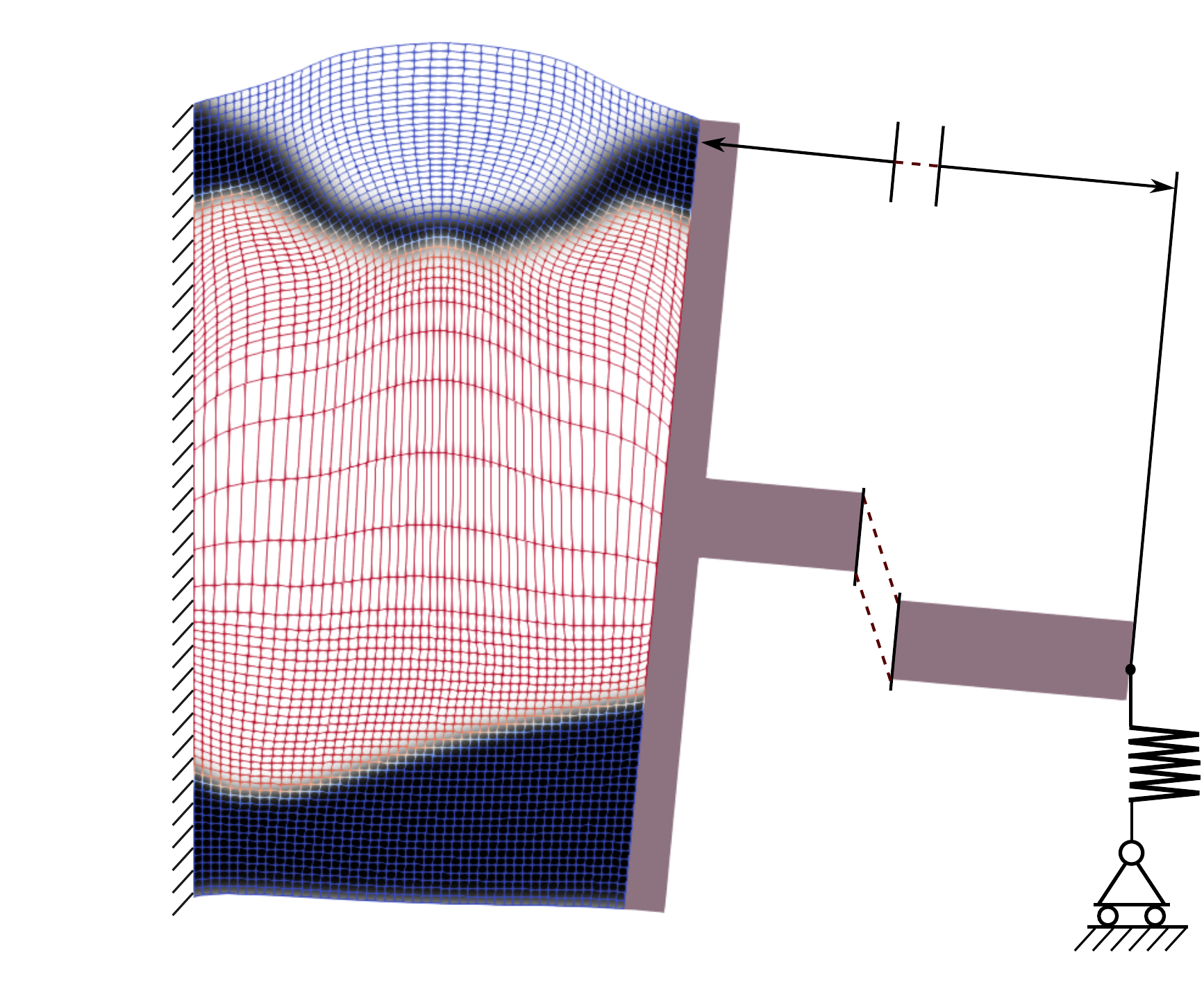}
    \put(68,82){$L_\mathrm{arm}$}
  \end{overpic}
\end{tabular}
\\[-1mm]
\includegraphics[width=0.65\textwidth,trim=0cm 0cm 0cm 0.8cm,clip]
                {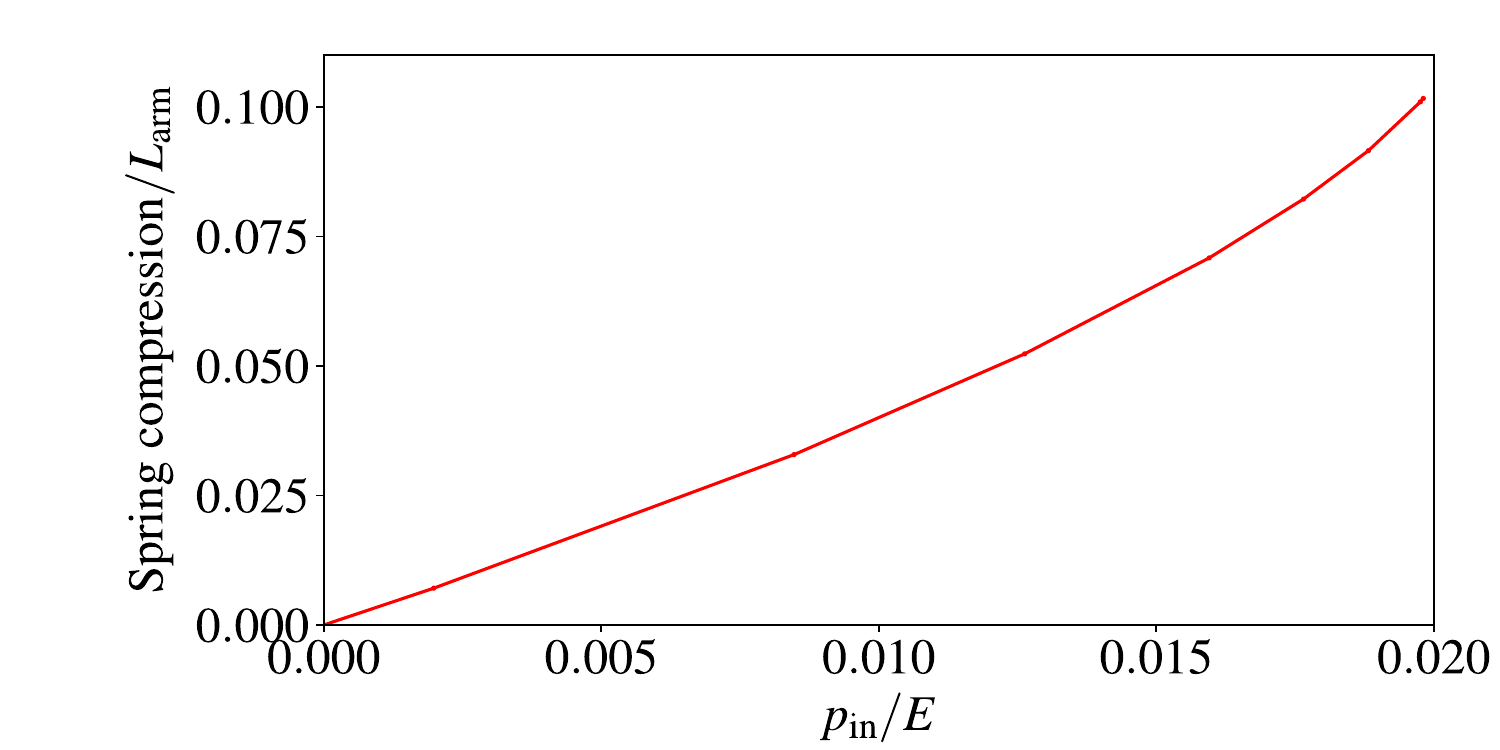}
\caption{Soft actuator design optimized for $p_\mathrm{in}\!=\!0.02 E$ and $\Psi_\mathrm{lim}\!=\!\tfrac{1}{2}\,0.17^2E$, shown in undeformed and deformed configurations. The graph shows spring compression as a function of actuation pressure in a post-simulation between $p_\mathrm{in}\!=\!0$ and $p_\mathrm{in}\!=\!0.02E$.}
\label{fig:opt_2D_design}
\end{figure}

Figure~\ref{fig:opt_2D_design} presents some additional details for one selected design out of the 22 cases of the parametric study, in particular for the design optimized for $p_\mathrm{in}\!=\!0.02E$ and $\Psi_\mathrm{lim}\!=\!\tfrac{1}{2}\,0.17^2E$.
The optimized design is shown in undeformed and deformed state, including the actuated rigid arm, in order to give a better visual impression of the benchmark result.
The graph at the bottom of Figure~\ref{fig:opt_2D_design}, shows how the spring compression increases as a function of the source pressure, when the latter is incremented in a post-simulation from zero to the target pressure, that the structure has been optimized for.
The system response is moderately nonlinear.
In general, it is expected that designs obtained with larger allowable strain levels, will also exhibit stronger nonlinear responses.

\subsubsection*{Convergence history}
\begin{figure}[!t]
\centering
\begin{overpic}[abs,scale=0.6,trim=1cm -8cm 0cm -6.5cm]
               {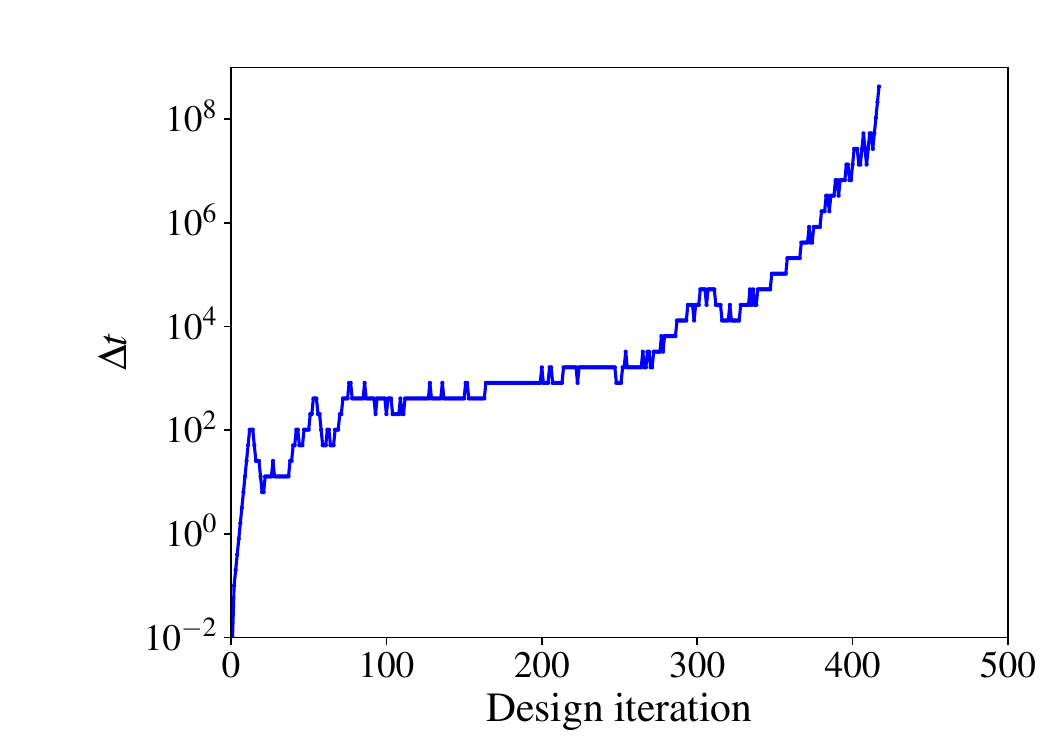}
  \put(-20,345){
    \includegraphics[abs,scale=0.16,trim=25cm 1cm 29cm 3cm,clip]
                    {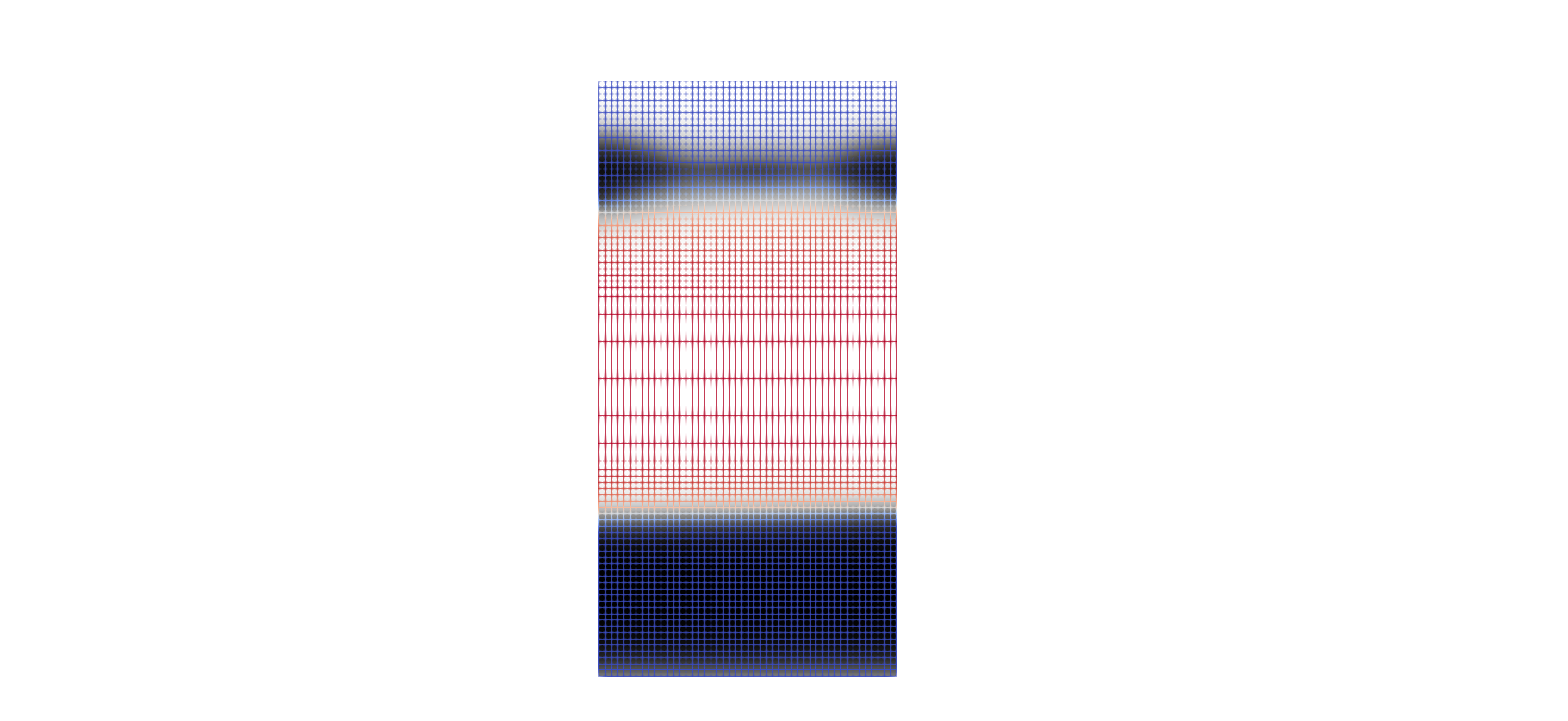}
  }
  \put(60,345){
    \includegraphics[abs,scale=0.16,trim=25cm 1cm 29cm 3cm,clip]
                    {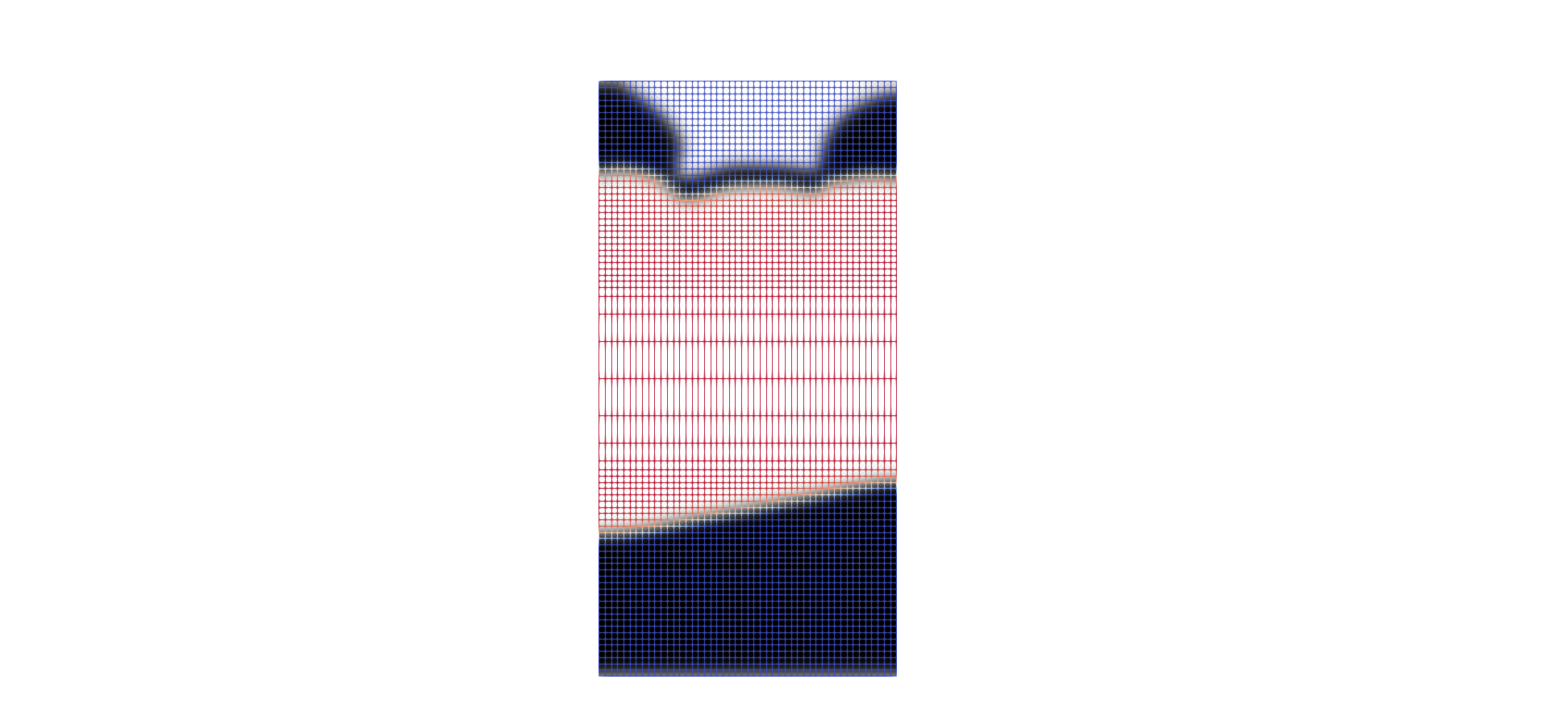}
  }
  \put(140,345){
    \includegraphics[abs,scale=0.16,trim=25cm 1cm 29cm 3cm,clip]
                    {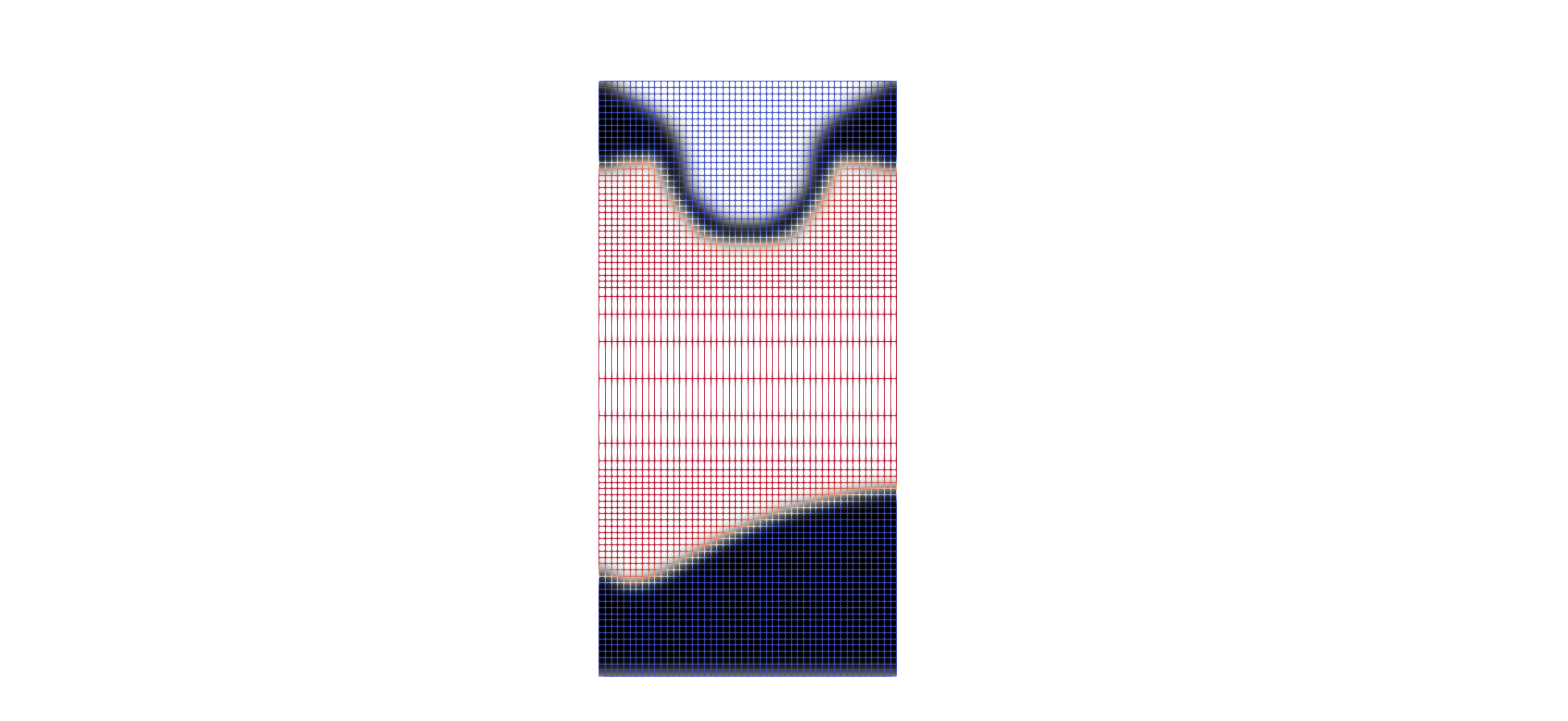}
  }
  \put(220,345){
    \includegraphics[abs,scale=0.16,trim=25cm 1cm 29cm 3cm,clip]
                    {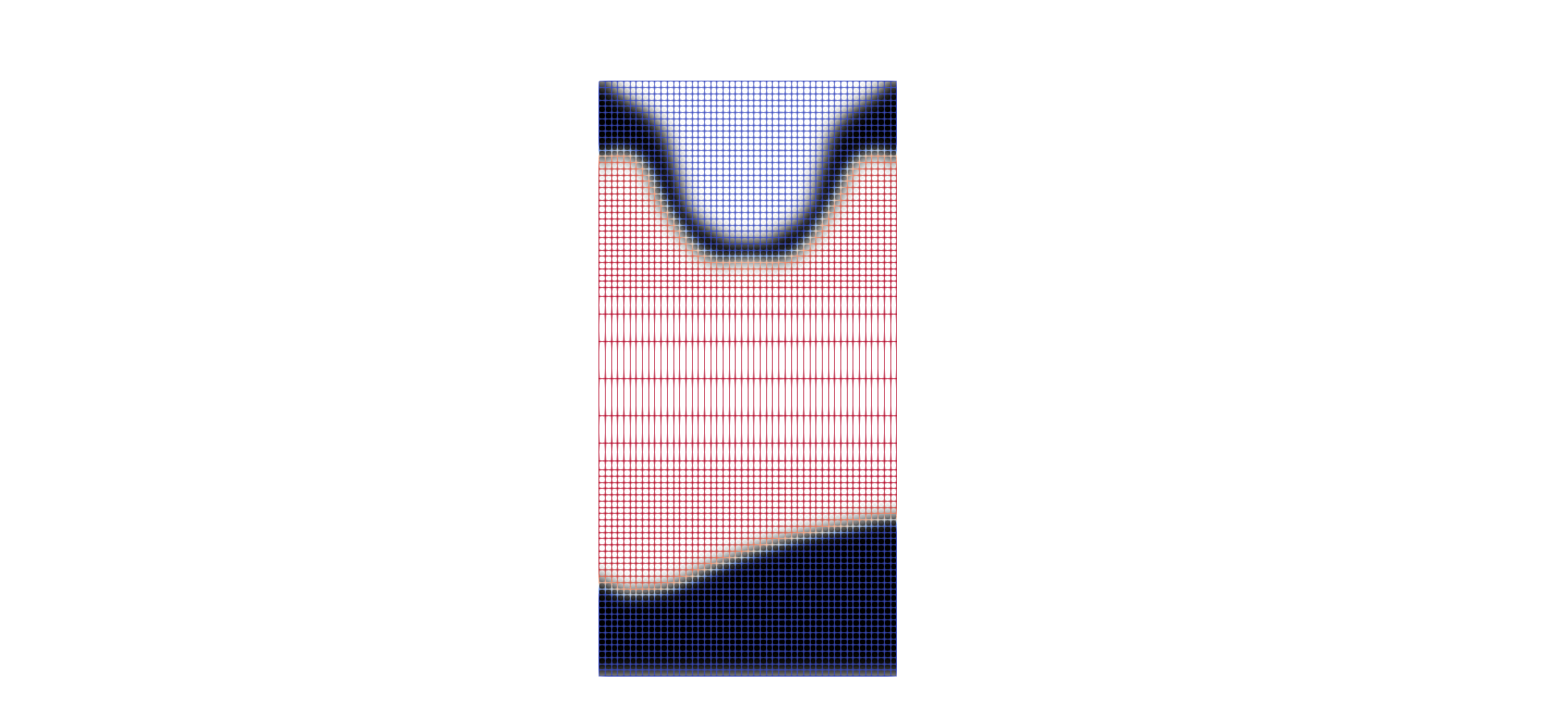}
  }
  \put(-20,0){
    \includegraphics[abs,scale=0.16,trim=25cm 1cm 26cm 1cm,clip]
                    {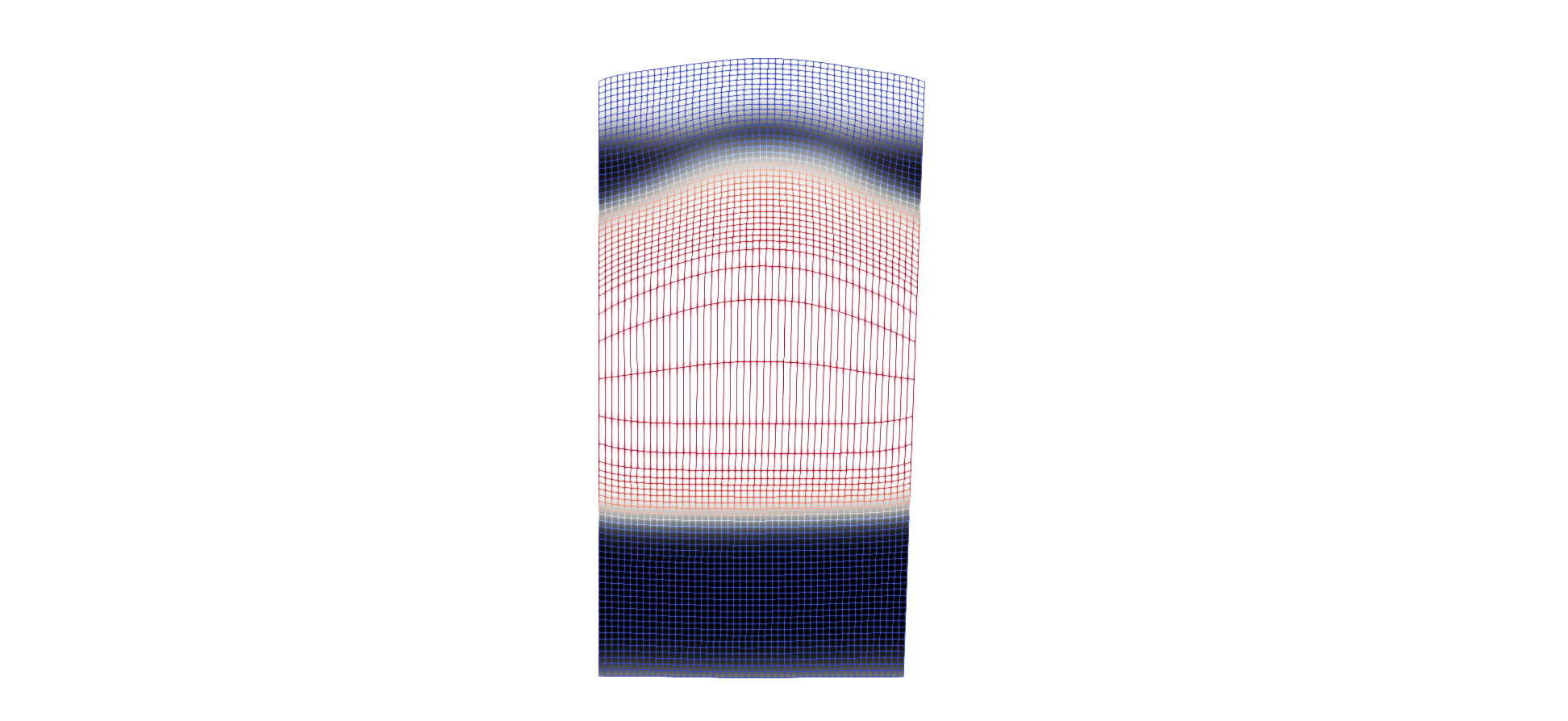}
  }
  \put(60,0){
    \includegraphics[abs,scale=0.16,trim=25cm 1cm 26cm 1cm,clip]
                    {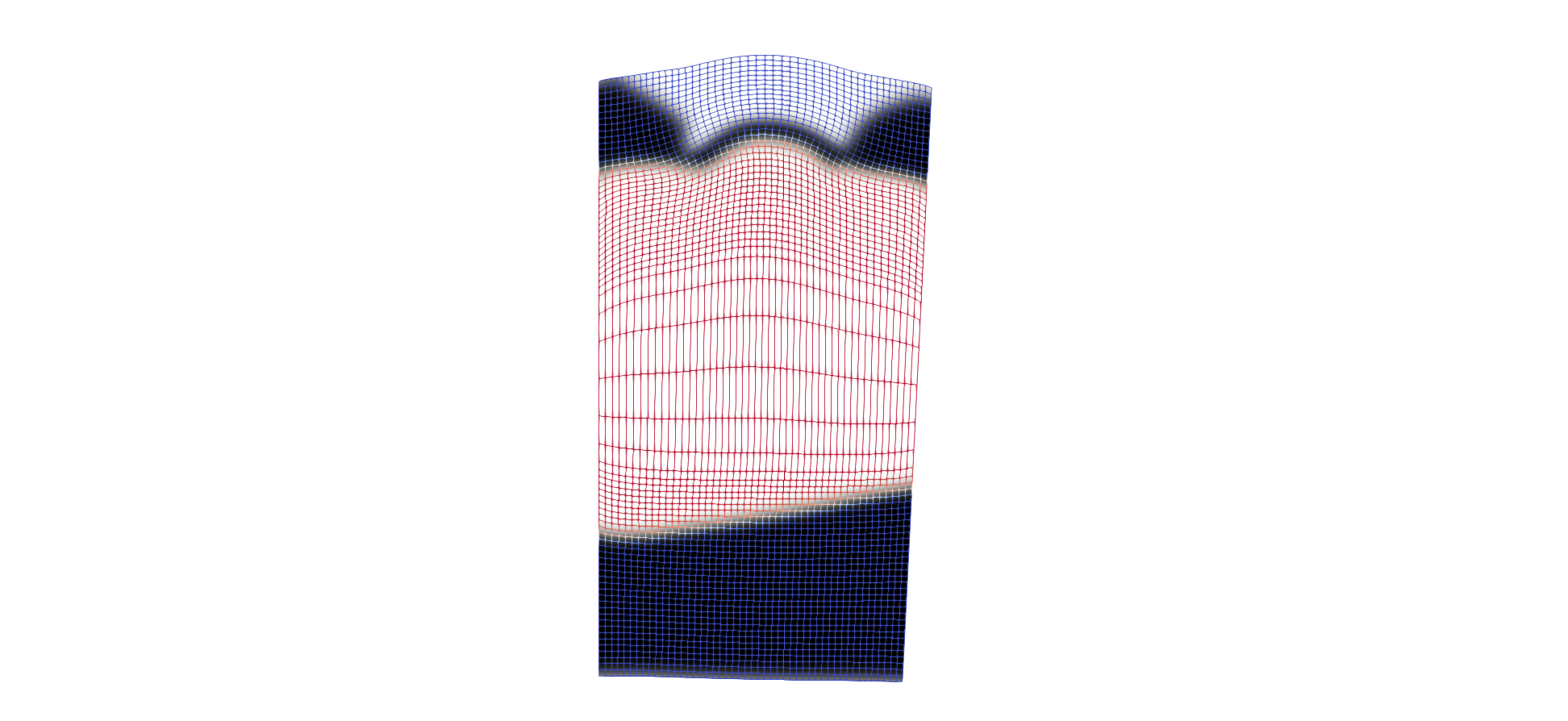}
  }
  \put(140,0){
    \includegraphics[abs,scale=0.16,trim=25cm 1cm 26cm 1cm,clip]
                    {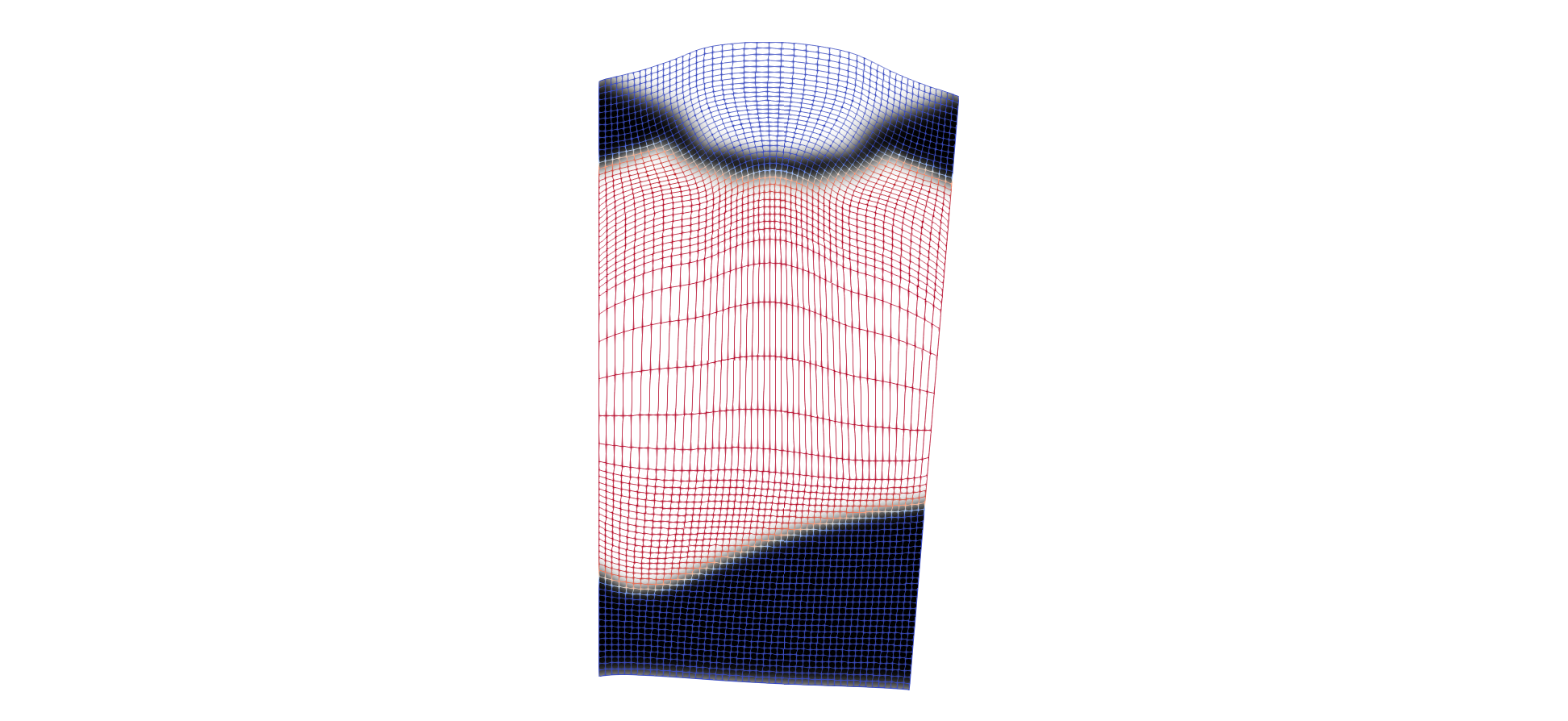}
  }
  \put(220,0){
    \includegraphics[abs,scale=0.16,trim=25cm 1cm 26cm 1cm,clip]
                    {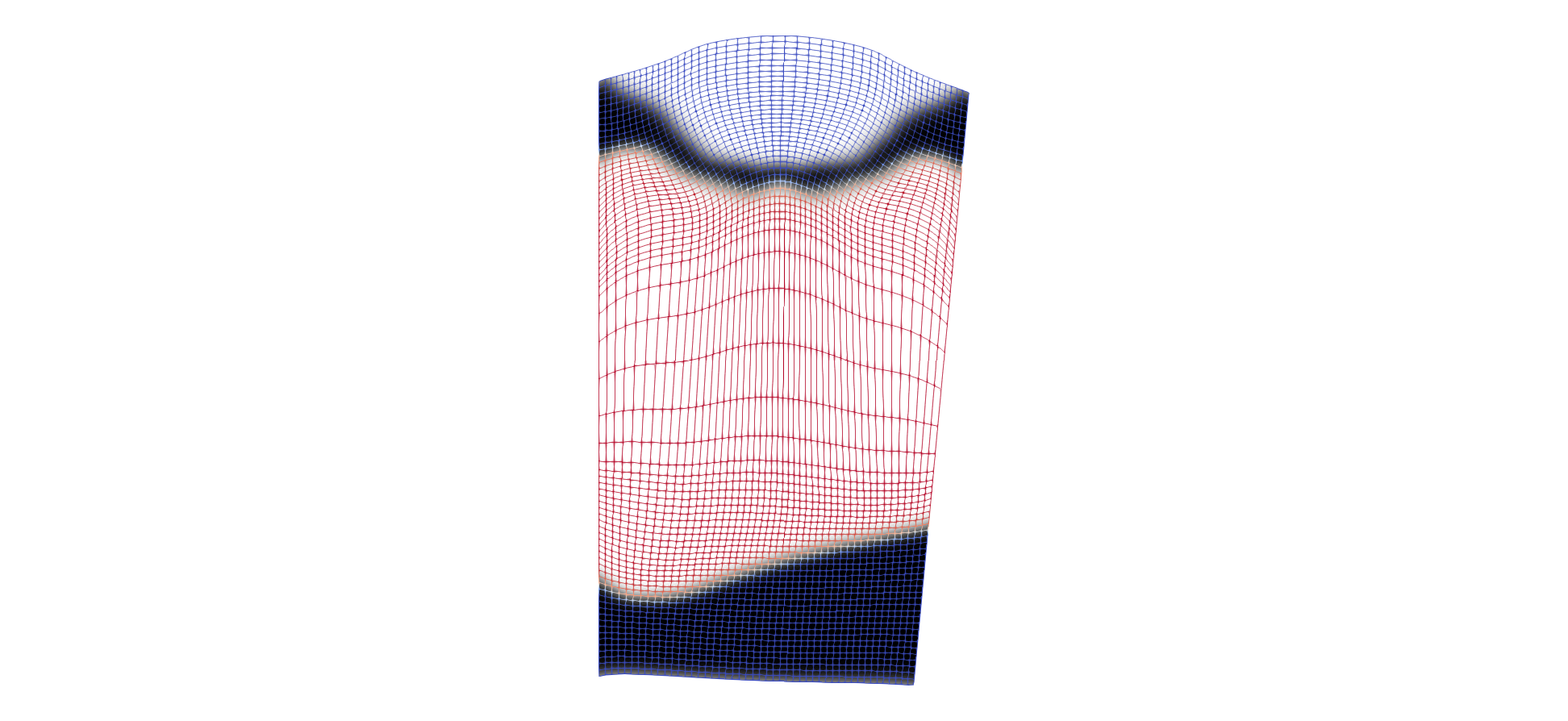}
  }
  \put(72.25,169){\color{green} \line(0,1){165}}
  \put(72.25,169){\color{green} \vector(-1,-1){40}}
  \put(72.25,334){\color{green} \vector(-2,1){22}}
  \put(94.5,169){\color{green} \line(0,1){165}}
  \put(94.5,149){\color{green} \vector(0,-1){22}}
  \put(94.5,334){\color{green} \vector(0,1){10}}
  \put(139.5,169){\color{green} \line(0,1){165}}
  \put(151,141){\color{green} \vector(1,-1){12}}
  \put(139.5,334){\color{green} \vector(3,1){25}}
  \put(207,169){\color{green} \line(0,1){165}}
  \put(207,169){\color{green} \vector(1,-1){40}}
  \put(207,334){\color{green} \vector(3,1){25}}
\end{overpic}
\caption{Evolution of the design pseudo-time step $\Delta t$ during the optimization and snapshots of the design at different optimization increments (50, 100, 200, 350).}
\label{fig:convergence}
\end{figure}

For the same representative case discussed above, Figure~\ref{fig:convergence} provides further details about the optimization convergence history.
The graph shows, on a logarithmic scale, how the pseudo-time step $\Delta t$ evolves during the 417 performed design iterations.
The higher $\Delta t$, the weaker the damping applied to the optimality condition for the respective design increment.
For most of the optimization history, $\Delta t$ remains within a limited range.
After approximately 350 design iterations, the current solution is already so close to the optimality point, that the overall system of nonlinear equations requires no more than a few Newton-Raphson iterations to converge.
Consequently, the algorithm from Figure~\ref{fig:dt_update_scheme} leads to a rapidly increasing $\Delta t$, until the temporary damping is practically eliminated from the optimality condition.

\begin{figure}[!t]
\centering
\begin{overpic}[scale=0.55,trim=0cm 0cm -3cm 0cm]
               {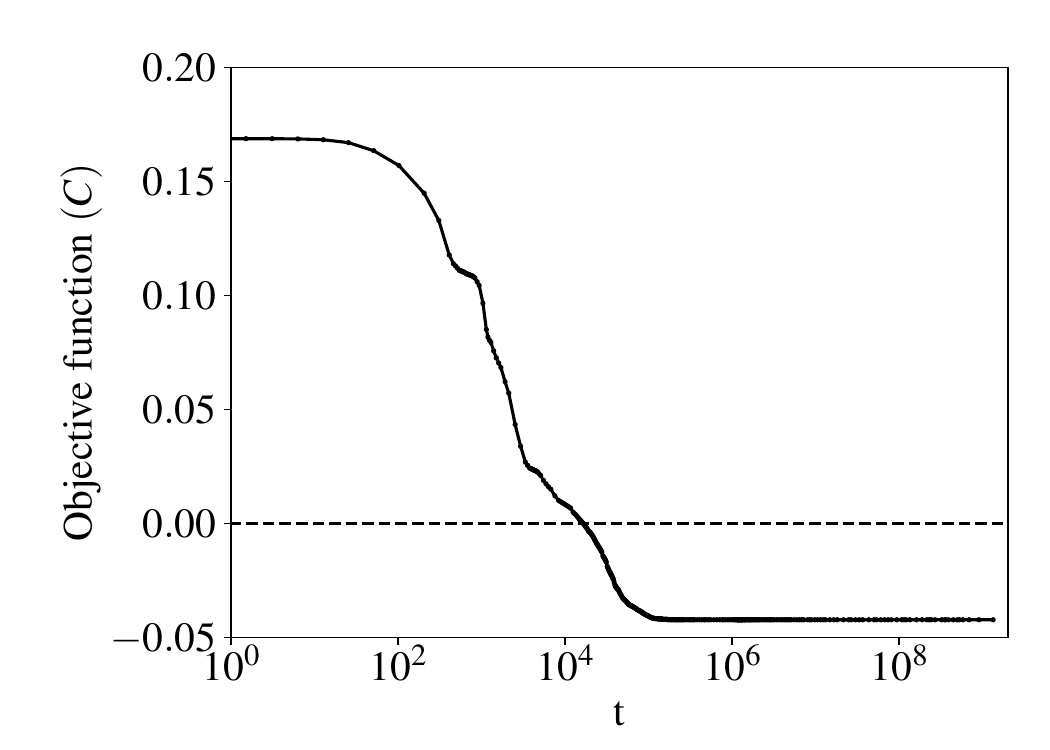}
\end{overpic}
\\[-6mm]
\setlength{\tabcolsep}{4pt}
\begin{tabular}{c c}
  \begin{overpic}[abs,scale=0.46,trim=0.5cm 0.1cm 0.4cm -0.7cm,clip]
                 {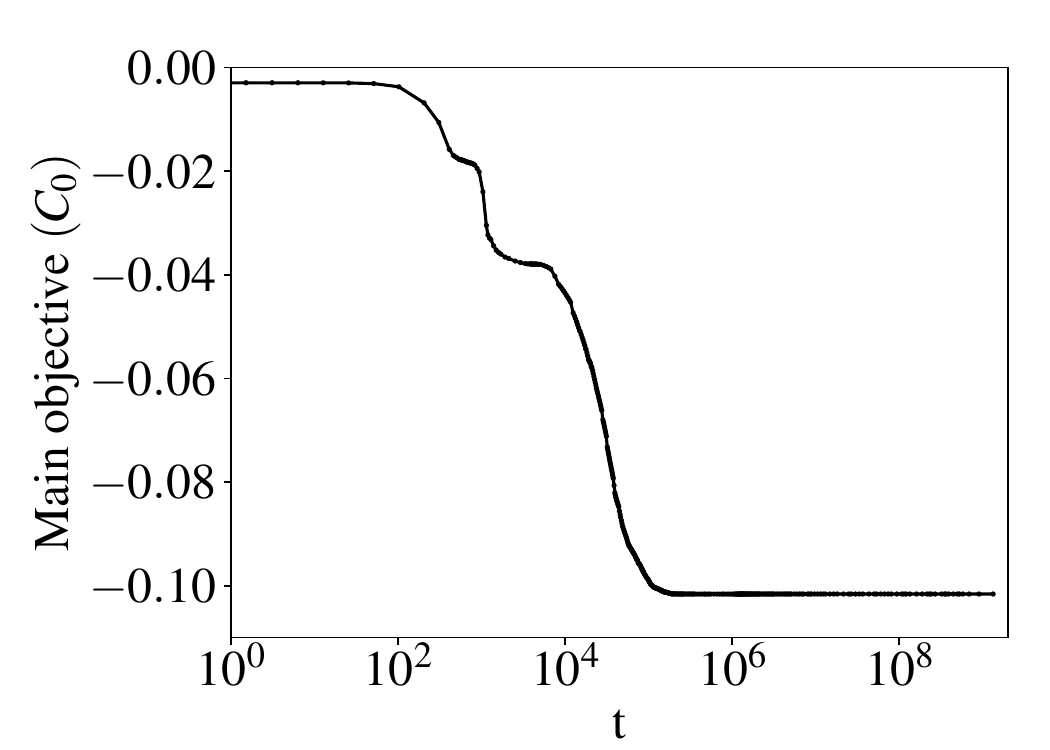}
  \end{overpic}
  &
  \begin{overpic}[abs,scale=0.46,trim=1.25cm 0.1cm 0cm -0.7cm,clip]
                 {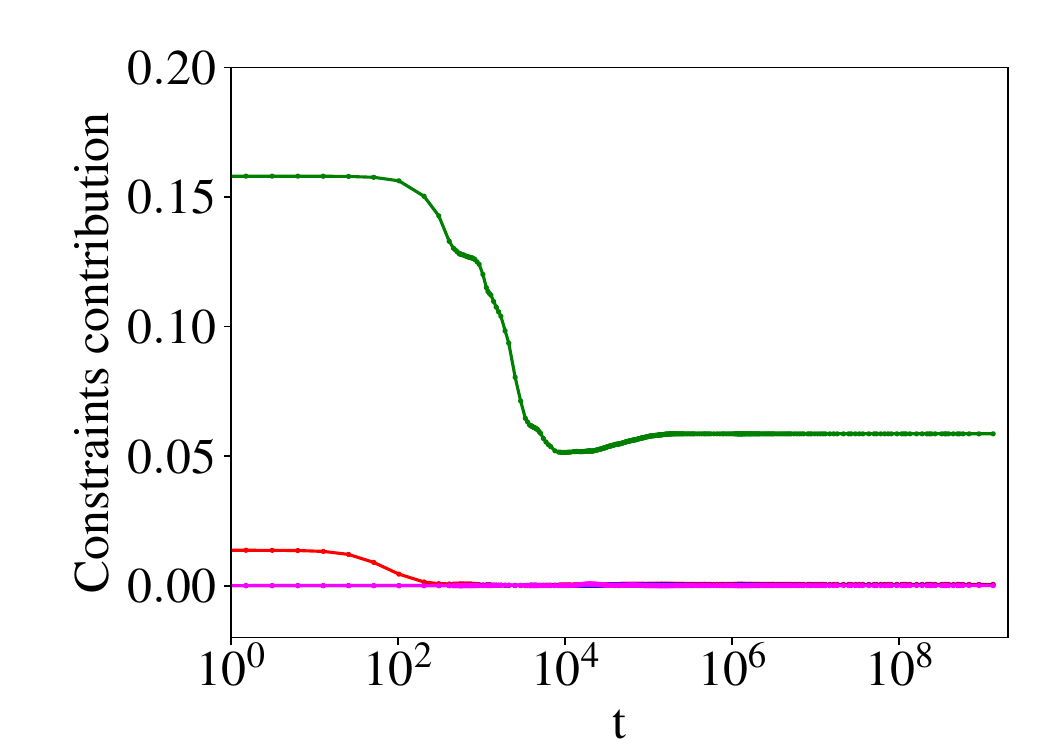}
    \put(100,80){
    \includegraphics[scale=0.3,trim=30mm 7mm 6mm 0cm,clip]
                    {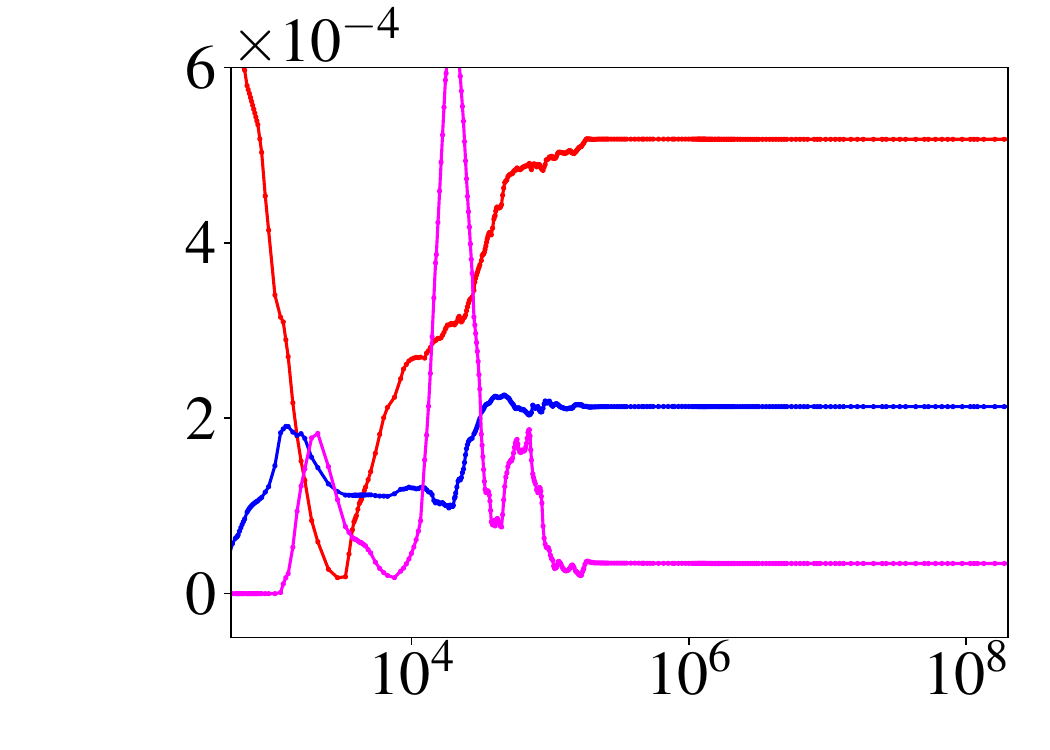}
    }
    \put(50,52){\textcolor{red}{$C_i$}}
    \put(82,62){\textcolor{darkgreen}{$C_A$}}
    \put(165,135){\textcolor{blue}{$C_\Psi$}}
    \put(165,110){\textcolor{magenta}{$C_p$}}
  \end{overpic}
\end{tabular}
\caption{Evolution of the total objective $C$, the main objective $C_0$, and the individual constraint contributions $C_A$, $C_i$, $C_p$, $C_\Psi$ to the total objective, as a function of pseudo-time.}
\label{fig:C_0_and_constraints}
\end{figure}

Figure~\ref{fig:C_0_and_constraints} presents the evolution of the objective function during the optimization, as a function of pseudo-time $t$, shown on a logarithmic scale.
The Hessian and gradient-based optimization algorithm ensures a monotonically decreasing total objective value until the optimality condition is achieved.
Furthermore, Figure~\ref{fig:C_0_and_constraints} also depicts the individual contributions in the total objective function, namely the main objective $C_0$, the surface area penalization $C_A$, and the penalized constraints $C_i$, $C_p$, and $C_\Psi$.
The primary contribution of $C_0$ arises from the spring compression, making it a dominating factor for the overall objective, together with the penalization $C_A$ of the total surface area of the design.
In terms of multi-objective optimization, one should imagine a typical Pareto front between $C_0$ and $C_A$.
As it will be demonstrated later, increasing the scaling of $C_A$ will lead to geometrically simpler designs at the cost of smaller spring compression.
On the contrary, increasing the scaling of all other contributions, $C_i$, $C_p$, and $C_\Psi$, has, after some point, a diminishing effect, because these penalized constraints are feasible.
The respective weights $c_i$, $c_p$, and $c_\Psi$, reported in Table~\ref{tab:model_params}, were chosen in a range where the respective Pareto fronts have flattened out.
Their exact values are therefore not as essential for the optimization problem as the scaling factor $c_A$.

The graph in the lower right corner of Figure~\ref{fig:C_0_and_constraints} shows that soon after the beginning of the optimization, all penalized constraints, $C_i$, $C_p$, and $C_\Psi$, are satisfied within good approximation.
The surface area penalization term $C_A$ is initially rather high, because of the larger portions of gray material in the non-converged design.
As the void-solid boundary becomes sharper, converging towards an interface thickness equal to $L_i$, the term $C_A$ begins to better represent the total surface area of the design.
Significant geometric changes in the design occur approximately until a pseudo-time $t\!\approx\!10^6$, reached after ca. 300 design iterations.
After that point, the curve for the total objective, as well as all curves for the individual contributions, flatten out, despite the large pseudo-time steps~$\Delta t$.

\subsubsection*{Effect of $c_A$, $k_\mathrm{sp}$, and $L_\mathrm{arm}$}
\begin{figure}[!h]
  \centering
  \raisebox{-0.5\height}{
    \begin{overpic}[scale=0.147,trim=36mm 0mm 38mm 21mm,clip]
                 {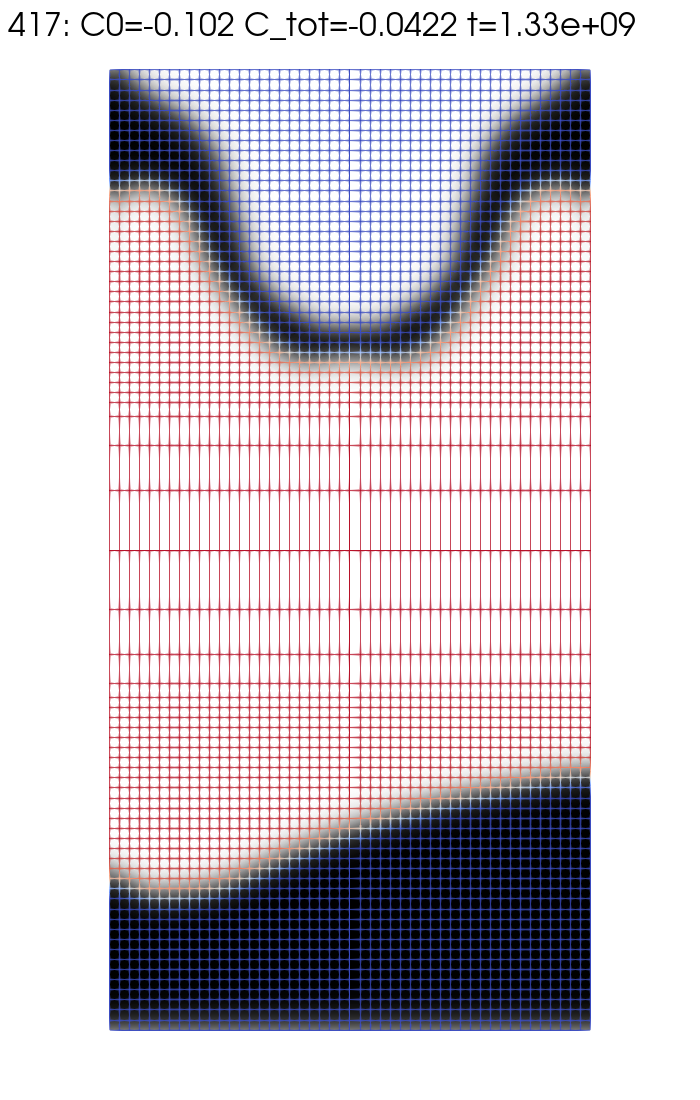}
  \put(10,0){$c_A\!=\!0.004$}
  \end{overpic}
  ~~
  \begin{overpic}[scale=0.147,trim=36mm 0mm 38mm 21mm,clip]
                 {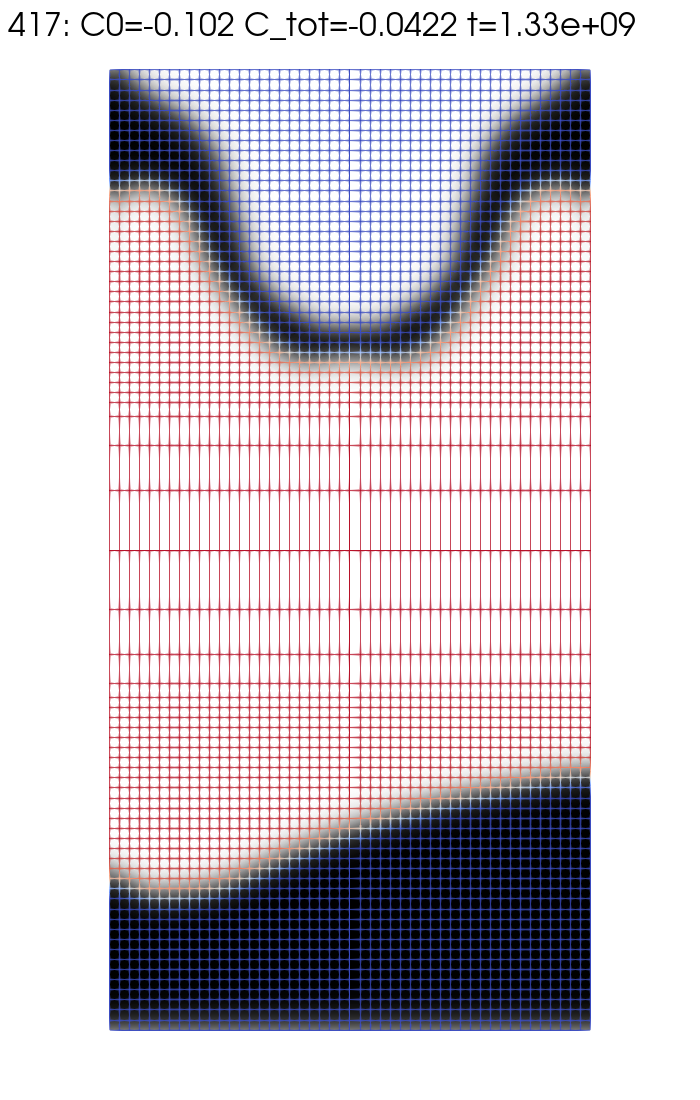}
  \put(10,0){$c_A\!=\!0.02$}
  \end{overpic}
  ~~
  \begin{overpic}[scale=0.147,trim=36mm 0mm 38mm 21mm,clip]
                 {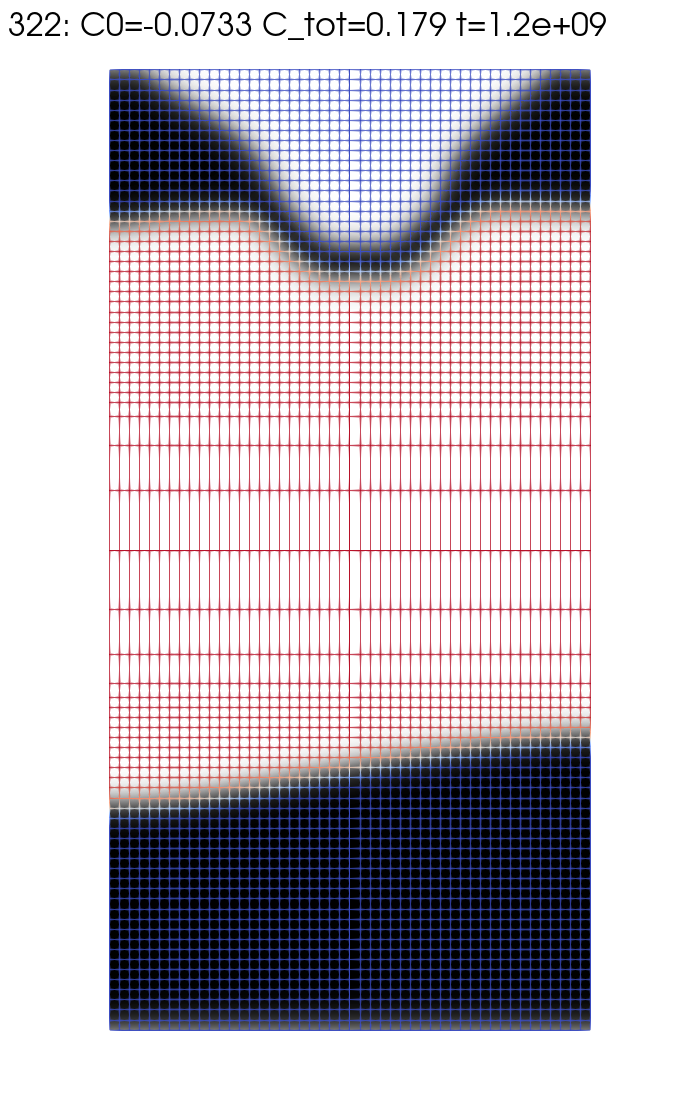}
  \put(10,0){$c_A\!=\!0.1$}
  \end{overpic}
  }
  ~
  \setlength{\tabcolsep}{5pt}
  \begin{tabular}{c|c|c|c}
    $c_A$ & $C_0$ & \parbox[c]{22mm}{\centering spring compression} &
                    \parbox[c]{13mm}{\centering surface area}\\[10pt]
    \hline
    & & \\[-10pt]
    0.04 & -0.1023 & \SI{6.14}{\milli\meter} & \SI{64.3}{\milli\meter}\\[3pt]
    0.02 & -0.1016 & \SI{6.10}{\milli\meter} & \SI{62.2}{\milli\meter}\\[3pt]
    0.1  & -0.0733 & \SI{4.40}{\milli\meter} & \SI{52.8}{\milli\meter}
  \end{tabular}
\caption{Effect of surface area penalization factor $c_A$ on the optimized design for $p_\mathrm{in}\!=\!0.02 E$ and $\Psi_\mathrm{lim}\!=\!\tfrac{1}{2}\,0.17^2E$.
For the 2D domain, surface area is reported in [\si{\milli\meter}] because it is per out-of-plane thickness.}
\label{fig:effect_c_A}
\end{figure}

Figure \ref{fig:effect_c_A} illustrates the effect of the estimated surface area penalization factor $c_A$ on the optimized design.
The design shown in the middle is the reference case discussed previously, obtained with the default value $c_A\!=\!0.02$.
The two designs shown to the left and right of the reference design were obtained with reduced $c_A\!=\!0.004$ and increased $c_A\!=\!0.1$, respectively.
Decreasing $c_A$ by a factor of five has a minimal effect, but increasing $c_A$ by a factor of five leads indeed, as expected, to a design with smaller surface area, at the cost of a larger $C_0$ value and smaller spring compression.
It is important to note that some minimal penalization of the total surface area of the design is mandatory for a well-posed optimization problem.
This is because, in the absence of any material volume constraint, clumps of disconnected solid material could be added arbitrarily, without any deterioration of the objective value.

\begin{figure}[!h]
\centering
\setlength{\tabcolsep}{3pt}
\begin{tabular}{c l l l}
  &
  $k_\mathrm{sp}\!=\!EH/2000$
  &
  $k_\mathrm{sp}\!=\!EH/400$
  & 
  $k_\mathrm{sp}\!=\!EH/80$\\
  \rotatebox[origin=c]{90}{undeformed}
  &
  \raisebox{-0.5\height}{
  \includegraphics[scale=0.147,trim=36mm 21mm 38mm 21mm,clip]
                  {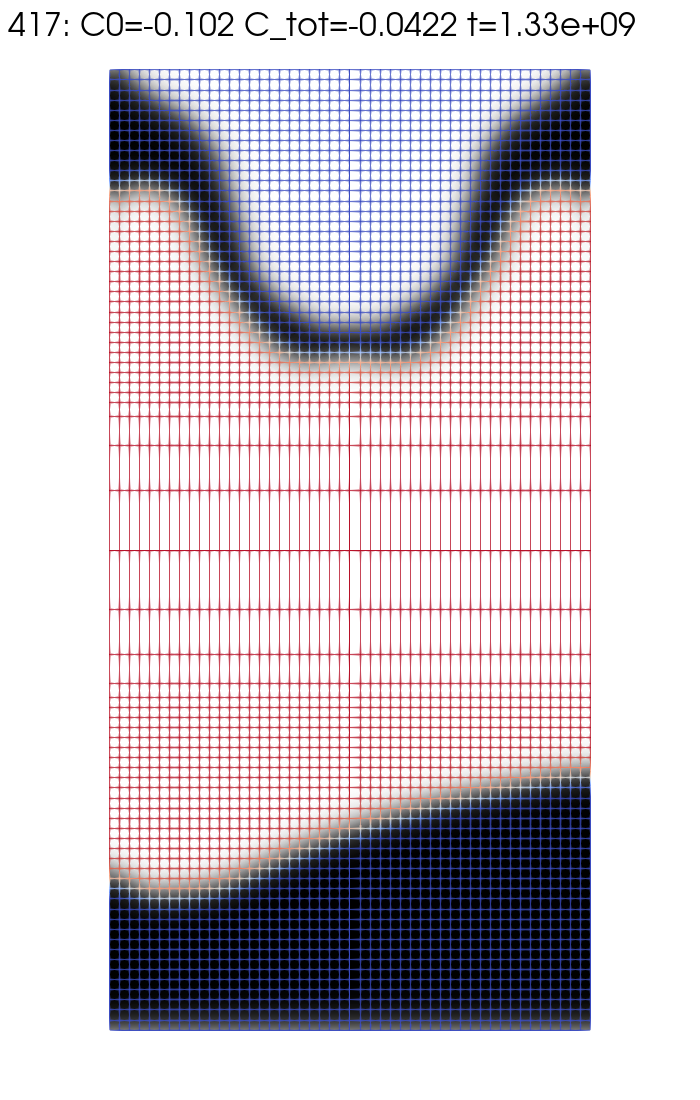}}
  &
  \raisebox{-0.5\height}{
  \includegraphics[scale=0.147,trim=36mm 21mm 38mm 21mm,clip]
                  {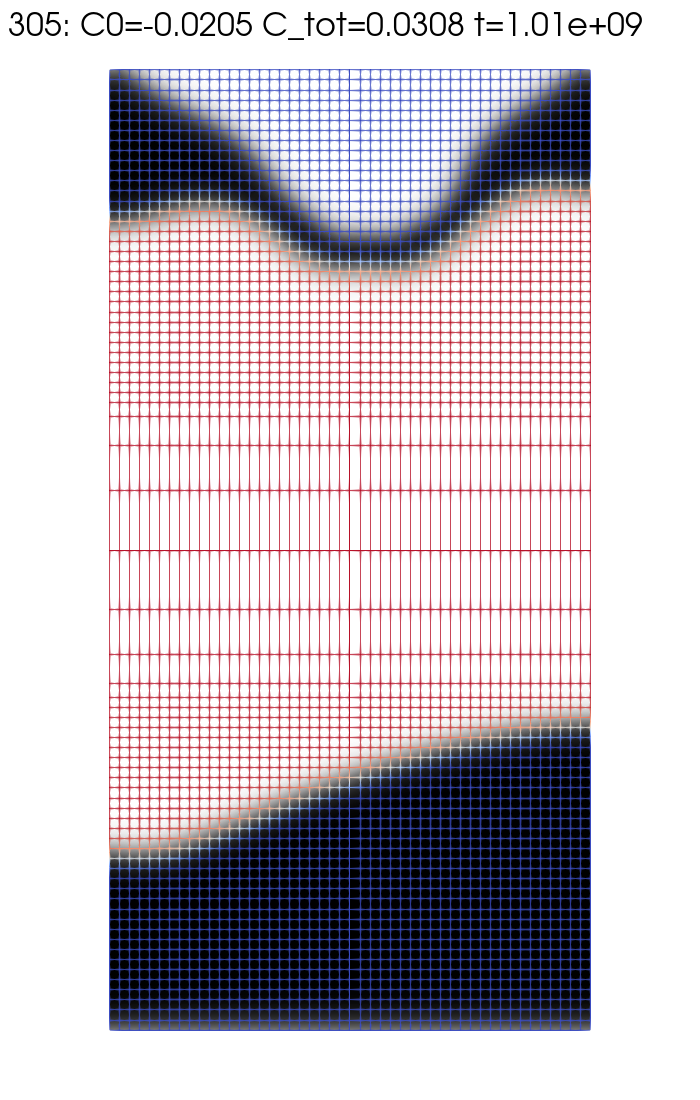}}
  &
  \raisebox{-0.5\height}{
  \includegraphics[scale=0.147,trim=36mm 21mm 38mm 21mm,clip]
                  {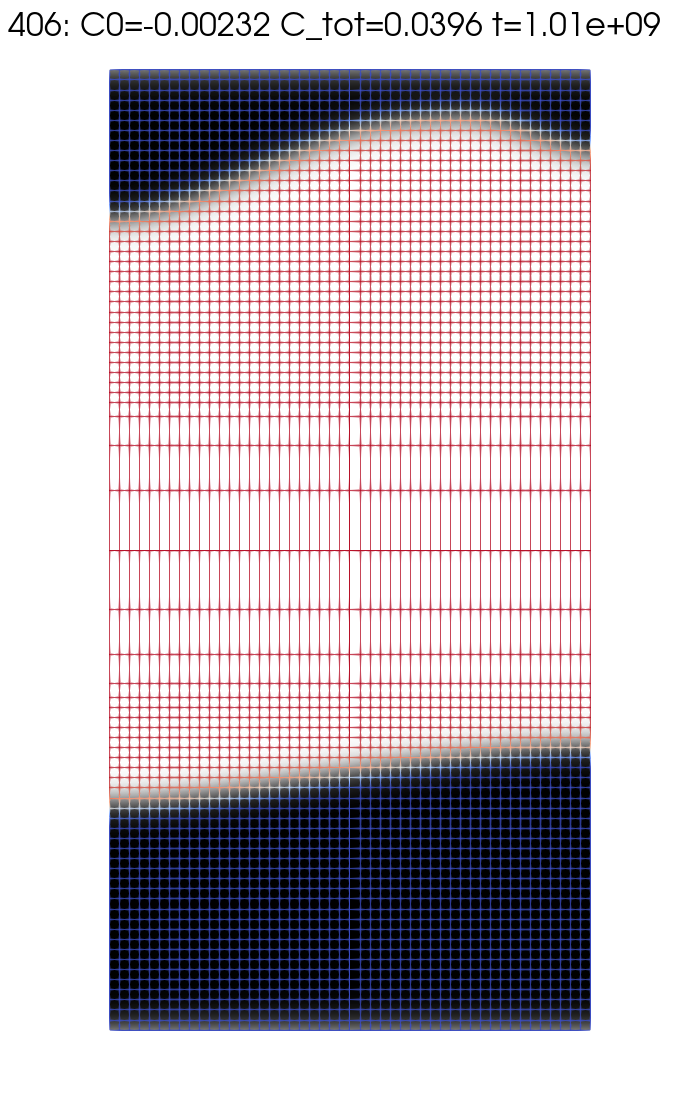}}
               
  \\
  \rotatebox[origin=c]{90}{deformed}
  &
  \raisebox{-0.5\height}{
  \includegraphics[scale=0.19,trim=2.8cm 5.0cm 5.2cm 3.5cm,clip]
                  {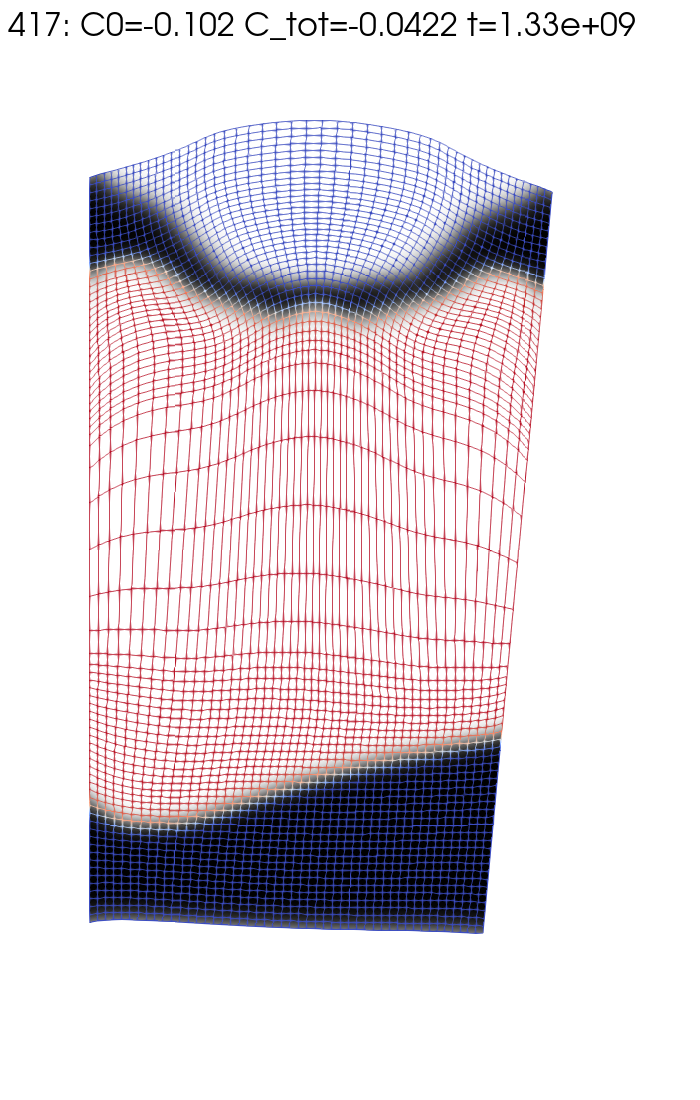}}
  &
  \raisebox{-0.5\height}{
  \includegraphics[scale=0.19,trim=2.8cm 5.0cm 5.2cm 3.5cm,clip]
                  {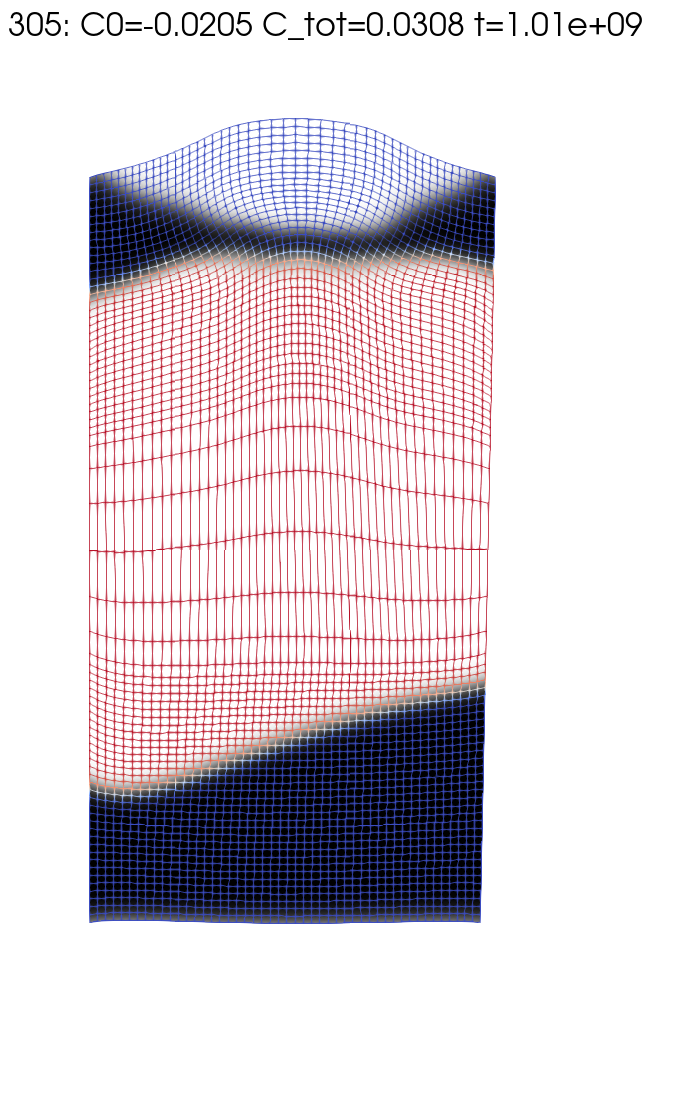}}
  &
  \raisebox{-0.5\height}{
  \includegraphics[scale=0.19,trim=2.8cm 5.0cm 5.2cm 3.5cm,clip]
                  {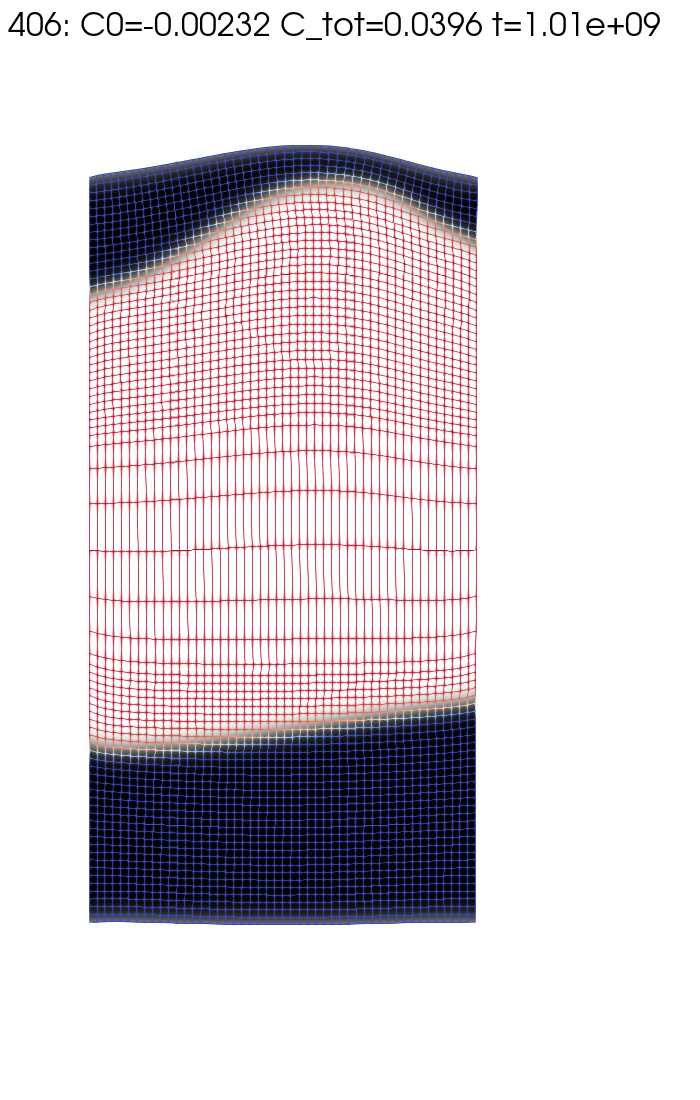}}
\end{tabular}
\begin{tabular}{c|c|c}
  $k_\mathrm{sp}$ & $C_0$ & \parbox[c]{22mm}{\centering spring compression}\\[10pt]
  \hline
  & & \\[-7pt]
  $\dfrac{EH}{2000}$ & -0.1016 & \SI{6.10}{\milli\meter}\\[14pt]
  $\dfrac{EH}{400}$  & -0.0205 & \SI{1.23}{\milli\meter}\\[14pt]
  $\dfrac{EH}{80}$   & -0.0023 & \SI{0.14}{\milli\meter}\\[14pt]
\end{tabular}
\caption{Effect of spring stiffness $k_\mathrm{sp}$ on the optimized actuator design.}
\label{fig:effect_k_sp}
\end{figure}

\begin{figure}[!h]
  \centering
  \raisebox{-0.5\height}{
  \begin{overpic}[scale=0.147,trim=36mm 0mm 38mm 21mm,clip]
                 {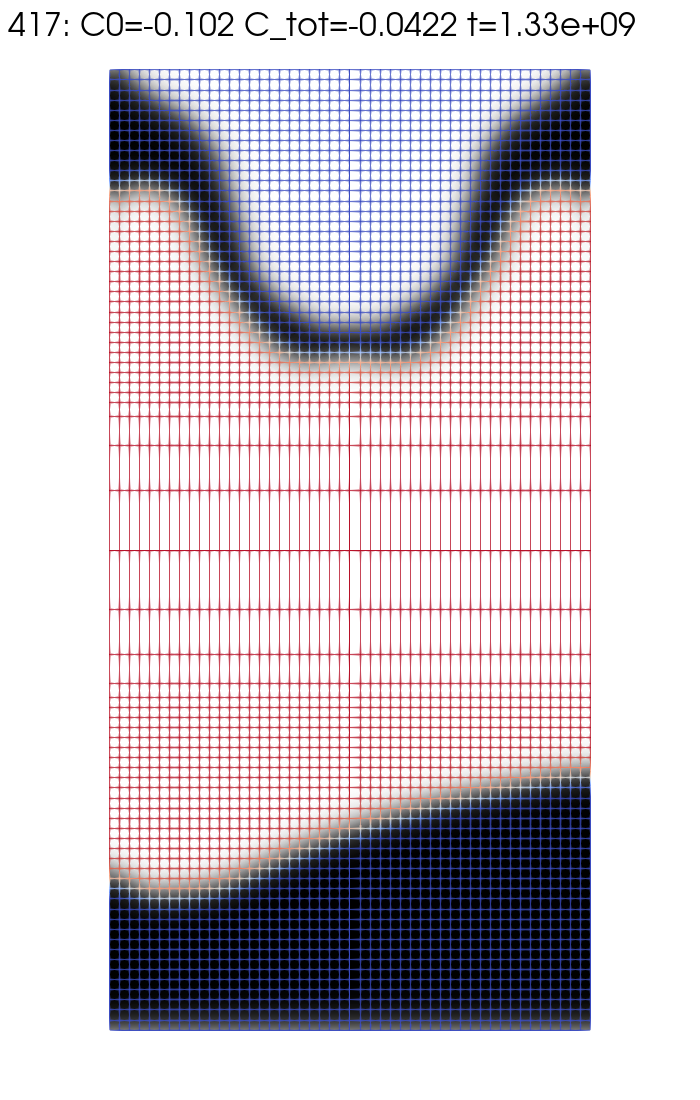}
  \put(5,0){$L_\mathrm{arm}\!=\!4 L$}
  \end{overpic}
  ~~
  \begin{overpic}[scale=0.147,trim=36mm 0mm 38mm 21mm,clip]
                 {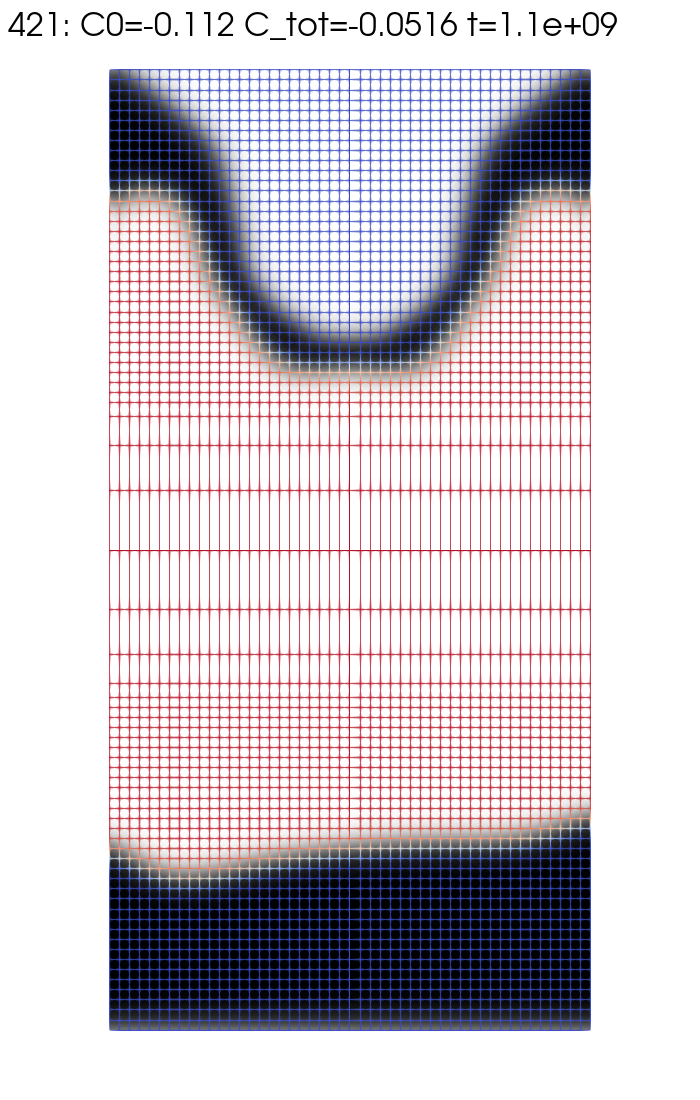}
  \put(4,0){$L_\mathrm{arm}\!=\!16 L$}
  \end{overpic}
  ~~
  \begin{overpic}[scale=0.147,trim=36mm 0mm 38mm 21mm,clip]
                 {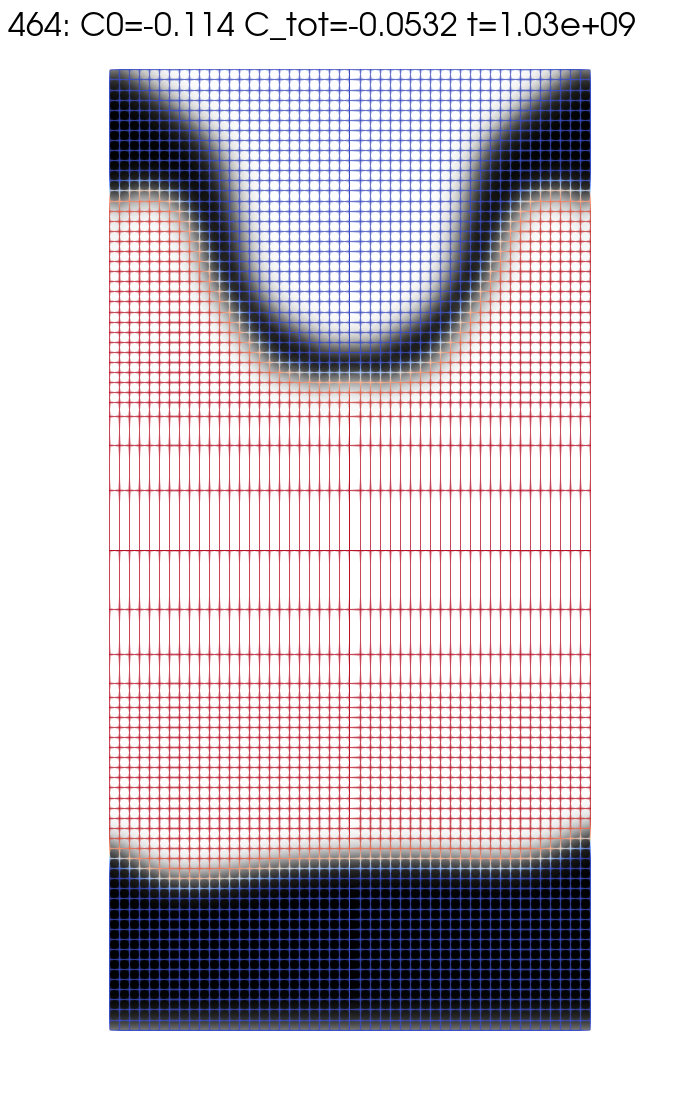}
  \put(4,0){$L_\mathrm{arm}\!=\!32 L$}
  \end{overpic}
  }
  ~
  \begin{tabular}{c|c|c|c}
    $L_\mathrm{arm}$ & $k_\mathrm{sp}$ & $C_0$
      & \parbox[c]{22mm}{\centering spring compression}\\[10pt]
    \hline
    & & \\[-7pt]
    $4L$  & $\dfrac{EH}{2000}$ &-0.1016 & \SI{6.10}{\milli\meter}\\[14pt]
    $16L$ & $\dfrac{EH}{32000}$ & -0.1112 & \SI{26.81}{\milli\meter}\\[14pt]
    $32L$ & $\dfrac{EH}{128000}$ & -0.1139 & \SI{54.69}{\milli\meter}\\[14pt]
  \end{tabular}
\caption{Effect of the rigid arm length $L_\mathrm{arm}$ on the optimised actuator design.}
\label{fig:effect_L_arm}
\end{figure}

The next study, presented in Figure \ref{fig:effect_k_sp}, addresses the impact of the spring stiffness $k_\mathrm{sp}$ on the optimized actuator design.
The design shown in the first column, in both undeformed and deformed configurations, is the representative case with $p_\mathrm{in}\!=\!0.02 E$, $\Psi_\mathrm{lim}\!=\!\tfrac{1}{2}\,0.17^2E$ and $c_A\!=\!0.02$, discussed extensively above.
The second and third columns show two new optimized designs, obtained for spring stiffness values increased by a factor of 5 and 25, respectively, relative to the reference case.
Increasing the spring stiffness leads to significant changes in the actuator design, showcasing again that the finite deformation capable framework provides optimized designs accounting for the specific compliance of the actuator's environment.
Unfortunately, in the case with the largest spring stiffness $k_\mathrm{sp}\!=\!EH/80$, the optimization procedure has switched to a different local minimum, that performs notably worse than the previous two cases.
Nevertheless, a cross-check between the remaining two cases, demonstrates that the design optimized for $k_\mathrm{sp}\!=\!EH/400$ indeed outperforms the original design optimized for $k_\mathrm{sp}\!=\!EH/2000$, when both designs are used in a post-simulation against the spring with stiffness $k_\mathrm{sp}\!=\!EH/400$.

Finally, Figure~\ref{fig:effect_L_arm} presents a study regarding the effect of the rigid arm length $L_\mathrm{arm}$ on the actuator design.
In this study, the arm length $L_\mathrm{arm}$ and the spring stiffness $k_\mathrm{sp}$ are changed simultaneously, so that the output bending stiffness, proportional to $L_\mathrm{arm}^2 k_\mathrm{sp}$, is kept constant.
This is in order to obtain different shear-to-bending load ratios, but without significantly affecting the bending load.
As the length $L_\mathrm{arm}$ increases, the obtained designs converge towards the case of a purely bending-dominated structure.

\section{Concluding remarks}
This work successfully established a new nonlinear density-based topology optimization framework for the computational design of pneumatically driven soft actuators subjected to large deformations.
The theory of porohyperelasticity was leveraged in order to transfer the actuation pressure to all surfaces of the structure that are connected to the pressurization source.
The proposed formulation fully addresses the case of design-dependent follower loads and is, to the best of the authors' knowledge, the first to incorporate the theory of porohyperelasticity within a framework of topology optimization.
To avoid pressure leakage through solid walls, the existing method of drainage in the solid was refined and incorporated in the proposed framework.
Additionally, a third-medium void regularization scheme, known from the third medium contact literature, turned out to be necessary in order to effectively prevent extreme mesh distortion around pressurized void regions.

A simple 2D benchmark problem was used to demonstrate the capabilities of the proposed framework.
The main objective of the optimization, was to maximize the compression of a spring by a rigid arm attached to the benchmark actuator.
However, in order to achieve a well-posed problem, a term penalizing the total surface area of the design, was essential to include in the overall objective.
Moreover, imposing an upper strain energy limit proved to be necessary in order to avoid excessively thin structure members.
A basic but systematic dimensional analysis allowed to express the overall objective function in dimensionless form, so that any scaling factors provided in this work are not problem specific.

Parametric studies were performed with regard to the actuation pressure, the strain energy density limit, the surface area penalization, the rigid arm length, and the spring stiffness.
For the 2D benchmark problem, the proposed topology optimization framework resulted in rather simple designs that could have also been obtained with shape optimization.
Nevertheless, the obtained load-dependent designs for a wide range of parameters, demonstrate the potential of the method, especially with the perspective of its future applications to 3D problems.
Such 3D applications of great interest are e.g. those concerning pneumatic actuators that interact with biological tissues, such as devices for laparoscopic surgery.

\section*{Acknowledgments}
This work was supported by a research grant (VIL50407) from VILLUM FONDEN.

\addcontentsline{toc}{section}{References}
\bibliographystyle{elsarticle-num}
\bibliography{references}

\appendix
\section*{Appendices}
\subsection*{A. Reference solution for the two-chamber structure}
\label{appendixA}
\renewcommand\thefigure{A.\arabic{figure}}
\setcounter{figure}{0}
\added{
Figure~\ref{fig:reference_solution} shows a comparison between the deformed porohyperelastic model with drainage reported in Figure~\ref{fig:leak_demo}c, and a reference solution obtained with a body fitted mesh.
The geometry of the body fitted mesh was extracted at the isoline $\chi\!=\!1$, which is equivalent to $\rho\!=\!0.731$, in order to account for the RAMP interpolation of the material stiffness.
The superimposed magenta contour in the Figure~\ref{fig:reference_solution} corresponds to the surface of the deformed body fitted mesh and is captured rather accurately by the fictitious domain porohyperelastic solution.
The source code for both cases compared in Figure~\ref{fig:reference_solution} is available at \url{https://data.mendeley.com/datasets/58cstgmrg9/1}.
}

\begin{figure}[!h]
\centering
\includegraphics[width=0.5\textwidth]{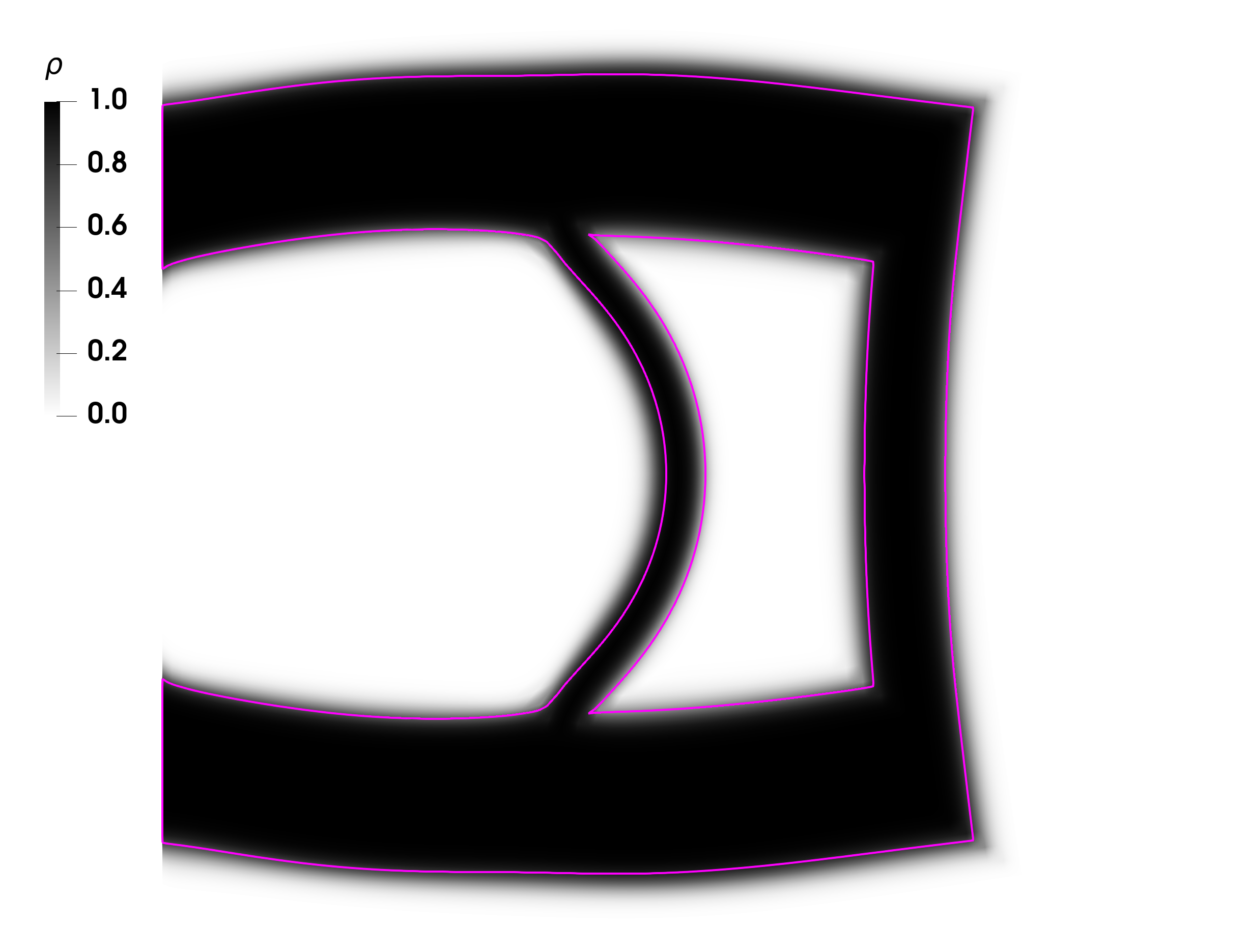}
\caption{\added{Comparison between porohyperelastic solution in the fictitious domain and reference solution with follower pressure load applied to body fitted mesh.}}
\label{fig:reference_solution}
\end{figure}

\subsection*{B. Remaining terms for the adjoint and optimality system}
\label{appendixB}
\setcounter{equation}{0}
\renewcommand\theequation{B.\arabic{equation}}

\added{Variations} with respect to the design variable $\chi$ (in addition to Eq.~\eqref{eq:design_damping_discr}):
\begin{equation}
C_{A,\chi}[\delta \chi]\!=\!
  \int_\Omega
  \dfrac{c_A}{|\Omega|^{1/2}} \dfrac{8}{L_i} \left(1 - 2\rho(\chi)\right)\rho'(\chi)\delta \chi
  ~d\Omega
\end{equation}
\begin{equation}
C_{i, \chi}[\delta \chi]\!=\!
  \int_\Omega
  c_i \left< \!\Norm{\nabla \chi}\!
             - \dfrac{8}{L_i} \right>^{\!5}
  \dfrac{\nabla \chi}
         {\Norm{\nabla \chi}} \cdot \nabla \delta \chi
 ~d\Omega
\end{equation}
\begin{align}
C_{\Psi,\chi}[\delta \chi]\!=\!
  \mathop{\mathlarger{\mathlarger{\int}}}_{\Omega}
  \Biggl<\dfrac{\Psi\!\left(\chi\!+\!\dfrac{8}{L_i}\ell_\Psi, \uu\right)}
              {\Psi_\mathrm{lim}}
        -1\Biggr>^{\!5}
  \Bigg(
  &K'\!\left(\!\chi\!+\!\dfrac{8}{L_i} \ell_\Psi\!\right)
       \dfrac{(\ln \detF)^2}{2}\nonumber \\[-10pt]
  &+G'\!\left(\!\chi\!+\!\dfrac{8}{L_i} \ell_\Psi\!\right)
  \dfrac{\detF^{-2/3} \Norm{\F}^2-3}{2} 
  \Bigg)\delta \chi~d\Omega
\end{align}
\begin{align}
\RR_{,\chi}\left[\Lambda_p,\LLambda_\uu\right]\left[\delta \chi \right]\!=\!
  \mathop{\mathlarger{\int}}_\Omega
  \bigg(&
    k'(\chi) \detF \left(\left(\F^T \F\right)^{-1} \nabla p \right)
    \cdot \nabla\Lambda_{p} 
    + Q_s \rho'(\chi)
    (p-p_\mathrm{out})\Lambda_p\nonumber\\[-10pt]
    &+
    \Big(K'(\chi) \ln\detF ~\II
         +G'(\chi) \detF^{-2/3} \Dev{\F\F^T}
    \Big)
    \!:\!
    \left(\nabla\LLambda_\uu \F^{-1}\right)
  \bigg) \delta \chi~d\Omega
\end{align}

\noindent
\added{Variations} with respect to the adjoint pressure variable $\Lambda_p$:
\begin{equation}
C_{p,p}[\dLambda_p]\!=\dfrac{1}{\Det{\Gamma_\mathrm{ext}}}
                      \!\int_{\Gamma_\mathrm{ext}}
                      \dfrac{c_p}{p_\mathrm{in}^2}\,p\,\dLambda_p
                      ~d\Gamma
\end{equation}
\begin{align}
\RR_{,p}\left[\Lambda_p,\LLambda_\uu\right]\left[\dLambda_p\right]\!=\!
  \mathop{\mathlarger{\int}}_\Omega
  &~k(\chi) \detF \left( \F^{-1}\F^{-T} \nabla \dLambda_p\right)
    \!\cdot\! \nabla\Lambda_p \nonumber\\[-10pt]
  &+\bigg(
      \Big(Q_{\mathrm{in}}(\XX) + Q_{\mathrm{out}}(\chi)\Big) \Lambda_p
      -\detF~\F^{-T} \!:\! \nabla\LLambda_\uu
    \bigg)\dLambda_{p}
  ~d\Omega
\end{align}

\noindent
\added{Variations} with respect to the adjoint displacement variable $\LLambda_\uu$:
\begin{align}
C_{\Psi,\uu}\left[\ddLambda_\uu\right]\!=\!\!
  \mathop{\mathlarger{\mathlarger{\int}}}_{\Omega}
  \!&
  \dfrac{c_\Psi}{\Psi_\mathrm{lim}}
  \Biggl<\dfrac{\Psi\!\left(\chi\!+\!\dfrac{8}{L_i}\ell_\Psi, \uu \right)}
               {\Psi_\mathrm{lim}}
         -1
  \Biggr>^{\!5}\nonumber\\
  +&\bigg(\!
    K\!\left(\!\chi\!+\!\dfrac{8}{L_i}\ell_\Psi\!\right) \ln\detF~\II
    +
    G\!\left(\!\chi\!+\!\dfrac{8}{L_i}\ell_\Psi\!\right) \detF^{-2/3} \Dev{\F\F^T}
  \!\bigg)
  \!:\! \left(\nabla\ddLambda_\uu \F^{-1}\right)
  ~d\Omega
\end{align}
\begin{align}
\RR_{,\uu}\left[\Lambda_p,\LLambda_\uu,\LLambda_\qq\right]
          \left[\ddLambda_\uu\right]\!=\!
  \mathop{\mathlarger{\int}}_\Omega
  k(\chi)
  \left(
  \left(\Big(\detF\,\F^{-1}\F^{-T}\Big)_{\!,\nabla\uu}
        \!:\!\nabla\ddLambda_\uu\right) \nabla p
  \right)
  \!\cdot\!\nabla\Lambda_p
  +\LLambda_\qq\!\cdot\!\ddLambda_\uu
  &\nonumber\\
  +\bigg(
    \Big(
      \big(K(\chi)\ln\detF\!-\!\detF p\big) \II
      +G(\chi) \detF^{-2/3} \Dev{\F\F^T}
    \Big)_{\!,\nabla\uu}
   \!:\!\nabla\ddLambda_\uu
   \bigg)
   \!:\! \left(\nabla\LLambda_\uu \F^{-1}\right)
  &\nonumber\\
  +c_r ~ \mathbb{H}\ddLambda_\uu \svdots \mathbb{H}\LLambda_\uu
  &~d\Omega
\end{align}

\noindent
\added{Variations} with respect to the adjoint displacement variable $\LLambda_\qq$:
\begin{equation}
\RR_{,\qq}\left[\LLambda_\uu,\LLambda_\TT\right]
          \left[\ddLambda_\qq \right]\!=\!
  \mathop{\mathlarger{\int}}_\Gamma
  \biggl(
  \LLambda_\uu
  -\begin{pmatrix}\Lambda_{T_x}\\\Lambda_{T_y}\end{pmatrix}
  +X_2\begin{pmatrix}\cos(T_\theta)\\\sin(T_\theta)\end{pmatrix}\Lambda_{T_\theta}
  \biggr)
  \cdot\ddLambda_\qq
  ~d\Gamma
\end{equation}

\noindent
\added{Variations} with respect to the adjoint rigid arm motion variables $\LLambda_\TT$:
\begin{equation}
C_{0,\TT}\left[\ddLambda_\TT\right]\!=\!
  \left(\dfrac{1000}{L_\mathrm{arm}^2} \left< T_y\right>
        +\dfrac{1}{L_\mathrm{arm}}
  \right)\dLambda_{T_y}
  +\cos(T_\theta) ~ \dLambda_{T_\theta}
\end{equation}

\begin{align}
\RR_{,\TT}\left[\LLambda_\qq,\LLambda_\TT\right]
          \left[\ddLambda_\TT\right]\!=
  &~~~~
  k_\mathrm{sp} 
  \left(
  \Lambda_{T_y}+L_\mathrm{arm}\cos(T_\theta)\Lambda_{T_\theta}
  \right)
  \left(
  \dLambda_{T_y}+L_\mathrm{arm}\cos(T_\theta)\dLambda_{T_\theta}
  \right)\nonumber\\
  &-
  k_\mathrm{sp} 
  \left(
  T_y+L_\mathrm{arm}\sin(T_\theta)
  \right)
  L_\mathrm{arm}\sin(T_\theta)\,\Lambda_{T_\theta}\,\dLambda_{T_\theta}
  \nonumber\\
  &+
  \mathop{\mathlarger{\mathlarger{\int}}}_{\Gamma}
  \left(
  -\begin{pmatrix}\dLambda_{T_x}\\\dLambda_{T_y}\end{pmatrix}
  +X_2\begin{pmatrix}\cos(T_\theta) \\\sin(T_\theta)\end{pmatrix} 
  \delta \Lambda_{T_\theta}
  \right)\cdot\LLambda_\qq
  \nonumber\\[-8pt]
  &~~~~~~~~~~~~~~~~~~~~~~~~+
  X_2\begin{pmatrix}-\sin(T_\theta)\\\cos(T_\theta)\end{pmatrix}
  \cdot \qq\,
  \Lambda_{T_\theta}\,\dLambda_{T_\theta}
  ~d\Gamma
\end{align}

\end{document}